\documentclass[a4paper]{article}

\usepackage[T1]{fontenc}
\usepackage[a4paper,margin=2.5cm]{geometry}
\usepackage{calc}
\usepackage{parskip}
\usepackage{booktabs}
\usepackage{xcolor}
\usepackage{graphicx}
\usepackage{bm}

\usepackage[inline]{enumitem}
\setlist[itemize,1]{wide=0pt,leftmargin=*}
\setlist[enumerate,1]{wide=0pt,leftmargin=*,label={(\alph*)}}

\usepackage{caption}
\usepackage[skip=0.5ex]{subcaption}

\usepackage[colorlinks,allcolors=darkgray,breaklinks=true]{hyperref}
\usepackage[acronym,nogroupskip]{glossaries}
\usepackage[capitalise,noabbrev]{cleveref}

\usepackage{needspace}

\usepackage[sort&compress,super]{natbib}
\usepackage[resetlabels]{multibib}
\newcites{si}{References}
\newcommand\biblabeldigits[1]{%
\ifnum#1<10 #1.\hphantom{0}\else #1.\fi}
\makeatletter
\renewcommand\@biblabel[1]{\biblabeldigits{#1}}
\makeatother

\makeglossaries
\setacronymstyle{long-short}

\newacronym{dft}{DFT}{density-functional theory} 
\newacronym{hd}{HD}{Hellinger distance}
\newacronym{llpt}{LLPT}{liquid-liquid phase transition} 
\newacronym{mae}{MAE}{mean absolute error}
\newacronym{maxE}{MaxE}{maximum absolute error}
\newacronym{md}{MD}{molecular dynamics} 
\newacronym{mlp}{MLP}{machine-learning interatomic potential} 
\newacronym{pbe}{PBE}{Perdew-Burke-Ernzerhof} 
\newacronym{rdf}{RDF}{radial distribution function}
\newacronym{rmse}{RMSE}{root-mean-squared error}
\newacronym{wdh}{WDH}{warm dense hydrogen} 

\newcommand{\dataset}{\texttt{h-llpt-24}} 

\begin{document}

\author{Thomas Bischoff$^1$, Bastian J{\"a}ckl$^1$, Matthias Rupp$^2$}
\date{%
	$^1$~Department of Computer and Information Science, University of Konstanz, Konstanz, Germany
	$^2$~Luxembourg Institute of Science and Technology (LIST), Esch-sur-Alzette, Luxembourg\\[3ex]
}
\title{Hydrogen under Pressure as a Benchmark\\ for Machine-Learning Interatomic Potentials}
\maketitle


\begin{abstract}\noindent
Machine-learning interatomic potentials (MLPs) are fast, data-driven surrogate models of atomistic systems' potential energy surfaces that can accelerate ab-initio molecular dynamics (MD) simulations by several orders of magnitude.
The performance of MLPs is commonly measured as the prediction error in energies and forces on data not used in their training.
While low prediction errors on a test set are necessary, they do not guarantee good performance in MD simulations.
The latter requires physically motivated performance measures obtained from running accelerated simulations.
However, the adoption of such measures has been limited by the effort and domain knowledge required to calculate and interpret them.

To overcome this limitation, we present a benchmark that automatically quantifies the performance of MLPs in MD simulations of a liquid-liquid phase transition in hydrogen under pressure, a challenging benchmark system.
The benchmark's \dataset{} dataset provides reference geometries, energies, forces, and stresses from density functional theory MD simulations at different temperatures and mass densities.
The benchmark's Python code automatically runs MLP-accelerated MD simulations and calculates, quantitatively compares and visualizes pressures, stable molecular fractions, diffusion coefficients, and radial distribution functions.
Employing this benchmark, we show that several state-of-the-art MLPs fail to reproduce the liquid-liquid phase transition.
\end{abstract}


\clearpage
\tableofcontents

\vfill
\printglossary[type=\acronymtype]
\clearpage


\section{Introduction}


\Gls{md} simulations of atomistic systems are a cornerstone of computational physics, chemistry, and materials science, but are limited by either accuracy or computational cost.
The trade-off between accuracy and cost is determined by the method used to compute the system's potential energy, whose negative gradient are the forces that propagate atoms in the simulation.

Typical choices include classical force fields, characterized by fixed functional forms, and quantum-mechanical (\textit{ab initio}) approaches.
Force fields are computationally efficient but limited in accuracy, transferability, and by parametrization effort.
Ab initio approaches are accurate and transferable but have high computational costs.
Consequently, \gls{md} simulations are limited in either the phenomena (e.g., bond breaking and formation) or the system sizes and time scales they can model.


\Glspl{mlp} \cite{uctm2021q} are data-driven approximations of ab-initio potential energy surfaces that exploit correlations between the atom positions and the potential energy of a system.
Essentially, a flexible functional form such as a neural network is parametrised (``trained'') on a dataset of ab initio calculations.
The parametrised model then calculates the forces during a simulation.
\glspl{mlp} are typically several orders of magnitude faster to evaluate than the ab initio reference method.~\cite{zcwo2020q,xrh2023q} 
This greatly accelerates \gls{md} simulations, enabling running more simulations, simulating larger systems, and increasing simulation times, resulting in improved configurational sampling, property estimates, and the ability to model otherwise inaccessible phenomena.


\gls{mlp}-accelerated \gls{md} simulations should reproduce the results of corresponding ab-initio \gls{md} simulations. 
More specifically, the discrepancies in derived macroscopic quantities, such as radial distribution functions and diffusion coefficients, should be as small as possible. 
The \gls{mlp}-\gls{md} should also reproduce any monotony, limits, and asymptotic behaviour of such quantities.
This is necessary to consistently describe physical phenomena such as proton exchange and phase transitions. 
Validation and performance assessment of \glspl{mlp} for \gls{md} should therefore base on \gls{mlp}-accelerated \gls{md} simulations. 

In practice, \gls{mlp} performance is commonly evaluated as the average error in force predictions on a test dataset (data not used during training but from the same distribution as the training data), without involving any \gls{md} simulations.
However, force accuracy on a test set is necessary but insufficient for successful \gls{md} simulations. \cite{lqb2021q,sggm2022q,fwgj2023q,mgd2023q}
Consequently, \glspl{mlp} with reported state-of-the-art test-set accuracy can fail catastrophically in actual \gls{md} simulations.

The inadequacy of test-set force errors as the sole performance indicator for \gls{md} simulations is increasingly recognized in the community.
Despite this, the uptake of assessment methods based on \gls{mlp}-\gls{md} simulations is slow, 
possibly because these require more human and computational effort as well as expert domain knowledge of the simulated system and the derived properties.


In this work, we introduce an \gls{md}-based benchmark of \gls{mlp} performance that is straightforward to use and does not require expert domain knowledge.
For this we employ simulation of a liquid-liquid phase transition in hydrogen under pressure as a challenging application.
The benchmark provides the \dataset{} dataset with geometries, energies, forces, and stresses from reference MD simulations at the density functional level of theory for different temperatures and mass densities.
This data serves as training, validation, and testing data for \gls{mlp} parametrisation.

Once parametrised, the benchmark's Python code automatically runs \gls{mlp}-accelerated \gls{md} simulations and calculates and visualises pressures, stable molecular fractions, diffusion coefficients, and radial distribution functions as functions of cell density.
These MD-derived properties characterise the liquid-liquid phase transition and are quantitatively compared with those from the reference simulations.
Outputs include automatically generated figures and tables.
We employ the benchmark to show that several state-of-the-art MLPs fail to reproduce the liquid-liquid phase transition.

The benchmark
(a) focuses on learning a single potential energy surface, as opposed to generalising across multiple atomistic systems simultaneously;~\cite{rtml2012q}
(b) focuses on learning the map from local atomic geometry to energy; in particular, it does not measure an \gls{mlp}'s ability to model different chemical elements;
(c) provides physical system-specific derived properties as performance metrics, beyond generic performance metrics;
(d) resolves differences between \glspl{mlp} well due to the high difficulty of the learning task.


\section{Background}\label{sec:background}


\textbf{{Hydrogen under pressure}}.
The term warm dense matter refers to atomistic systems in a regime of temperature and density between the condensed phase and hot plasma.
\Gls{wdh} exhibits complex behaviour, including multiple solid and liquid phases and the transitions between them. \cite{bvsd2024}
These are relevant, among others, for modelling gas planets, \cite{hmr2020} high-pressure physics, hydrogen-based inertial confinement fusion, and the design of hydrogen-rich materials.

The phase diagram of \gls{wdh} has been intensely debated in the literature for almost a century. 
There are multiple reasons for disagreements:
Experimental synthesis and characterisation of \gls{wdh} are challenging due to the high temperatures and pressures involved.~\cite{f2018b}
Challenges of computational approaches include \gls{wdh}['s] many-body quantum nature, high computational costs, and limits of the approximations used, for example, strong dependency on the exchange-correlation functional \cite{gwma2019,bvsd2024}.

This work focuses on a \gls{llpt} between molecular and atomic liquid hydrogen, using \gls{dft} with the \gls{pbe}~\cite{pbe1996} functional as the ground truth.
In this, our aim is not further insight into the \gls{llpt} itself.
Instead, we use modelling of the \gls{llpt} at the \gls{dft}/\gls{pbe} level of theory as a hard and therefore discriminating benchmark for \glspl{mlp}.

While the nature of the \gls{llpt} in \gls{wdh} remains under debate, \cite{cmpc2020q,khht2021,cmpc2021q} computational studies \cite{gwma2019, s2003b, mpsc2010, lhr2010, vtmb2007, dpc2006, mhs2018, khht2021,bvsd2024} indicate a first-order transition.
This is also the case for the \gls{dft}/\gls{pbe} method used in this work.
Such a transition is characterised by 
a sudden discontinuous change in observables, including pressure, stable molecular fraction, diffusion coefficient, and radial distribution function as a function of density.
These observables can be estimated from \gls{md} simulations of \gls{wdh}.


\textbf{Molecular dynamics simulations}\cite{t2023}
enable observing the evolution of an atomistic system over time and, based on this, calculating derived properties of that system.
For every time step of the simulation, each atom's position is updated based on its mass, velocity, and the force acting on it.
As the force is the negative gradient of the potential energy with respect to atom position, this requires a potential energy function, or potential, that calculates the potential energy of a system given the position and chemical species of its atoms.

The computational cost of an \gls{md} simulation scales with the number of atoms and time steps. 
Extended systems are often approximated by simulating a smaller part, the unit cell, with periodic boundary conditions.
While this saves computational cost, the unit cell cannot be too small, as this would lead to finite-size effects.

For long simulations of large systems, it is therefore important that the potential is efficient to evaluate.
Traditional empirical potentials have fixed functional forms that allow for efficient evaluation but limit accuracy.
Ab-initio potentials solve the quantum-mechanical equations governing the system, leading to accurate forces but high computational costs.
Data-driven potentials offer a compromise, see \cref{sec:background}.

Some thermodynamic variables are held constant during an \gls{md} simulation to imitate real-world conditions. 
In the canonical ensemble ($NVT$), the number of atoms~$N$, the volume~$V$ of the unit cell and the temperature~$T$ are held constant.
The latter is controlled via a thermostat that adjusts the kinetic energy of the atoms.


\textbf{Machine-learning interatomic potentials}\cite{uctm2021q}
are computationally efficient, data-driven approximations of ab-initio potential-energy surfaces.
Their computational costs typically fall between those of traditional empirical potentials and those of ab-initio potentials.
Their approximation error determines how well they can reproduce the physics in an \gls{md} simulation and, thus, the accuracy of the derived observables.
Many types of \glspl{mlp} have been developed.
They differ, among other aspects, in the interactions they model, how they featurise these, and the regression algorithm used.

Global \glspl{mlp} consider all interactions between atoms; for this reason, they usually scale superlinearly with system size.
To scale linearly, most \glspl{mlp} consider only finite subsets of interactions.
Typically, such local \glspl{mlp} for each atom consider only interactions with neighbouring atoms within a given distance.
Semilocal \glspl{mlp} build up longer-range correlations iteratively from local ones, often through message-passing mechanisms.
Interactions between atoms are numerically represented (``featurised'') for regression.
Corresponding basis functions are either explicitly given or learned from the training data.
Popular regression algorithms include linear regression, kernel regression, and neural networks.


\section{Related work}\label{sec:related}


Benchmarks for \glspl{mlp} typically report prediction errors for energies and forces on test sets, that is, data sampled from the same distribution as the training data but not used for training the \gls{mlp}.
As prediction errors are straightforward to compute, these benchmarks provide reference data but not program code to calculate or visualise performance metrics.
Popular datasets for benchmarking \glspl{mlp} include QM7 \cite{rtml2012q,hmdt2021q}, QM9 \cite{rdrl2014q}, MD17 \cite{ctsm2017q,cl2020q}, and OC20 \cite{cdzu2021q,cdzu2021bq}.
See Vita et~al.\cite{vfmt2023q} for a recent collection of such datasets.

The need for validation of \glspl{mlp} beyond test-set prediction errors has been recognised in the community for several years.
Only recently, however, have systematic efforts been made to address this problem:

Stocker et~al.{}\cite{sggm2022q} investigate the robustness of the GemNet\cite{gbg2021q} \gls{mlp}, trained on the QM7-x\cite{hmdt2021q} dataset of small organic molecules, in long dynamics simulations (280\,ns total simulation time).
They find that ``low test set errors are not sufficient for obtaining stable dynamics and that severe pathologies sometimes only become apparent after hundreds of ps of dynamics.''
Their study focuses on stability and does not investigate properties derived from simulations.

Fu et~al.~\cite{fwgj2023q} benchmark nine \glspl{mlp}
on water, small organic molecules, alanine dipeptide, and Li$_{6.75}$P$_3$S$_{11}$.
They find that ``force accuracy alone does not suffice for effective simulation.''
For alanine dipeptide, the only \gls{mlp} stable over five nanoseconds of simulation time failed to reproduce its free energy surface.
Their benchmark provides simulation protocols, an open-source codebase,\cite{fwgj2023code} and quantitative metrics for derived properties,
specifically interatomic distance distributions, radial distribution functions, diffusion coefficients, free energies, and simulation stability.
The benchmark in this work includes two of these generic metrics, radial distribution functions and diffusion coefficients, but also provides application-specific metrics related to the \gls{llpt} of hydrogen under pressure.

Fonseca et~al.\cite{fpt2023q} provide the cross-platform software package \texttt{FFAST} (Force Field Analysis Software and Tools) to benchmark, analyse, and visualise \gls{mlp} performance and limitations.
They demonstrate their software for the Nequip \cite{bmsk2022q} and MACE \cite{bkoc2022q} \glspl{mlp} applied to stachyose and docosahexaenoic acid.
Their approach is geared towards molecules, that is, multi-element systems where covalent bonding dominates and does not change much.
Consequently, many analysis features do not apply to our setting, such as element-resolved error distributions or identification of specific atoms with higher prediction errors.
Where applicable, we adapt features, such as error distribution plots.

Morrow et~al.\cite{mgd2023q} discuss best practices for validating \glspl{mlp} but do not provide specific benchmarks.
We follow their guidance where appropriate.
In particular, we evaluate \glspl{mlp} only with respect to the ground-truth ab-initio method that generated the training data,
as comparing with experimental results would evaluate the suitability of the reference method instead of the suitability of the \gls{mlp}.
They conclude that it would be ``desirable to create openly available `packaged' tests that can be run directly from an openly available and easily accessible code, say, a Python script or automated workflow,`` which this benchmark provides.

In active learning, \cite{s2012b} an \gls{mlp} is dynamically (``on the fly''\cite{capd2004q}) retrained during an accelerated \gls{md} simulation whenever it encounters a system configuration outside its domain of applicability.
Since reference calculations are only done when required for the \gls{md}, active learning is a more data-efficient alternative to static \glspl{mlp} trained on a fixed training set and to extending the training set via unbiased sampling from \gls{md} trajectories.
In this work, we do not employ active learning as it would 
(a) introduce additional requirements for \glspl{mlp}, such as predictive uncertainties; (b) introduce additional factors influencing \gls{mlp} performance, such as the active learning algorithm; and (c) substantially complicate benchmarking by tightly coupling training to simulations.


\enlargethispage*{2\baselineskip}

\section{Dataset}\label{sec:dataset}

The benchmark is based on the \dataset{} dataset introduced below.
This dataset contains 612 \gls{dft} reference \gls{md} simulations of 10\,k steps each.
Part of this data is provided for training, validating, and testing \glspl{mlp}.
All data (after equilibration) was used to calculate reference values for the \gls{md}-derived properties in Section~\ref{sec:properties}.

We constructed the \dataset{} dataset explicitly to benchmark \glspl{mlp}.
Its complexity does not come from interactions of multiple chemical elements but solely from the non-linearity of interactions between its particles.
In particular, the effects of partially bound electrons have to be detected indirectly from the hydrogen nuclei's positions.
As such, this setting primarily tests an \gls{mlp}'s ability to learn complex high-dimensional functions.

\begin{figure}[bthp]
    \begin{minipage}[b]{11.3cm}
        \includegraphics[width=\linewidth]{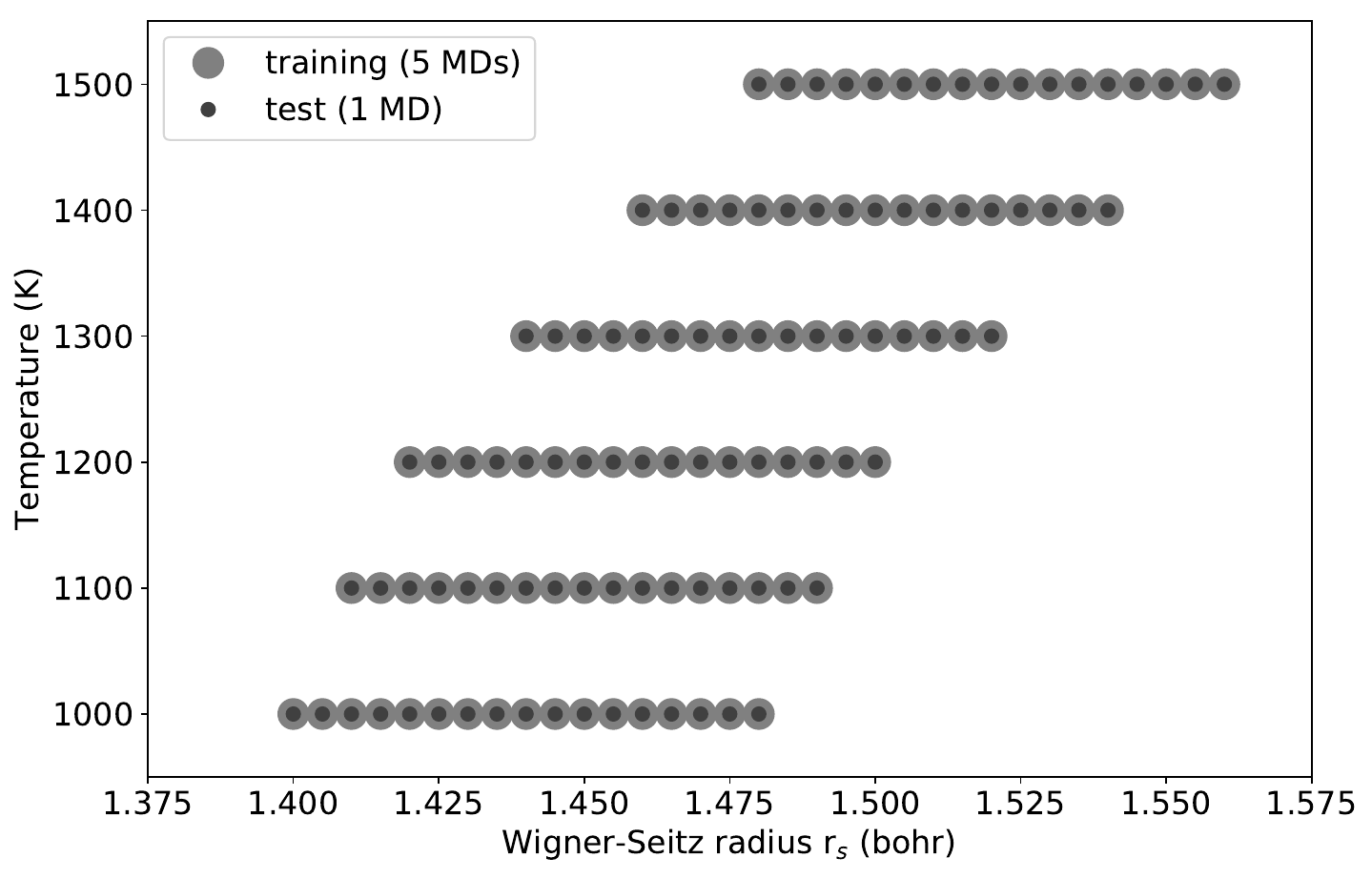}
    \end{minipage}%
    \hfill%
    \begin{minipage}[b]{\linewidth-11.3cm-2\tabcolsep}
        \caption{%
            \emph{Dataset coverage.}
            Each disk represents a combination of Wigner-Seitz radius and temperature for which six \gls{dft} \gls{md} simulations were performed, five for training of \glspl{mlp} and one for testing.
            \smallskip
        }
        \label{fig:dataset_overview}
    \end{minipage}
\end{figure}

To generate the \dataset{} dataset, we ran ab-initio \gls{md} simulations of 128 hydrogen atoms in a cubic unit cell under periodic boundary conditions at different temperatures and densities (\cref{fig:dataset_overview}, \cref{si:tab:rsindices}).
For each of the six temperatures~$T$, we chose 17 equally-spaced Wigner-Seitz radii~$r_s$, covering the regime of the \gls{llpt}.
We ran six \gls{md} simulations for each combination $(T,r_s)$, resulting in a total of $6 \cdot 17 \cdot 6 = 612$ \gls{md} simulations.
These simulations were performed in the $NVT$ ensemble, using \gls{dft} with the \gls{pbe}\cite{pbe1996} functional as the potential, and were carried out using the \texttt{Quantum ESPRESSO}\cite{gbuw2009,gawb2017} software.
To decorrelate configurations, we subsampled each simulation's trajectory, taking two configurations from early in the 0.5\,ps equilibration phase and 12 configurations at regularly-spaced intervals of 40\,fs after equilibration.

This resulted in 8\,568 configurations, which we split into a training set of 7\,140 and a test set of 1\,428 configurations. 
The training configurations were sampled from five of the six repetitions; the test configurations were sampled from the remaining trajectory.
Training and test data are thus sampled independently from the same distribution, satisfying the underlying assumption of machine-learning models that data are independent and identically distributed.

If a validation set is desired for \gls{mlp} training, e.g., for early stopping or hyperparameter optimisation, the training set can be split further into a (smaller) training set and a validation set following the same approach:
One of the five trajectories is used for validation, and the other four are used for training.
Similarly, the trajectory designations can also be used for five-fold cross-validation.

\begin{figure}[phbt]
    \includegraphics[width=0.32\linewidth]{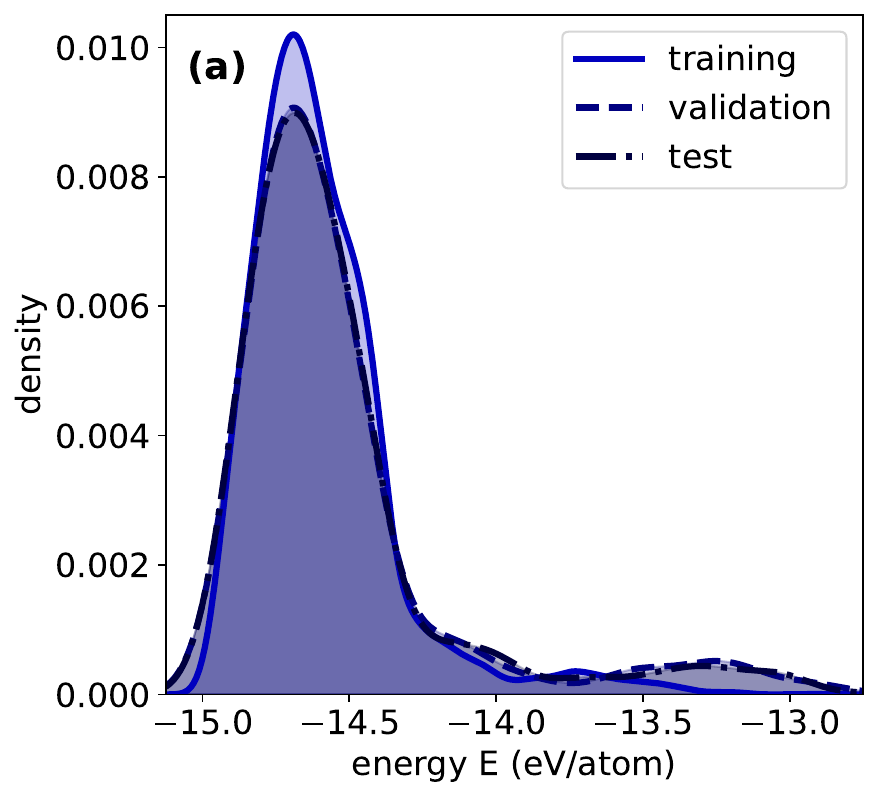}\hfill%
    \includegraphics[width=0.32\linewidth]{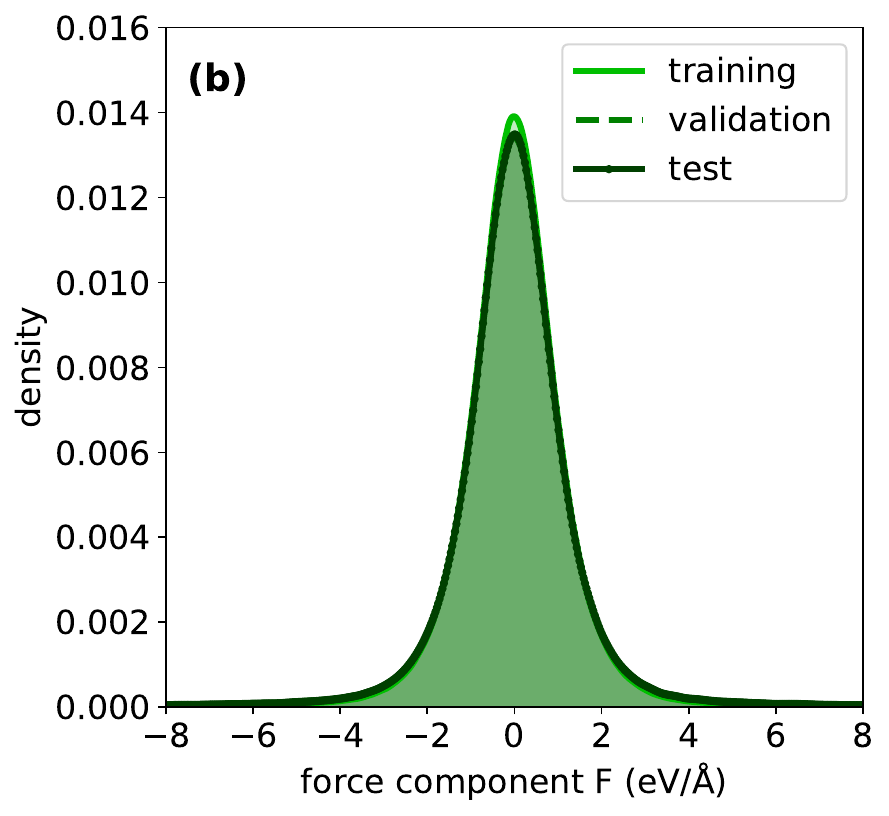}\hfill%
    \includegraphics[width=0.32\linewidth]{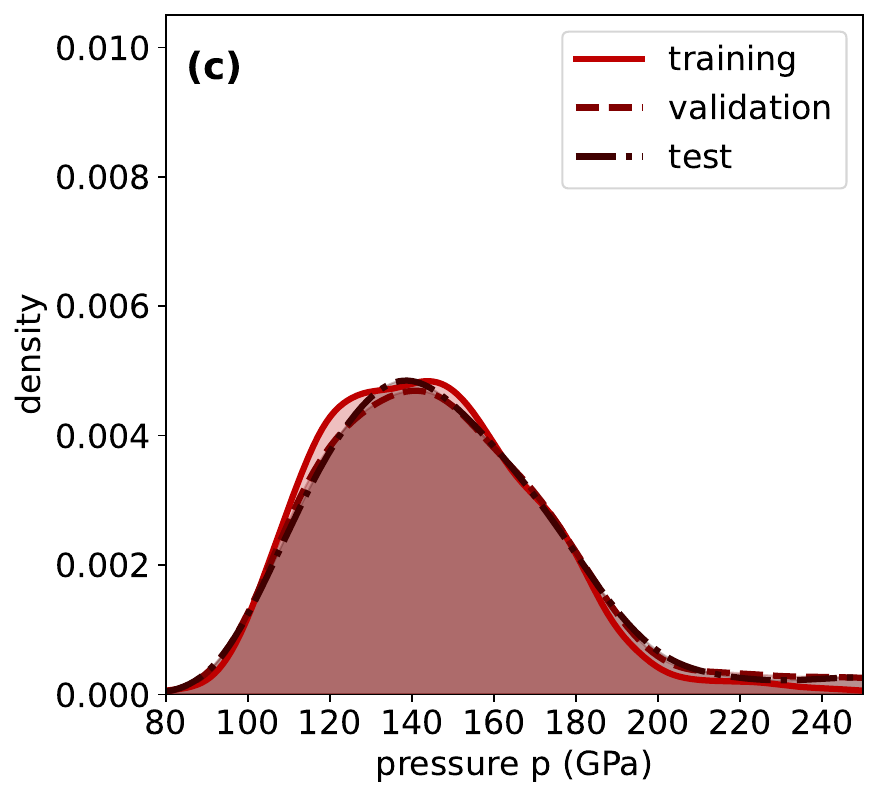}

    \caption{%
        \emph{Distribution of energies, forces, and pressures} in the \dataset{} dataset.
        Shown are smoothed histograms of energies (left), forces (middle), and pressures (right) in the training (solid), validation (dashed), and test (dotted) subsets.
    }
    \label{fig:dataset_distribution}
\end{figure}

For each configuration, the dataset provides lattice vectors (\AA), atom positions (\AA), total energy (eV), forces (eV/\AA), stress tensor (eV/\AA$^3$), pressure (GPa), Wigner-Seitz radius (bohr), temperature (K), and trajectory number (unitless).
Figure~\ref{fig:dataset_distribution} presents the distribution of energies, forces, and pressures in the training, validation (\texttt{md=4}), and test sets.
For further computational details, see \cref{si:sec:methods}.

\section{Performance measures} \label{sec:properties}

The benchmark uses four \gls{md}-derived properties to quantify the performance of \glspl{mlp} in characterising the \gls{llpt} of \gls{wdh}.
This requires running and analysing \gls{md} simulations.
To facilitate this, it provides Python scripts that automatically run multiple \gls{md} simulations, calculate the derived properties including uncertainties (error bars), and visualise them.
When comparing with the \gls{dft}-\gls{md} ground truth, these uncertainties are taken into account via the Hellinger distance.

\begin{figure}[ptbh]
    \includegraphics[width=0.44\linewidth,trim={5pt 5pt 5pt 5pt},clip]{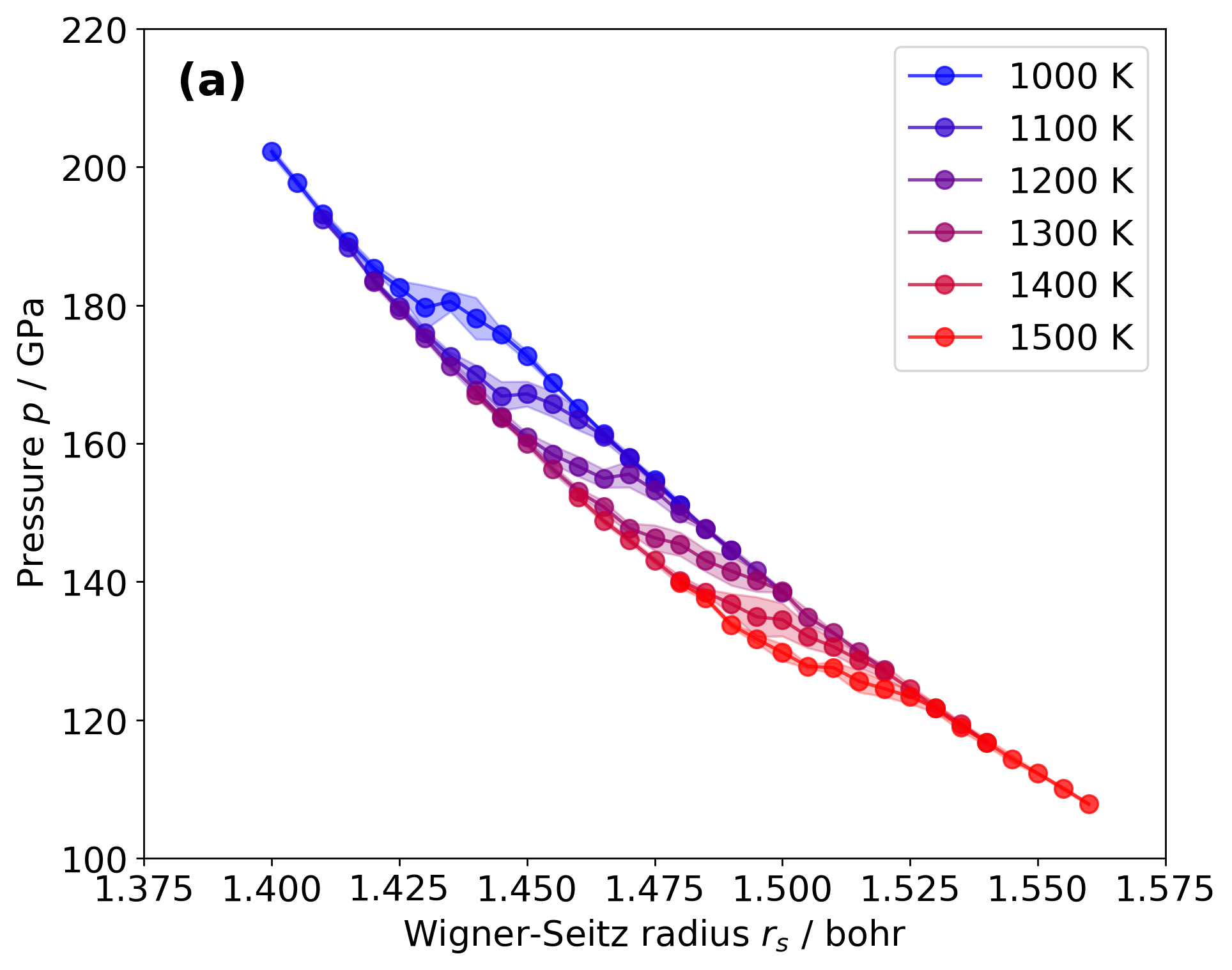}%
    \quad\!\!%
    \includegraphics[width=0.44\linewidth,trim={5pt 5pt 5pt 5pt},clip]{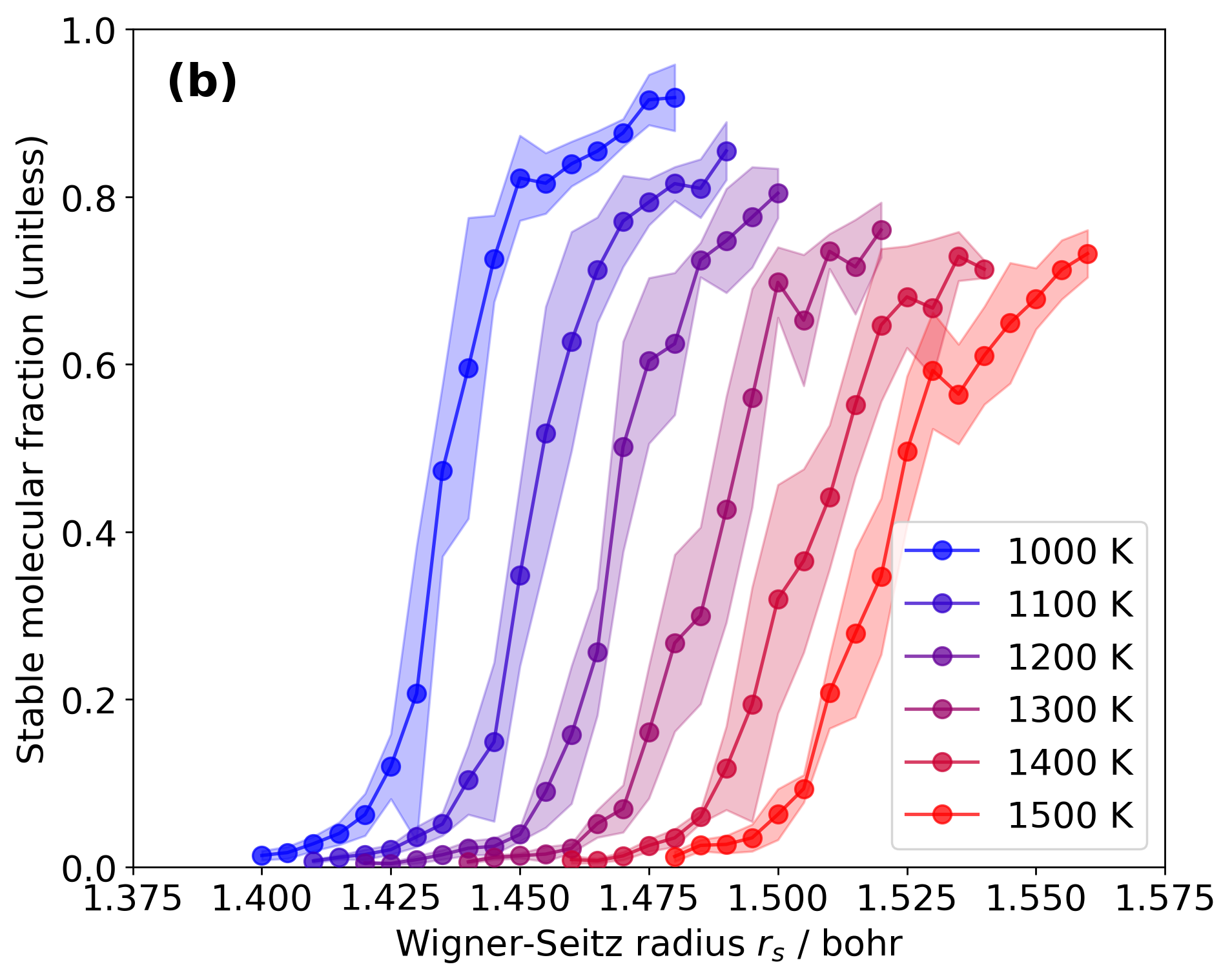}
    
    \includegraphics[width=0.44\linewidth,trim={5pt 5pt 5pt 5pt},clip]{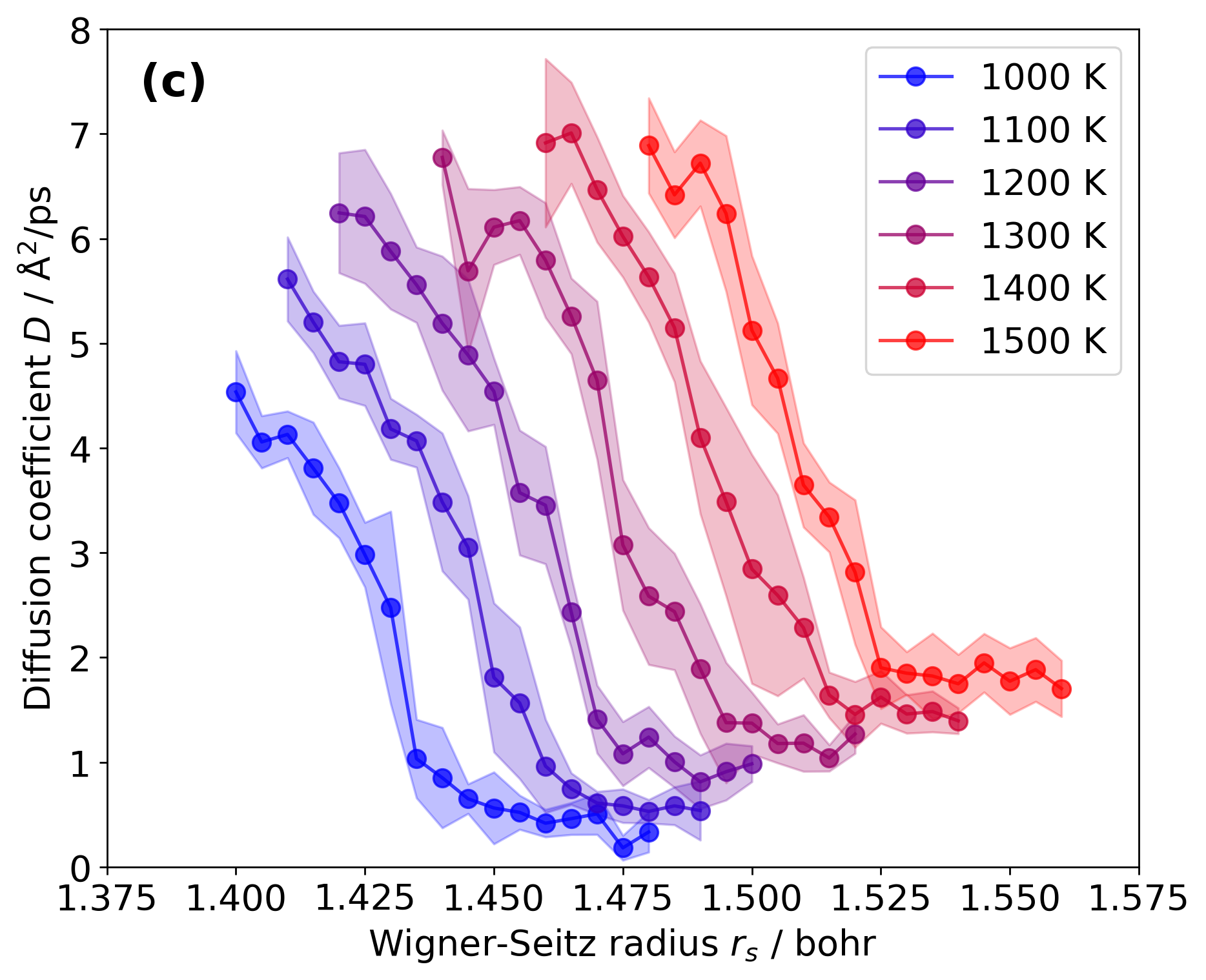}
    \includegraphics[width=0.55\linewidth,trim={5pt 5pt 5pt 5pt},clip]{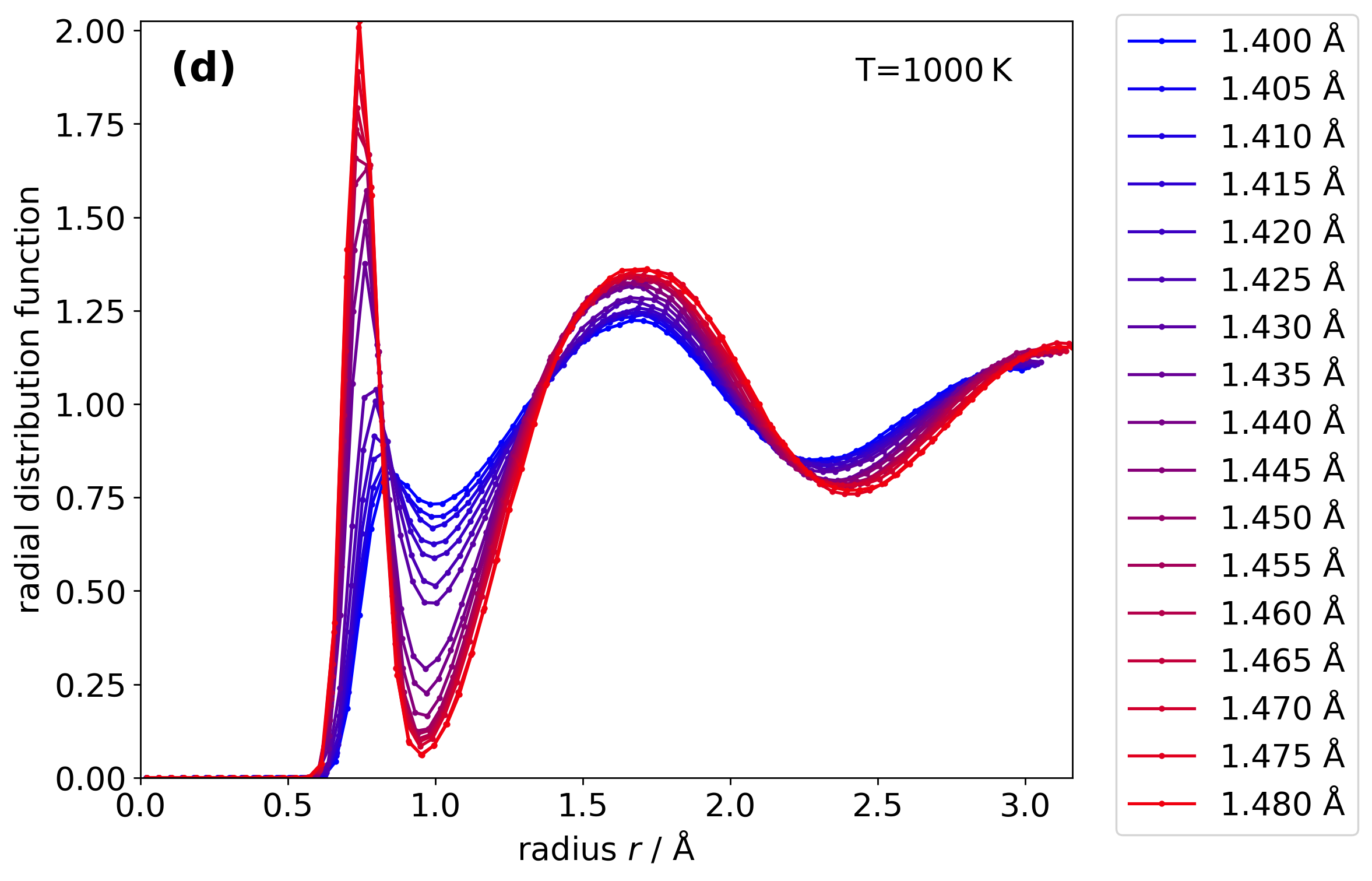}
    
    \caption{%
        \emph{Derived properties} based on the reference \gls{dft}/\gls{pbe} \gls{md} simulations of the \dataset{} dataset. 
        Shown are (a) pressure, (b) stable molecular fraction, and (c) diffusion coefficient as a function of the Wigner-Seitz radius~$r_s$ and for different temperatures (color-coded), as well as (d) radial distribution functions at 1000\,K for different Wigner-Seitz radii (color-coded).
    }
    \label{fig:performancemeasures}
\end{figure}

\textbf{Pressure curves} show isotropic pressure as a function of the Wigner-Seitz radius~$r_s$ at different temperatures (Figure~\ref{fig:performancemeasures}a).
These curves relate the state variables temperature, pressure, and density, effectively corresponding to equations of state.
For pure atomic and molecular phases, pressure decreases monotonically with~$r_s$.
Near the \gls{llpt}, pressure remains almost constant, leading to a characteristic plateau.
This effect decreases for higher temperatures.
While the existence, explanation, and position of the actual plateaus depend on the ab-initio method and have been controversially debated, results reported in the literature for \gls{dft}-\gls{pbe} are consistent. \cite{lhr2010,mpsc2010,mhs2018,gwma2019,cmpc2020q,hktc2020,khht2021,ttns2022q,nypc2023q}.
The benchmark measures the ability of an \gls{mlp} to reproduce the \gls{dft}-\gls{pbe} results.

\textbf{Stable molecular fractions}~$\mu$ indicate the fraction of hydrogen atoms that form an $H_2$ dimer for an extended time span.
Specifically, we consider two hydrogen atoms to be a stable dimer if they remain within 0.95\,{\AA} of each other for at least 75\,fs.
These thresholds reflect the typical time and length scales in \gls{wdh}.\cite{gwma2019} 
Moreover, we find that the inflection point of the isothermal $\mu$ over $r_s$ curve is not sensitive to the choice of thresholds.
The benchmark calculates~$\mu$ for each trajectory frame and reports the average.
In the atomic liquid phase, $\mu$ is (close to) zero, whereas in the molecular liquid phase, $\mu$ is (close to) one.
The transition between the two phases is visible as a sudden change in~$\mu$ (Figure~\ref{fig:performancemeasures}b). 

The (self-)\textbf{diffusion coefficient}~$D$ quantifies the mobility of the hydrogen atoms.
The benchmark code first computes the $3 \times 3$ diffusion tensor~$\bm{D}$ from the squared displacements of the atomic positions via the Einstein equation\cite{mmre2020}
\begin{equation}\label{equ:einstein}
    \bm{D} = \frac{1}{2} \lim_{t\rightarrow \infty} \frac{\partial}{\partial t} \frac{1}{N} \sum_{i=1}^N  \bigl( \bm{r_i}(t) - \bm{r_i}(0) \bigr)^2, 
\end{equation}
where $N$ is the number of atoms in the unit cell and $\bm{r_i}(t)$ is the position of atom~$i$ at time~$t$.
Averaging the main diagonal then yields $D = \frac{1}{3} ( \bm{D}_{1,1} + \bm{D}_{2,2} + \bm{D}_{3,3})$.
For homogeneous systems, the tensor's main diagonal elements should be similar, and the off-diagonal elements should be small. 

Equation~\ref{equ:einstein} converges for large enough~$N$ and~$t$.
While in this benchmark, $N$ and~$t$ are too small for convergence, they are large enough to facilitate the comparison between \gls{dft} and \glspl{mlp}, and to indicate the observed \gls{llpt}:
The atomic phase is characterised by highly mobile hydrogen atoms, and thus large $D$, whereas the molecular phase exhibits a more viscous behaviour, and thus smaller $D$.
The transition between the two phases is visible as a sudden change in~$D$ (Figure~\ref{fig:performancemeasures}c).

The \textbf{\gls{rdf}}~$g(r)$, also called pair correlation function, describes the average density of hydrogen atoms as a function of distance from a hydrogen atom.
For a homogeneous and isotropic system, the \gls{rdf} is measured relative to the system's overall density $N/V$, where $N$ is the number of atoms and $V$ is the volume of the unit cell.
The benchmark computes the \gls{rdf} as a histogram by counting the atoms in spherical shells around a central atom and dividing by the shells' volumes $\frac{4}{3} \pi (r+\Delta)^3 - \frac{4}{3} \pi r^3$, where~$r$ is the distance at which the shell starts and $\Delta$ is its thickness. 
Together,
\[
    g(r,\Delta) = \biggl( \frac{1}{N} \sum_{i,j=1}^N \left[ r \leq ||\mathbf{r_j} - \mathbf{r_i}|| \leq r + \Delta \right] \biggr) \Bigl/ \; \frac{N}{V} \frac{4}{3} \pi \bigl( (r+\Delta)^3 - r^3 \bigr) ,
\]
where $[\ldots]$ are Iverson brackets, $\mathbf{r_i}$ is the position of atom~i, and $||\cdot||$ is Euclidean distance under periodic boundary conditions.
The molecular liquid shows a peak at the hydrogen-hydrogen bonding distance, whereas the atomic liquid does not (Figure~\ref{fig:performancemeasures}d).
For larger distances, the \gls{rdf} converges to the ideal gas limit.

\begin{figure}[b]
    \includegraphics[width=0.33\linewidth-\tabcolsep]{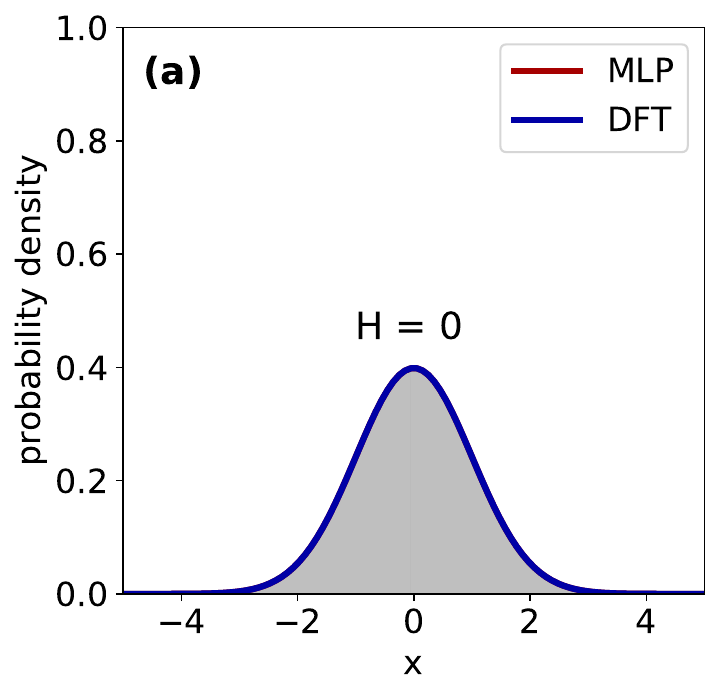}%
    \hfill%
    \includegraphics[width=0.33\linewidth-\tabcolsep]{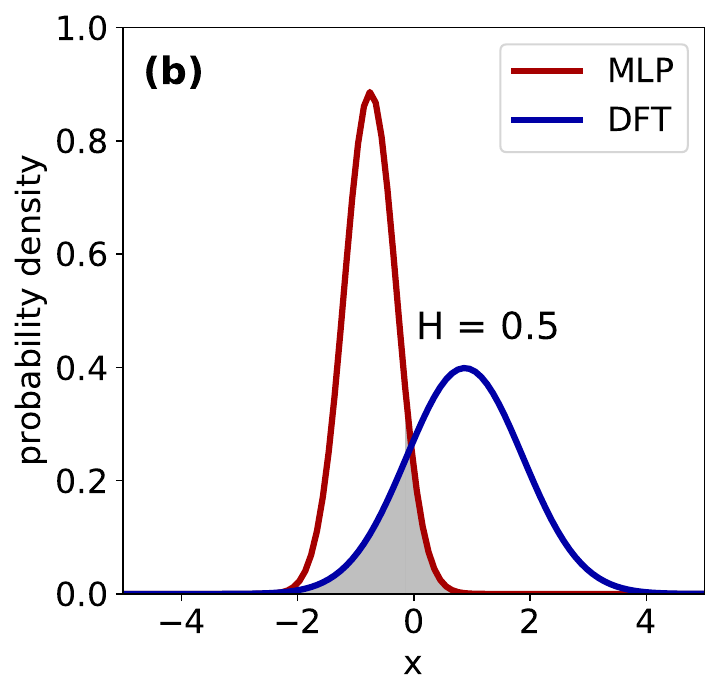}%
    \hfill%
    \includegraphics[width=0.33\linewidth-\tabcolsep]{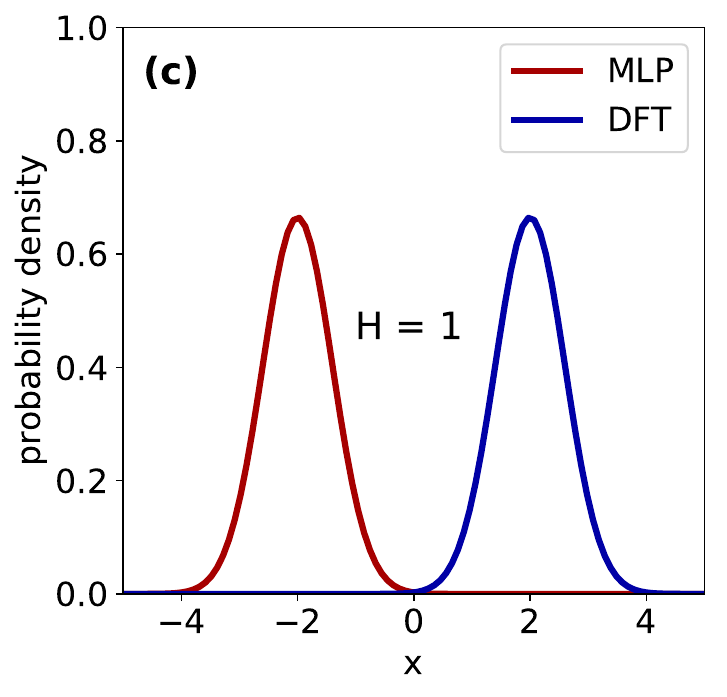}
    \caption{%
        \emph{Illustration of the Hellinger distance.}
        Shown are (a) perfect, (b) intermediate, and (c) vanishing overlap of two one-dimensional distributions (MLP, DFT). 
        The intersection is shown in gray.
    }
    \label{fig:hellingerdist}
\end{figure}

The benchmark uses the \textbf{\gls{hd}}~\cite{h1909} to quantify the discrepancies between properties derived from reference \gls{dft} and \gls{mlp}-accelerated \gls{md} simulations. 
This statistical measure quantifies the similarity between two probability distributions.
It is closely related to the Bhattacharyya coefficient~\cite{b1946} but is a metric in the mathematical sense:
It is non-negative, symmetric, and obeys the triangle inequality. 
Furthermore, it is bounded between 0 and 1, with 0 (1) representing perfect (vanishing) overlap between two distributions.
Figure~\ref{fig:hellingerdist} presents a graphical illustration of the \gls{hd}.

The benchmark assumes that property values derived from multiple \gls{md} simulations are normally distributed.
The \gls{hd} of two scalar, normally distributed random variables $f_\text{DFT}\sim \mathcal{N}(\mu_\text{DFT},\sigma_\text{DFT})$ and $f_\text{MLP}\sim \mathcal{N}(\mu_\text{MLP},\sigma_\text{MLP})$ is
\begin{equation}
    \text{HD}(f_\text{MLP},f_\text{DFT}) = 
    \sqrt{1 - 
    \sqrt{\frac{2\,\sigma_\text{MLP}\,\sigma_\text{DFT}}{\sigma_\text{MLP}^2+\sigma_\text{DFT}^2}}\,
    \exp \bigg( -\frac{1}{4}\frac{(\mu_\text{MLP}-\mu_\text{DFT})^2}{\sigma_\text{MLP}^2+\sigma_\text{DFT}^2} \bigg)}.
\end{equation}
For multivariate quantities, the benchmark averages over the \glspl{hd} between components.

\textbf{Test-set prediction errors}.
We also provide the error in predicted energies and forces on a dedicated test dataset sampled from the same distribution as the training data (Section~\ref{sec:dataset}).
These test-set errors can provide insights into an \gls{mlp} and its failure modes, e.g., via bias and outlier analysis.
However, test-set errors are not necessarily indicative of the accuracy of \gls{md}-derived properties (\cref{sec:related}).

For test-set predictions of energies and (component-wise) forces,  we report the \gls{mae}, \gls{rmse}, and maximum absolute error, both as-is and as fractions of their respective ranges, as well as $\log(1-R^2)$, where $R$ is Pearson's correlation coefficient, a measure of linear correlation.
For simplicity, we refrain from adjusting the range for outlying values and from dividing by standard deviation instead of range.

For visual analysis the benchmark provides scatter plots of predicted versus ground-truth values.
These plots also show the distribution of predicted and ground-truth \gls{dft} energies and force components, as well as $\log$-$\log$ histograms of prediction errors.
See \cref{fig:scatter:ufthreepacemace,si:fig:scatterplots} for examples.


\section{Benchmark}\label{sec:benchmark}

\textbf{Benchmarking} the performance of an \gls{mlp} in characterising the \gls{llpt} of \gls{wdh} consists of these steps:
\begin{enumerate}[label={(\arabic*)},itemsep=0pt,leftmargin=*]
    \item\label{alg:bmtrain} Training the \gls{mlp} on the \dataset{} dataset, using only the training data part.
    \item\label{alg:bmrunmd} Running \gls{md} simulations with the trained \gls{mlp}.
    \item\label{alg:bmprops} Calculating derived properties (Section~\ref{sec:properties}) based on the generated trajectories.
    \item\label{alg:bmcompr} Comparing these with the properties obtained from the \gls{dft} reference simulations via the \gls{hd}.
\end{enumerate}

Step~\ref{alg:bmtrain} depends on the specifics of the \gls{mlp} and is, therefore, not performed by the benchmark code. 
All other steps, \ref{alg:bmrunmd}, \ref{alg:bmprops}, and \ref{alg:bmcompr}, do not depend on the specifics of the \gls{mlp} and are fully automatised by the benchmark code.

Step~\ref{alg:bmrunmd} runs multiple \gls{md} simulations in the $NVT$ ensemble with the same settings as in Section~\ref{sec:dataset} but using the \gls{mlp} instead of the \gls{dft} potential energy surface.
Specifically, for each of the six temperatures and 17 Wigner-Seitz radii, 12 independent \gls{md} simulations are performed, resulting in a total of $6 \cdot 17 \cdot 12 = 1\,224$ \gls{md} simulations.
Of the 10\,k time steps in each simulation, the first 2\,500 time steps are discarded to allow for equilibration.
The remaining 7\,500 time steps are used to compute performance measures.

The benchmark uses the LAMMPS (Large-scale Atomic/Molecular Massively Parallel Simulator) code~\cite{tatp2022} to run the \gls{md} simulations.
Consequently, the benchmarked \gls{mlp} must have LAMMPS bindings.

The benchmark's \textbf{code} contains functionality to
\begin{itemize}
    \item configure the benchmark, create hydrogen starting configurations, input files, and directory structure (\texttt{create\char`_\ldots} routines);
    \item run \gls{md} simulations (\texttt{run\char`_simulations} routine);
    \item extract information and \gls{md}-derived properties (\texttt{get\char`_\ldots} routines);
    \item create figures (\texttt{figure\char`_\ldots} routines) and tables (\texttt{table\char`_\ldots} routines).
\end{itemize}
The benchmark is written in the open-source programming language Python (Python Software Foundation, \href{https://www.python.org/}{python.org}).
The benchmark repository (\cref{sec:datacode}) provides two Jupyter (NumFOCUS Foundation, \href{https://jupyter.org/}{jupyter.org}) notebooks.
One demonstrates benchmarking of an \gls{mlp} and can serve as a starting point;
the other generates \cref{fig:performancemeasures,si:fig:scatterplots} and \cref{tab:resultsmdprop,tab:resultstestset} in this work.


\section{Results and discussion}

We benchmark several interatomic potentials of increasing complexity, ranging from traditional empirical potentials to a state-of-the-art message-passing neural network.
We selected these potentials for diversity, covering local and semilocal models, different representations of atomic environments, and multiple regression algorithms.
Together, these potentials cover the whole range of the trade-off between computational efficiency and predictive accuracy.
Specifically, we benchmark the following potentials:

\begin{enumerate}[]
    \item%
        The two-body Yukawa \cite{y1935} and the three-body Tersoff \cite{t1988} potential, two established traditional empirical potentials.
        These are not \glspl{mlp} but serve as baseline models.
        Both were reparametrised for the \dataset{} dataset. 
        These are the fastest potentials in the selection and have the fewest parameters.

    \item%
        Ultra-fast potentials \cite{xrh2023q} in a two-body and a three-body parametrisation.
        These \glspl{mlp} represent effective two- and three-body terms in a cubic B-spline basis, with coefficients determined via regularised linear regression.
        They are as fast as traditional empirical potentials but more flexible.
    
    \item%
        PACE (performant ACE) \cite{lood2021q} and MACE (multi-ACE) \cite{bkoc2022q}, two \glspl{mlp} based on the atomic cluster expansion (ACE), \cite{d2019q} a complete representation of local atom environments.
        PACE uses semi-linear regression; MACE is a message-passing neural network with equivariant higher-order messages.
\end{enumerate}

\begin{figure}

    \includegraphics[width=0.25\linewidth]{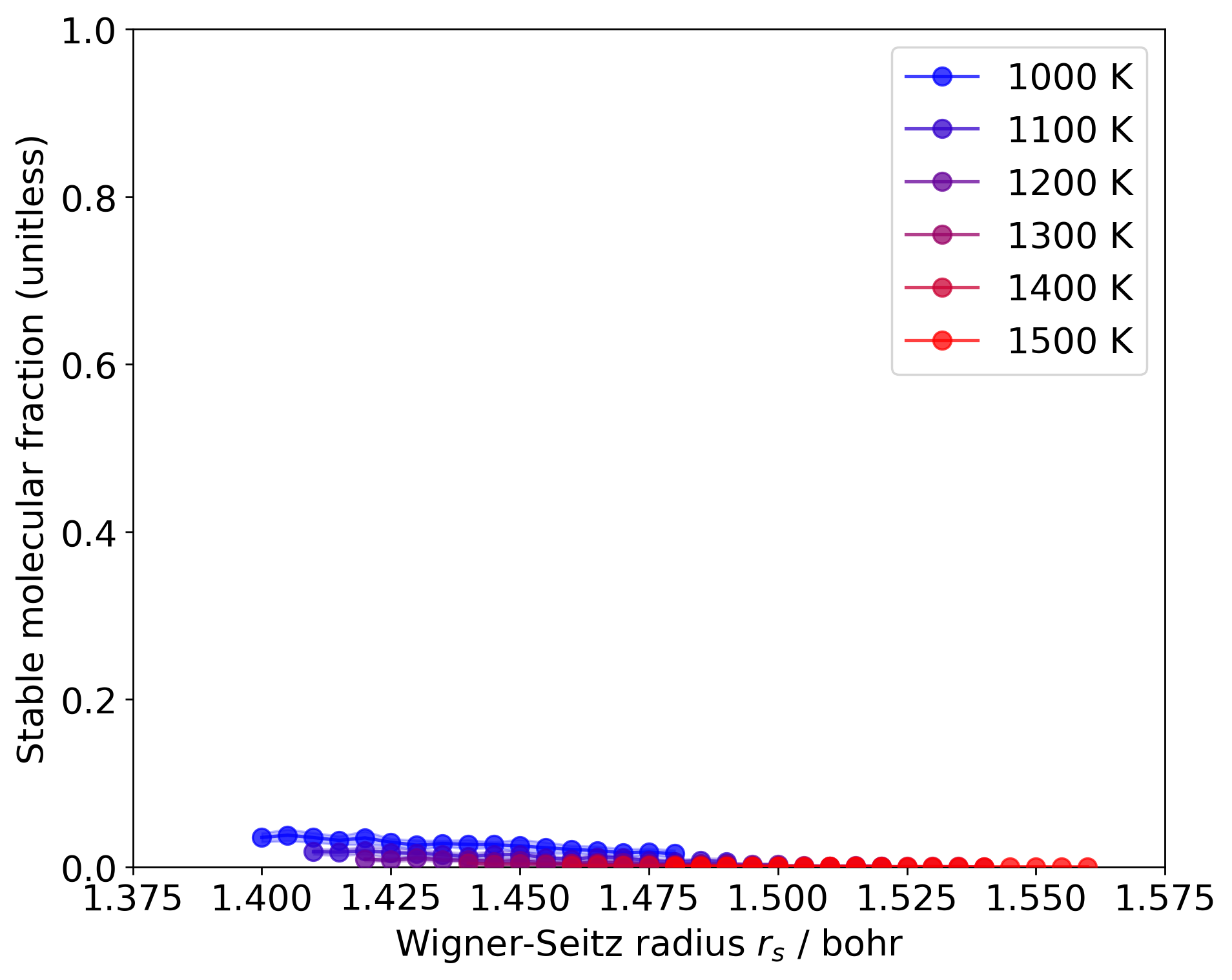}%
    \includegraphics[width=0.25\linewidth]{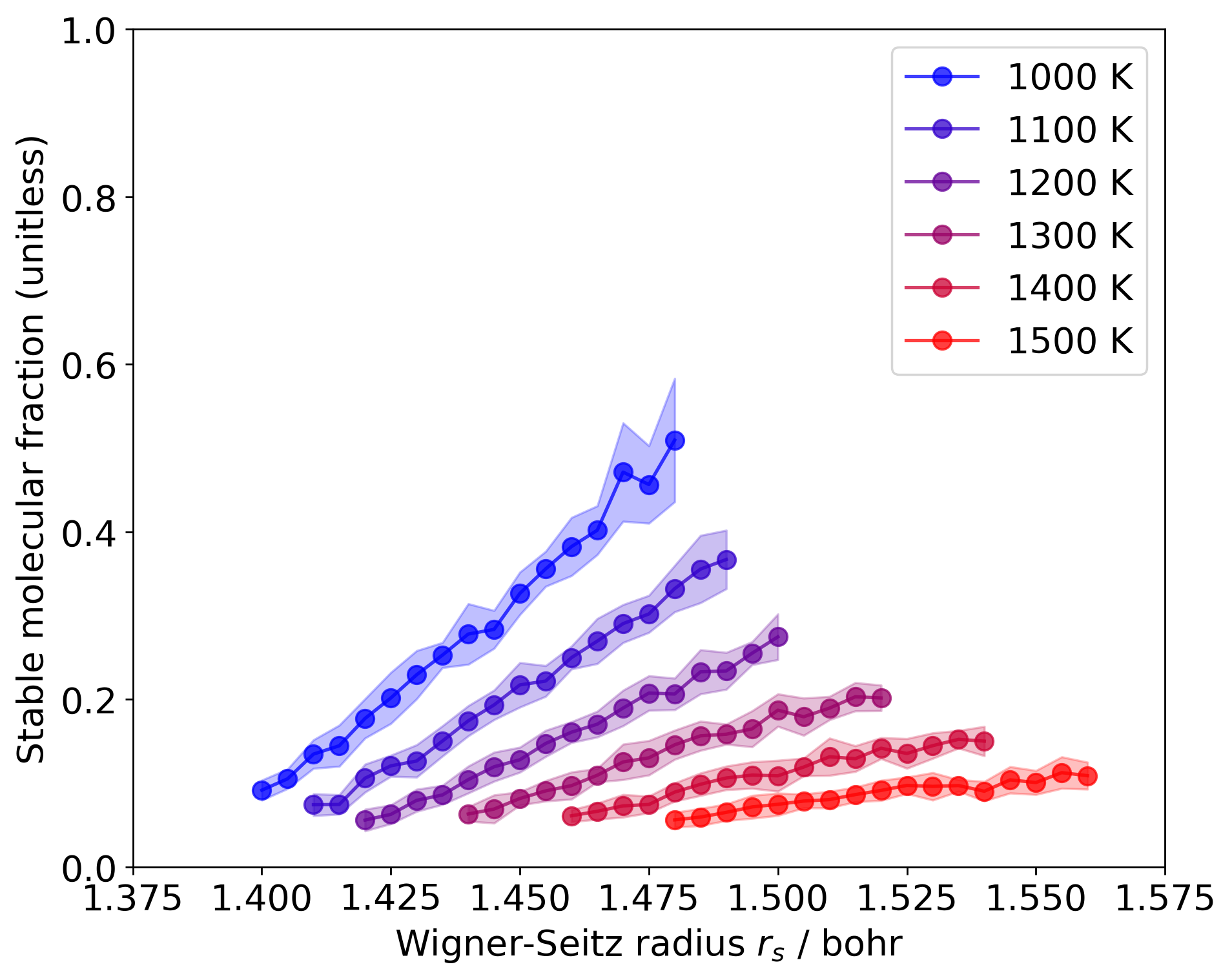}%
    \includegraphics[width=0.25\linewidth]{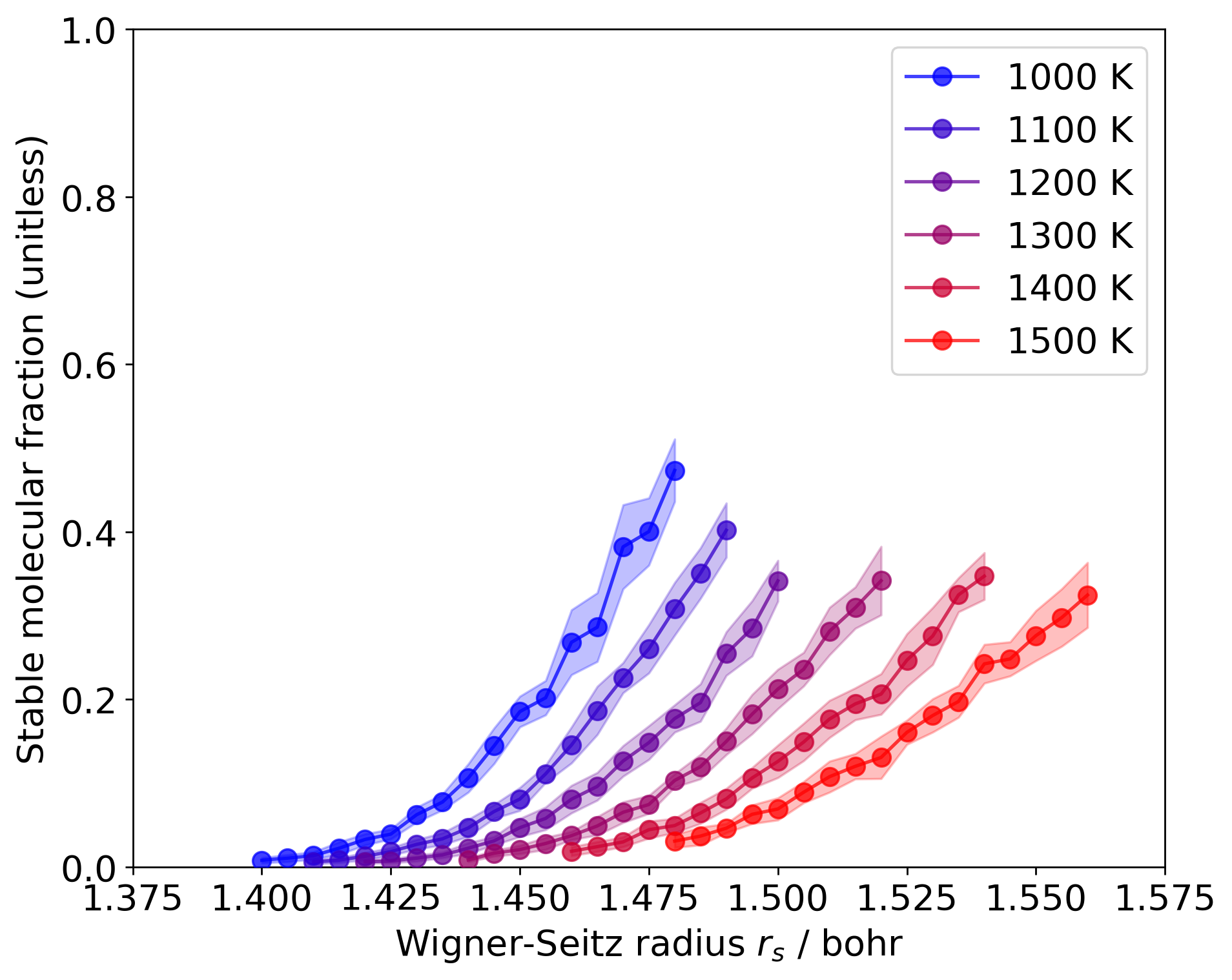}%
    \includegraphics[width=0.25\linewidth]{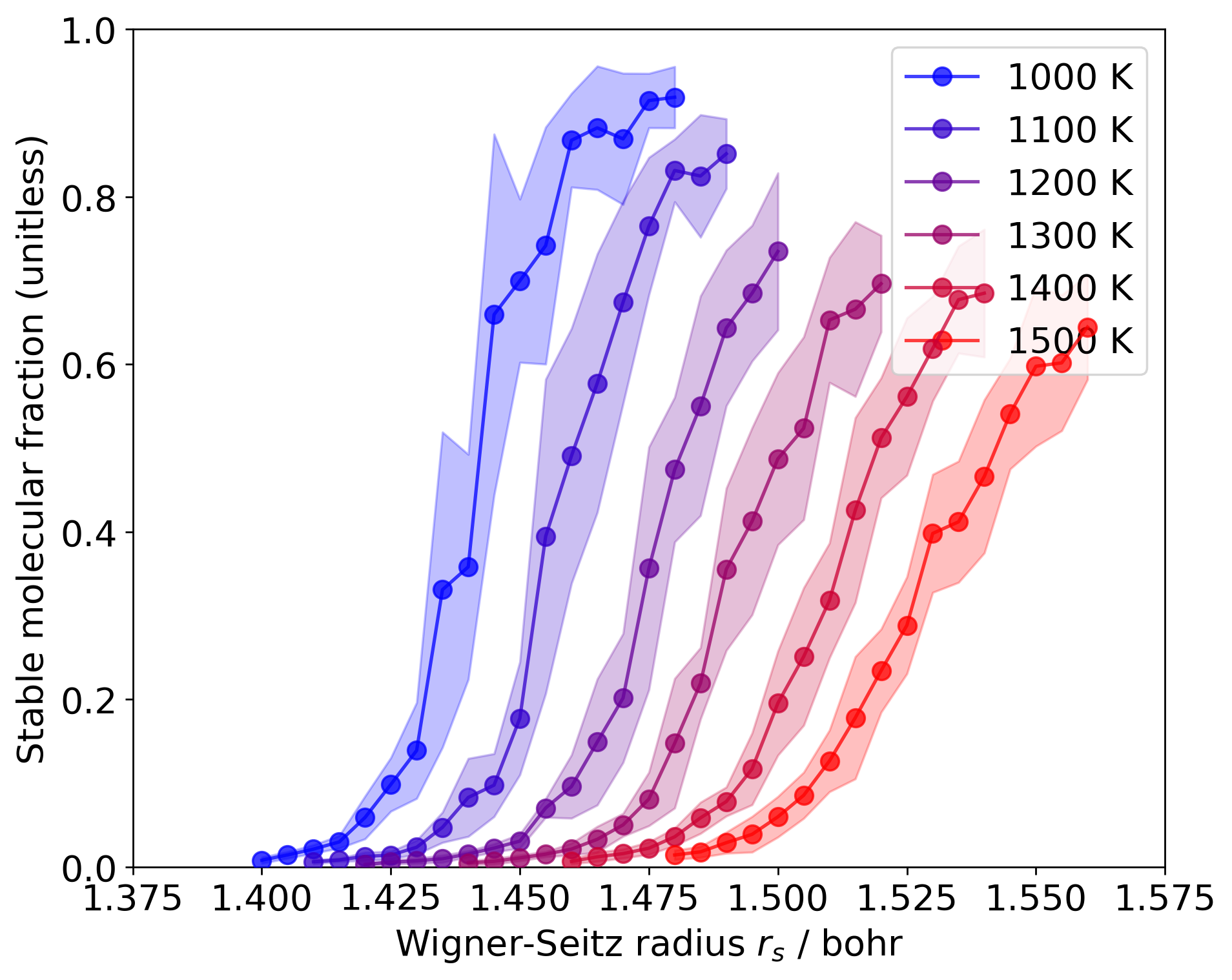}

    \includegraphics[width=0.25\linewidth]{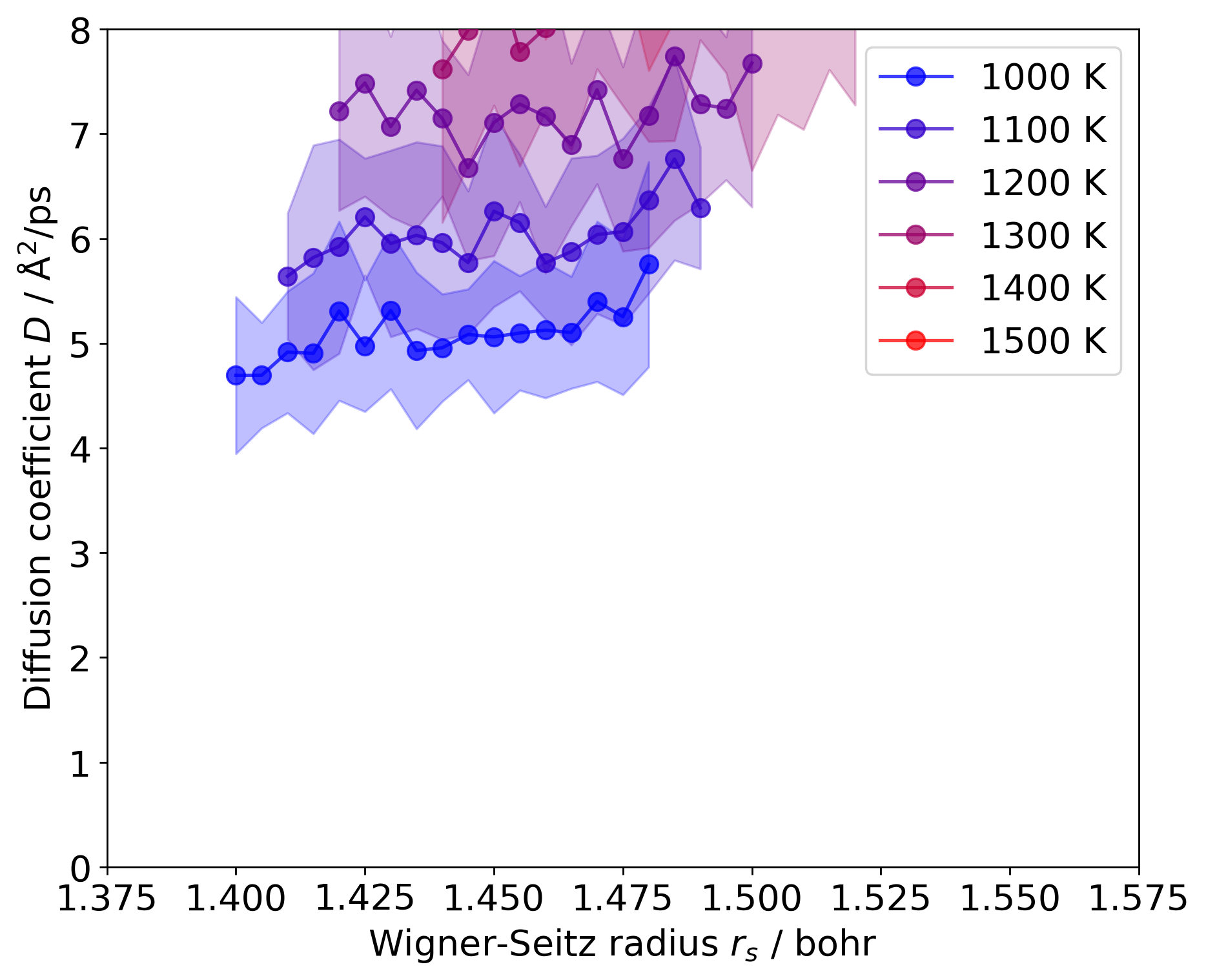}%
    \includegraphics[width=0.25\linewidth]{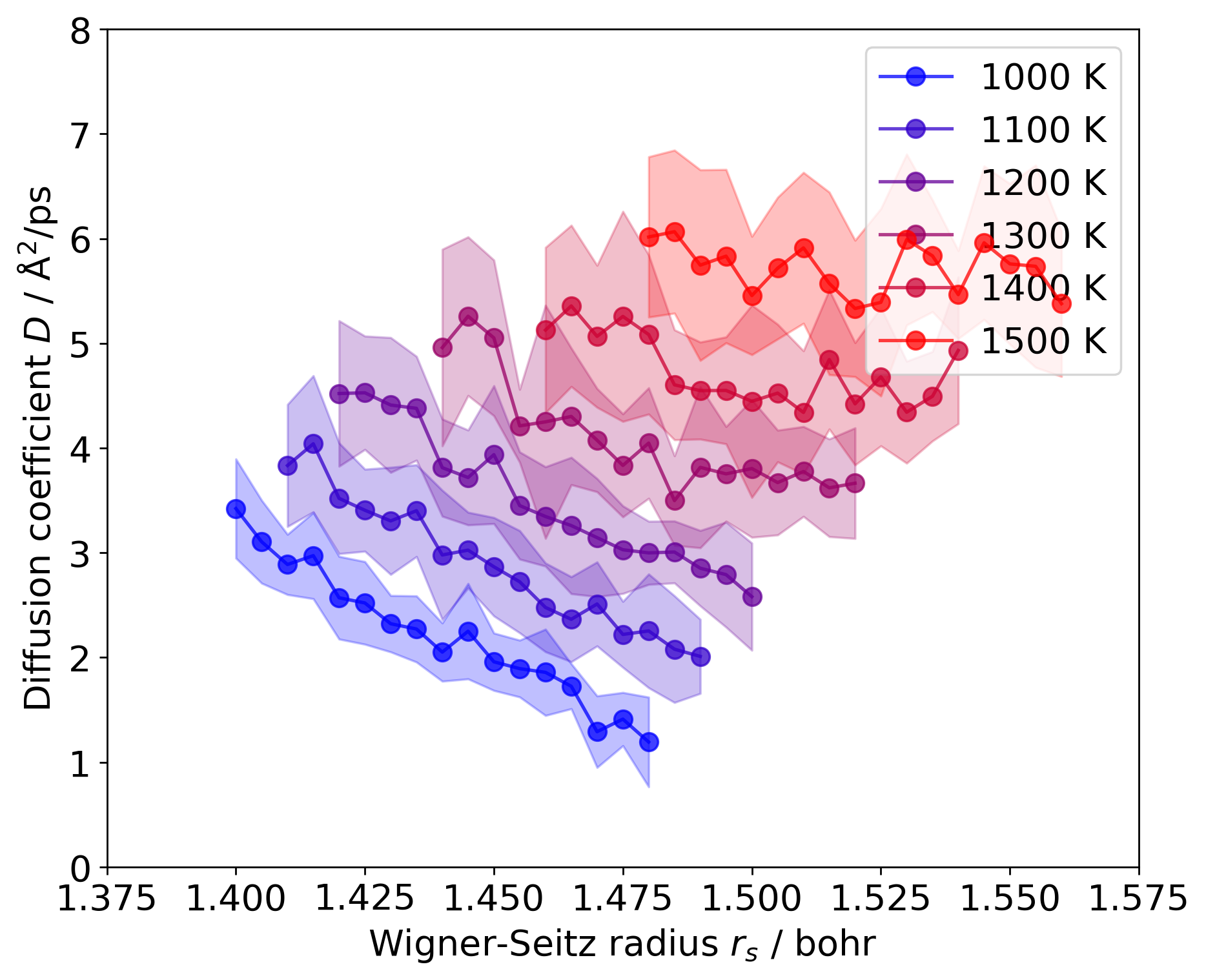}%
    \includegraphics[width=0.25\linewidth]{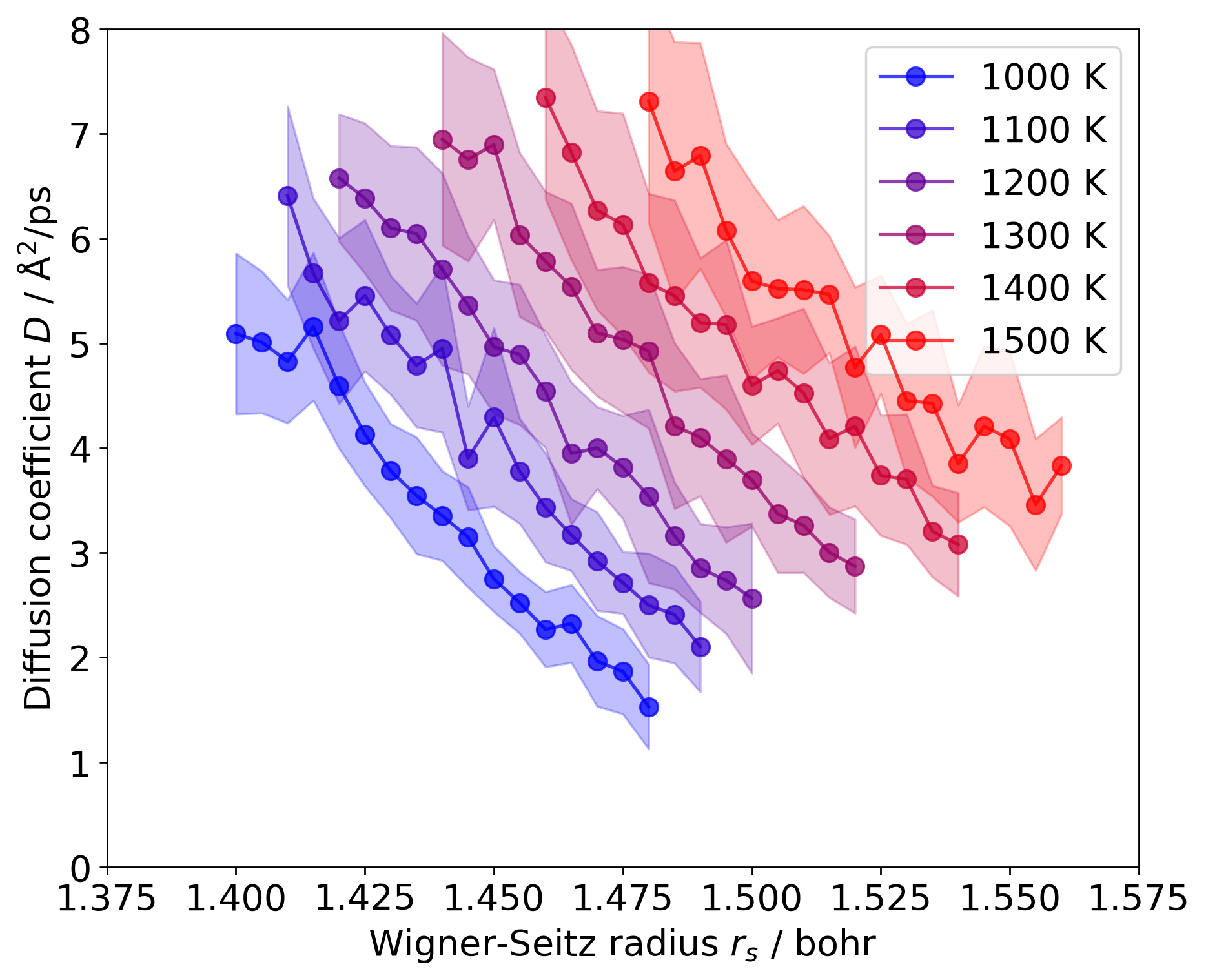}%
    \includegraphics[width=0.25\linewidth]{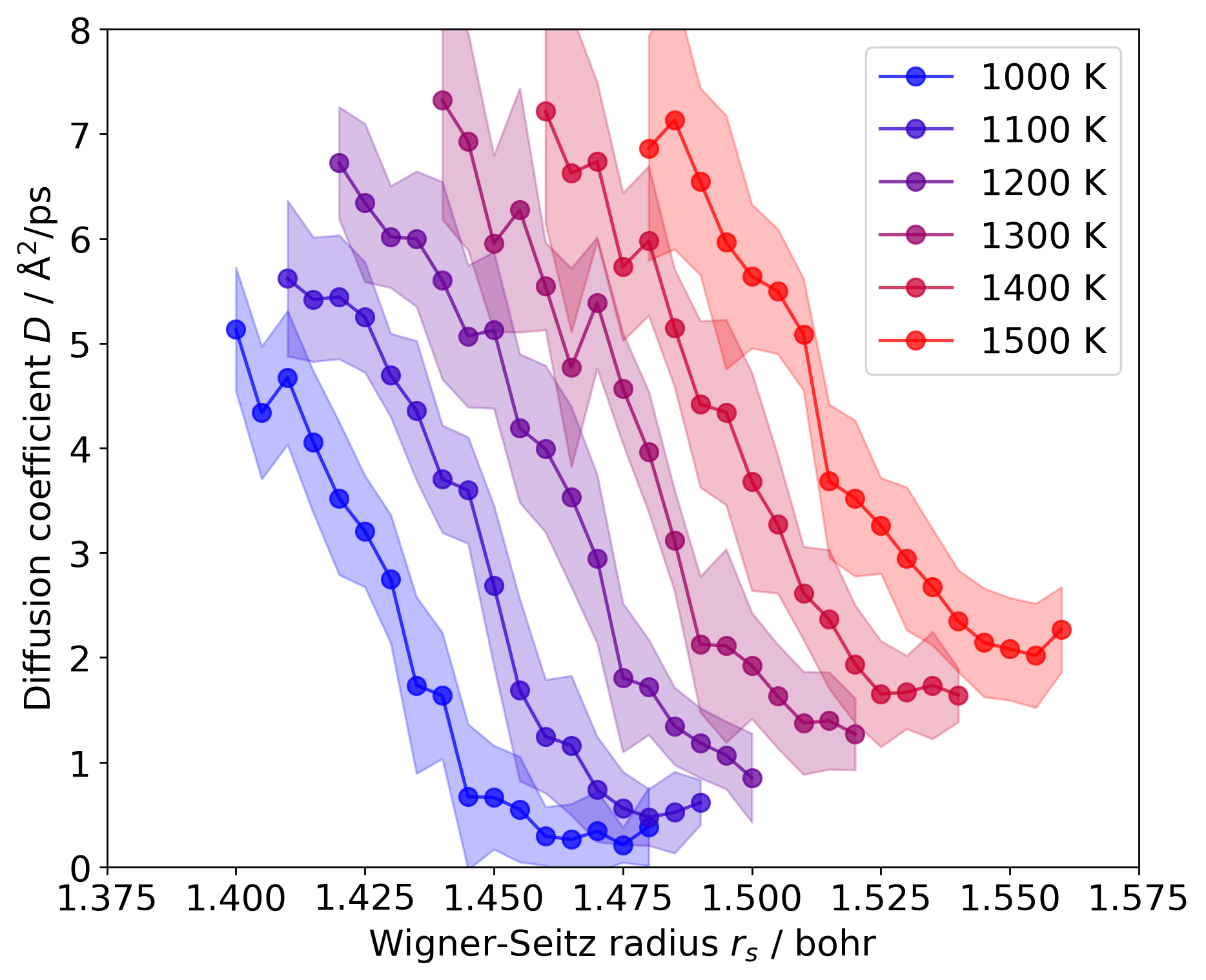}

    \mbox{}\hfill UFP2\hfill\hfill UFP3\hfill\hfill PACE\hfill\hfill MACE\hfill\mbox{}
    
    \caption{%
        \emph{Derived properties} from \gls{mlp}-accelerated \gls{md} simulations.
        Shown are stable molecular fractions (top row) and diffusion coefficients (bottom row) for UFP2, UFP3, PACE, and MACE (left to right).
    }
    \label{fig:selproperties}
\end{figure}

Table~\ref{tab:resultsmdprop} presents the main benchmark results. 
For the \glspl{mlp} above and the four \gls{md}-derived properties of Section~\ref{sec:properties}---pressure~$p$, stable molecular fraction~$\mu$, (self)diffusion coefficient~$D$, and \gls{rdf}~$g$---we report average \glspl{hd} between reference \gls{dft}/\gls{pbe} (six repetitions) and \gls{mlp}-accelerated (12 repetitions) \gls{md} simulations.
\Cref{fig:selproperties} shows stable molecular fractions and diffusion coefficients for the UFP2, UFP3, PACE, and MACE \glspl{mlp}.
Property and scatter plots for all \glspl{mlp} can be found in \cref{si:fig:properties,si:fig:scatterplots}.

The Yukawa and the Tersoff potential fail to reproduce the \gls{md}-derived properties.
Since the UFP2 and UFP3 potentials perform markedly better, we attribute this failure to the fixed functional forms of the traditional empirical potentials.
In the following, we will not discuss the Yukawa and Tersoff potentials further.

UFP3 performs as well as or better than UFP2 for all properties, consistent with its increased fitting capacity from its three-body terms. 
UFP3 is also the first \gls{mlp} to start qualitatively capturing the properties of \gls{wdh} (\cref{fig:selproperties}).

PACE improves substantially over UFP3 in \gls{hd} and visual assessment but does not yet closely match the properties from the reference \gls{md} simulations.
MACE, in turn, improves substantially over PACE in \gls{hd}, with the largest improvement in \cref{tab:resultsmdprop}, and visual assessment.
It is the only benchmarked \gls{mlp} whose \gls{md}-derived properties quantitatively agree with the \gls{dft}/\gls{pbe} reference.

Overall, we observe that $\text{HD}(p) > \text{HD}(\mu) > \text{HD}(D) > \text{HD}(g)$ (except for $\mu$ and $D$ for the Tersoff potential, which we attribute to noise), indicating that pressures are harder to reproduce than stable molecular fractions than diffusion coefficients than \glspl{rdf}.
The performance of \glspl{mlp} improves with their complexity (and computational requirements).

\begin{table}[bthp]
    \begin{minipage}[t]{0.51\linewidth}
    \begin{tabular}[t]{@{}lcccccc@{}}
        \toprule
        & & \multicolumn{5}{c}{HDs between MD Properties} \\
        \cmidrule(lr){3-7}
        MLP & Rank & $p$ & $\mu$ & $D$ & $g$ & Ø \\
        \midrule
        Yukawa  & 5.5 & 1.00 & 1.00 & 0.99 & 0.79 & 0.95 \\
        Tersoff & 5.5 & 1.00 & 0.95 & 0.99 & 0.91 & 0.96 \\
        UFP2    & 4 & 1.00 & 0.87 & 0.85 & 0.66 & 0.85 \\
        UFP3    & 3 & 0.99 & 0.88 & 0.65 & 0.59 & 0.78 \\
        PACE    & 2 & 0.84 & 0.67 & 0.60 & 0.31 & 0.61 \\
        MACE    & 1 & \bf 0.27 & \bf 0.22 & \bf 0.19 & \bf 0.10 & \bf 0.20 \\
        \bottomrule
    \end{tabular}
    \end{minipage}%
    \hfill%
    \begin{minipage}[t]{\linewidth-0.51\linewidth-2\tabcolsep}
        \caption{%
            \textit{Performance comparison} of \glspl{mlp}. 
            Shown are average \glspl{hd} for four \gls{md}-derived properties (pressure~$p$, stable molecular fraction~$\mu$, (self)diffusion coefficient~$D$, \gls{rdf}~$g$) between properties from \gls{dft} reference and \gls{mlp}-accelerated \gls{md} simulations.
            Best performance in bold.
        }
        \label{tab:resultsmdprop}
    
        \begingroup
        \scriptsize 
        MLP = machine-learning potential, 
        HD = Hellinger distance, 
        MD = molecular dynamics,
        DFT = density functional theory.
        \endgroup
    \end{minipage}
    
\end{table}

\begin{figure}
    \includegraphics[width=0.33\linewidth,trim={160mm 5pt 5pt 5pt},clip]{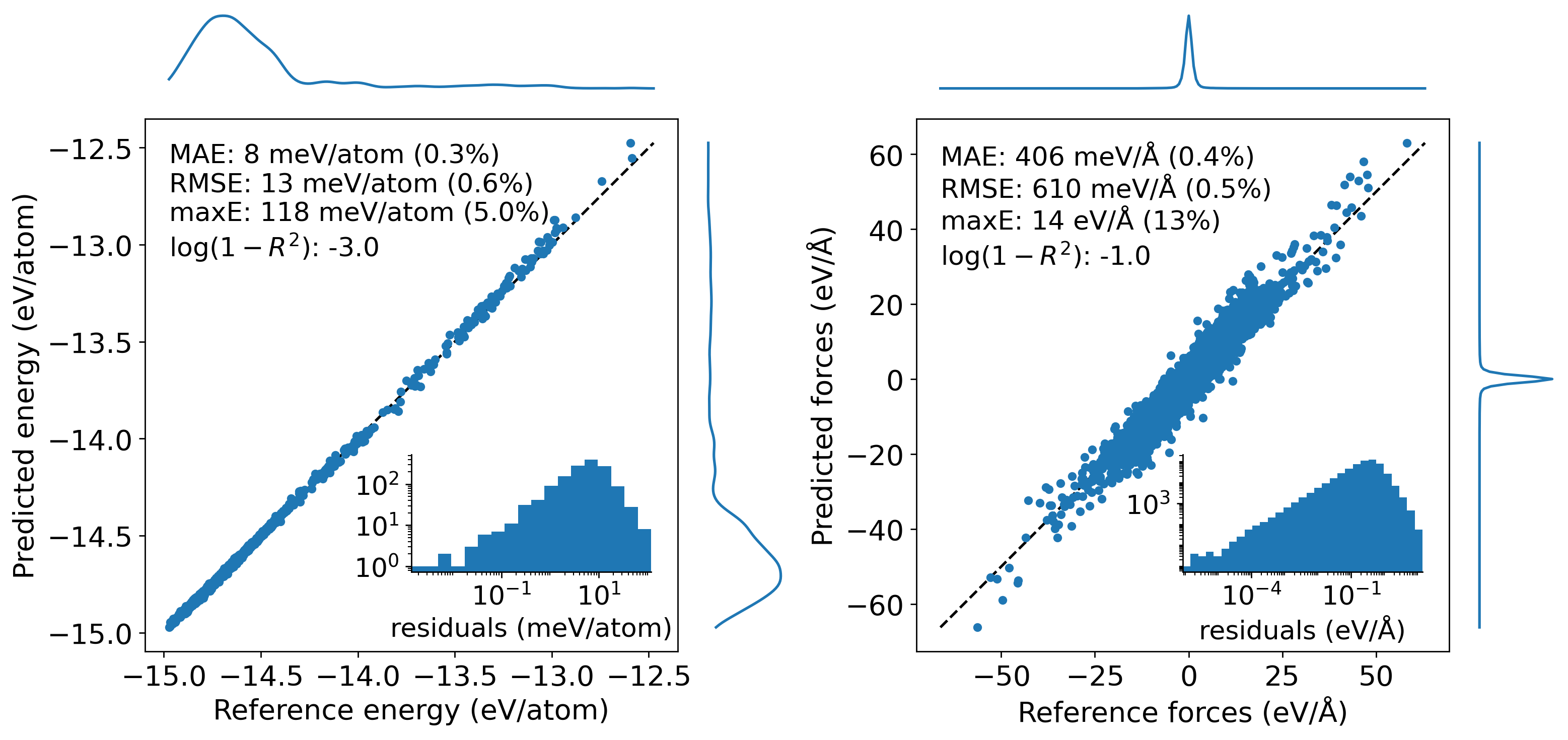}%
    \includegraphics[width=0.33\linewidth,trim={160mm 5pt 5pt 5pt},clip]{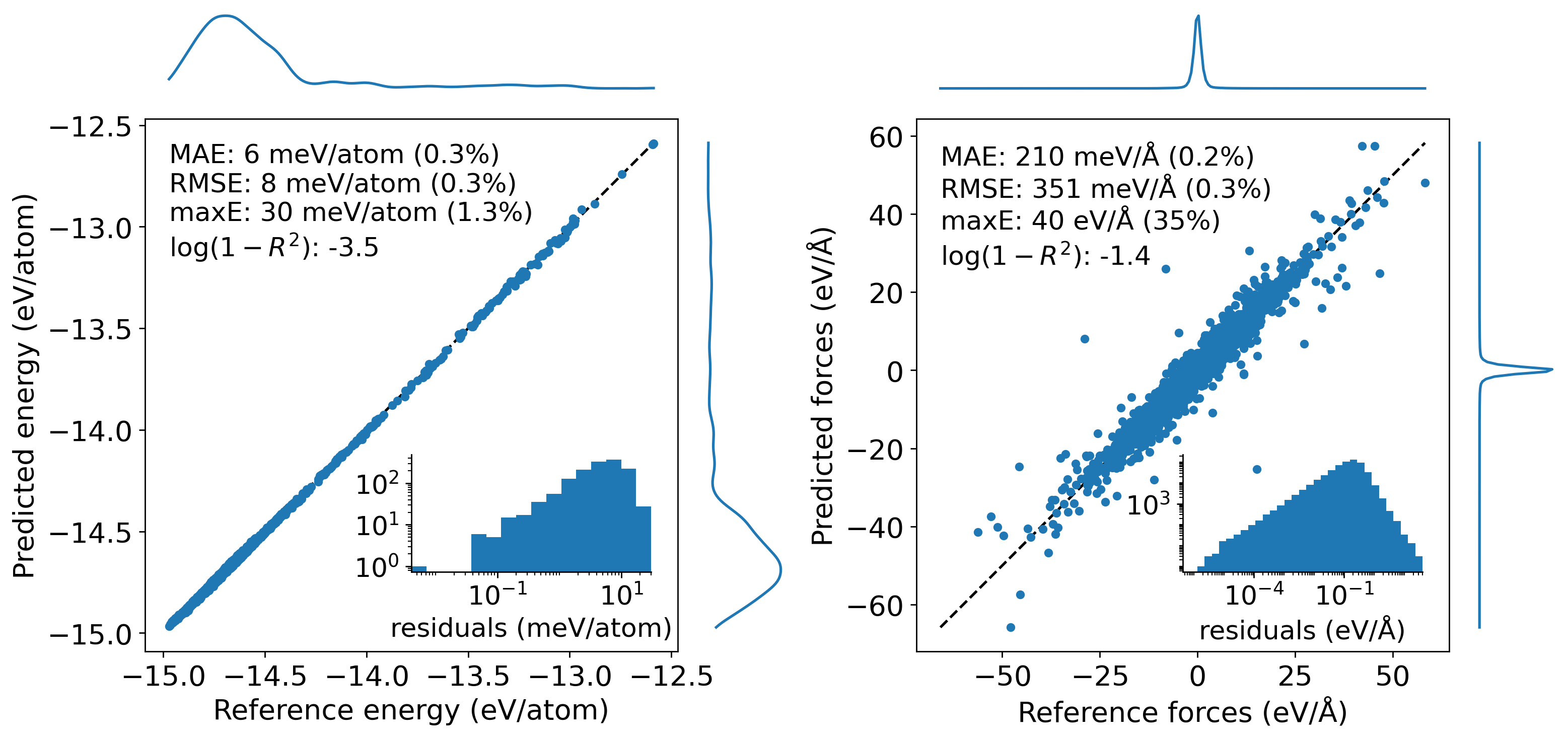}%
    \includegraphics[width=0.33\linewidth,trim={160mm 5pt 5pt 5pt},clip]{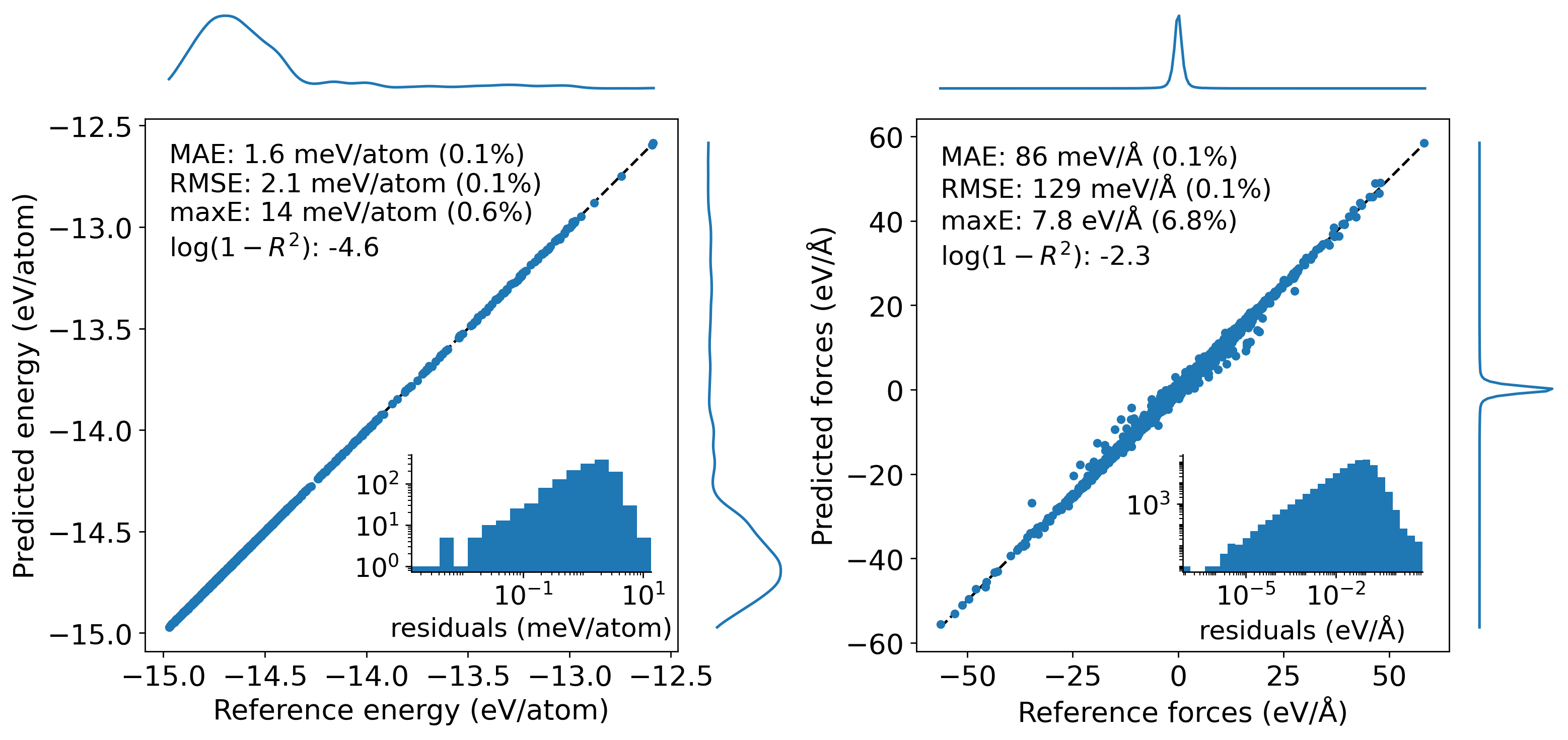}%
    
    \caption{%
        \emph{Test-set prediction errors for forces} (component-wise) of the UFP3 (left), PACE (middle), and MACE (right) \glspl{mlp}.
        Shown are predictions versus \gls{dft} reference (dots), error statistics (top left inside of plots), data distributions (top and right outside of plots), and a $\log$-$\log$ histogram of prediction errors (inset).
        See \cref{sec:properties} for details and \cref{si:fig:scatterplots} for scatter plots of energies and forces for all \glspl{mlp}.
    }
    \label{fig:scatter:ufthreepacemace}
\end{figure}

As an ancillary, Table~\ref{tab:resultstestset} presents test-set prediction-error statistics.
Specifically, it shows \gls{mae}, \gls{rmse}, and maximum error ($\max$), both as-is and as fractions of their respective ranges, and $\rho = \log(1-R^2)$, where $R$ is Pearson's correlation coefficient.
For all statistics, lower values are better.
For UFP3, PACE, and MACE, the three best-performing \glspl{mlp} in this benchmark, \cref{fig:scatter:ufthreepacemace} presents scatter plots of test-set force prediction errors, data distributions, and $\log$-$\log$ histograms of prediction errors.
For energy and force scatter plots of all \glspl{mlp}, see \cref{si:fig:scatterplots}.

Test-set prediction errors correlate with performance in \gls{md}-derived properties and lead to the same ranking of \glspl{mlp}. 
However, from the test-set errors alone, it would have been unclear a priori if and which \gls{mlp} would lead to property predictions in agreement with the \gls{dft}/\gls{pbe} reference.
In particular, the suitability of PACE would have been unclear from test-set errors alone.

\begin{table}[bthp]
    \centering
    \caption{%
        \textit{Test-set prediction error} statistics for selected \glspl{mlp}.
        Shown are, for energies and forces (com\-pon\-ent-wise), MAE, RMSE, maximum absolute error ($\max$), both as-is and as a fraction of their respective ranges, as well as $\rho = \log_{10}(1-R^2)$, where $R$ is Pearson's correlation coefficient.
        Best performance in bold.
    }
    \label{tab:resultstestset}

    \begingroup
    \scriptsize 
        MLP = machine-learning interatomic potential, 
        MAE = mean absolute error,
        RMSE = root-mean-squared error.
    \endgroup

    \bigskip

    \begin{tabular}{@{}l c@{}c c@{}c c@{\;}c c   c@{}cc@{}cc@{\;}cc@{}}
        \toprule
        & \multicolumn{7}{c}{E / meV/atom} & \multicolumn{7}{c}{F / meV/Å} \\
        \cmidrule(lr){2-8} \cmidrule(lr){9-15} 
        MLP & MAE & \% & RMSE & \% & $\max$ & \% & $\rho$ & MAE & \% & RMSE & \% & $\max$ & \% & $\rho$ \\
        \midrule
        Yukawa  & 112 & 4.69 & 193 & 8.11 & 850 &   36 & -0.79  & 964 & 0.84 & 1434 & 1.25 & 46\,k &  40 & -0.22 \\
        Tersoff &  90 & 3.76 & 170 & 7.11 & 923 &   39 & -1.78  & 820 & 0.72 & 1485 & 1.30 & 55\,k &  48 & -0.23 \\
        UFP2    &  21 & 0.88 &  31 & 1.30 & 139 & 5.83 & -2.19  & 646 & 0.56 &  912 & 0.80 & 15\,k &  13 & -0.64 \\
        UFP3    & 8.3 & 0.35 &  13 & 0.55 & 118 & 4.96 & -3.00  & 406 & 0.35 &  611 & 0.53 & 14\,k &  13 & -1.00 \\
        PACE    & 6.0 & 0.25 & 7.5 & 0.32 &  30 & 1.27 & -3.46  & 210 & 0.18 &  351 & 0.31 & 40\,k &  35 & -1.45 \\
        MACE    & \bf1.6 & \bf0.07 & \bf2.1 & \bf0.09 &  \bf14 & \bf0.57 & \bf-4.56  &  \bf86 & \bf0.08 &  \bf129 & \bf0.11 &  \bf8\,k &   \bf7 & \bf-2.31 \\
        \bottomrule
    \end{tabular}
\end{table}

\section{Conclusions}

This work provides the automated \dataset{} benchmark for \glspl{mlp} that characterises their ability to model the physics of the \gls{llpt} between atomic and molecular \gls{wdh}. 
The benchmark goes beyond test-set error statistics, which do not guarantee physical simulations, yet requires no expert domain knowledge.

The benchmark focuses on an \gls{mlp}'s ability to learn a complex high-dimensional potential-energy surface.
In line with this, the performance ranking of \glspl{mlp} in Table~\ref{tab:resultsmdprop} corresponds to the complexity of the benchmarked \glspl{mlp}.

A limitation of the benchmark is the restriction to a single \gls{md} package.
In particular, for experimental, less mature \glspl{mlp}, a LAMMPS interface might not be available.
Hence, supporting another widely used \gls{md} package, such as the Atomic Simulation Environment (ASE), \cite{amzj2017} would be desirable.

The benchmark could, in principle, be extended to active learning if the computational requirements of the used reference ab-initio method can be limited sufficiently to facilitate benchmarking.
For \gls{dft}, this could be attempted, e.g., via lower-rung functionals and smaller basis sets.

Other challenging aspects of \gls{mlp} development that could benefit from future automated application-driven benchmarks include the ability to model many chemical elements, e.g., high-entropy alloys, and the ability to model long-ranged interactions, e.g., van-der-Waals materials.


\section*{Acknowledgments}\label{sec:acknowledgments}
\addcontentsline{toc}{section}{\nameref{sec:acknowledgments}}

The authors thank 
Marcel F.{} Langer, Michele Ceriotti,
Richard G. Hennig, Stephen R. Xie, Ajinkya Hire, Hendrik Kra{\ss},
Kousuke Nakano, Andrea Tirelli, Giacomo Tenti, Michele Casula, 
Ralf Drautz, G{\'a}bor Cs{\'a}nyi
for constructive discussions
of the benchmark, \gls{wdh}, and specific \glspl{mlp}.

This project has received funding from the European Union's Horizon 2020 research and innovation programme under Grant Agreement No.~952165, European Center of Excellence in Exascale Computing TREX, Targeting Real chemical accuracy at the Exascale.
Calculations were performed at the bwUniCluster~2.0 as part of the high-performance-computing infrastructure of the state of Baden-W\"urttemberg.


\section*{Data and code availability}\label{sec:datacode}
\addcontentsline{toc}{section}{\nameref{sec:datacode}}

The benchmark code and \dataset{} dataset are available at \url{https://gitlab.com/qmml/h-benchmark} under the Apache license~2.0.
The raw data underlying the \dataset{} dataset are available at \url{https://zenodo.org/deposit/8160491}.


\needspace{5\baselineskip}
\addcontentsline{toc}{section}{References}


\providecommand{\BAN}{\,}\providecommand{\BAP}{.\,}\providecommand{\BANE}{}\providecommand{\BAPE}{.}



\appendix
\clearpage
\phantomsection
\addcontentsline{toc}{section}{\texorpdfstring{\rule[-.3\baselineskip]{0pt}{3\baselineskip}\large Supplementary information}{SupplementaryInformation}}

\setcounter{secnumdepth}{2}
\setcounter{figure}{0}
\renewcommand{\thefigure}{SI\hspace*{0.5pt}\arabic{figure}}
\setcounter{table}{0}
\renewcommand{\thetable}{SI\hspace*{0.5pt}\arabic{table}}

\begingroup
    \LARGE
    Supplementary Information:\\[0.5ex]
    Hydrogen under Pressure as a Benchmark\\ for Machine-Learning Interatomic Potentials

    \medskip

    \large
    Thomas Bischoff,$^1$ Bastian J{\"a}ckl,$^1$ Matthias Rupp$^2$

    \normalsize\raggedright
    $^1$~Department of Computer and Information Science, University of Konstanz, Konstanz, Germany\\
    $^2$~Luxembourg Institute of Science and Technology (LIST), Esch-sur-Alzette, Luxembourg
\endgroup

\bigskip


\section{Computational details}\label{si:sec:methods}


\subsection{Molecular dynamics simulations}\label{sec:methods_md}

For the reference \gls{md} simulations, \Gls{dft} calculations were carried out using the \texttt{Quantum-Espresso} \citesi{gbuw2009si} software. 
We used the semilocal \gls{pbe} \citesi{pbe1996si} functional together with a scalar-relativistic, ultrasoft projector-augmented wave (PAW) pseudopotential \citesi{d2014si} for hydrogen. 
We set the energy cutoff for the plane-wave basis to 60 Ry and used a shifted $4 \times 4 \times 4$ $\bm{k}$-point sampling. 
We smeared the occupations of the Kohn-Sham levels \citesi{ks1965si} according to the Fermi-Dirac distribution consistent with the ionic temperature applied in the \gls{md} simulations. 
We verified that the employed numerical parameters are consistent with those in the literature \citesi{gwma2019si, cmpc2020qsi, ttns2022qsi, nypc2023qsi}.

All \gls{md} simulations used cubic supercells.
We determined the cubic lattice constant~$a$ for each calculation according to the definition of the Wigner-Seitz radius~$r_s$. 
Specifically, we set $a = (4\pi r_s^3N/3)^{1/3}$, where $N=128$ is the number of hydrogen atoms in the unit cell. 
We verified that 128 hydrogen atoms are sufficient to identify the \gls{llpt}, in agreement with the literature. \citesi{gwma2019si, ttns2022qsi, khht2021si}.

\gls{md} simulations were carried out using the \texttt{LAMMPS} \citesi{tatp2022si} software in the canonical ensemble ($NVT$).
We set the time step to 0.2\,fs and controlled the temperature via a Bussi-Parrinello thermostat \citesi{bdp2007si}.
The time duration of each \gls{md} simulation was one ps, of which the first 0.5\,ps were used for equilibration, and the remaining 0.5\,ps were used for computing the thermodynamic quantities of interest. 

We verified that this simulation length is sufficient to obtain converged results for purely atomic or purely molecular hydrogen.
In proximity to the \gls{llpt}, we observed non-periodic oscillations between the two phases.
To ensure sufficient thermodynamic averaging for these cases, we performed six independent \gls{md} simulations (of the same duration) for each point in the investigated $(T,r_{s})$ phase space (\cref{fig:dataset_overview}). 

Individual \gls{md} simulations were initialised with distinct starting configurations composed of random atomic coordinates and velocities. 
The average and standard deviation over the six \gls{md} simulations were then considered as the value and statistical uncertainty of the \gls{md}-derived properties. 

Discarding equilibration, the average over six 0.5\,ps \gls{md} simulations is equivalent to the average over one three ps \gls{md} simulation.
However, in the former approach the \gls{md} simulations can be run in parallel. 
In addition, the six uncorrelated starting configurations prevent the hydrogen system from overproportionally remaining in either the atomic or the molecular phase.

The benchmark employs \gls{dft} with the \gls{pbe} functional and classically described nuclei to simulate hydrogen.
Such simulations provide a suitable description of the \gls{llpt} of \gls{wdh}. \citesi{lhr2010si,mpsc2010si,mhs2018si,gwma2019si,cmpc2020qsi,hktc2020si,khht2021si,ttns2022qsi,nypc2023qsi}
However, other electronic-structure methods, such as the local density approximation, the strongly constrained and appropriately normed (SCAN) \citesi{srp2015} functional, or quantum Monte Carlo approaches, yield quantitatively different descriptions of the \gls{llpt} \citesi{gwma2019si,hktc2020si,ttns2022qsi}.
The same holds for neglecting or incorporating the quantum-mechanical nature of the proton. \citesi{mpsc2010si,hktc2020si,nypc2023qsi}
Such effects from the choice of electronic-structure method, while important to better understand the physics of the \gls{llpt} in \gls{wdh}, are not relevant for this benchmark, whose sole purpose is to assess the ability of \glspl{mlp} to reproduce results from a given ground-truth electronic-structure method.

The \gls{mlp}-accelerated \gls{md} simulations follow the \gls{dft} \gls{md} simulations as closely as possible.
They are also carried out using the LAMMPS software and use the same time step, simulation length, equilibration phase, and thermostat.
However, enabled by the \glspl{mlp} computational efficiency, the benchmark runs twelve repetitions for each combination of temperature and Wigner-Seitz radius.


\subsection{Molecular dynamics-derived properties and post-processing}

\begin{figure}[htbp]
    \includegraphics[width=0.5\linewidth]{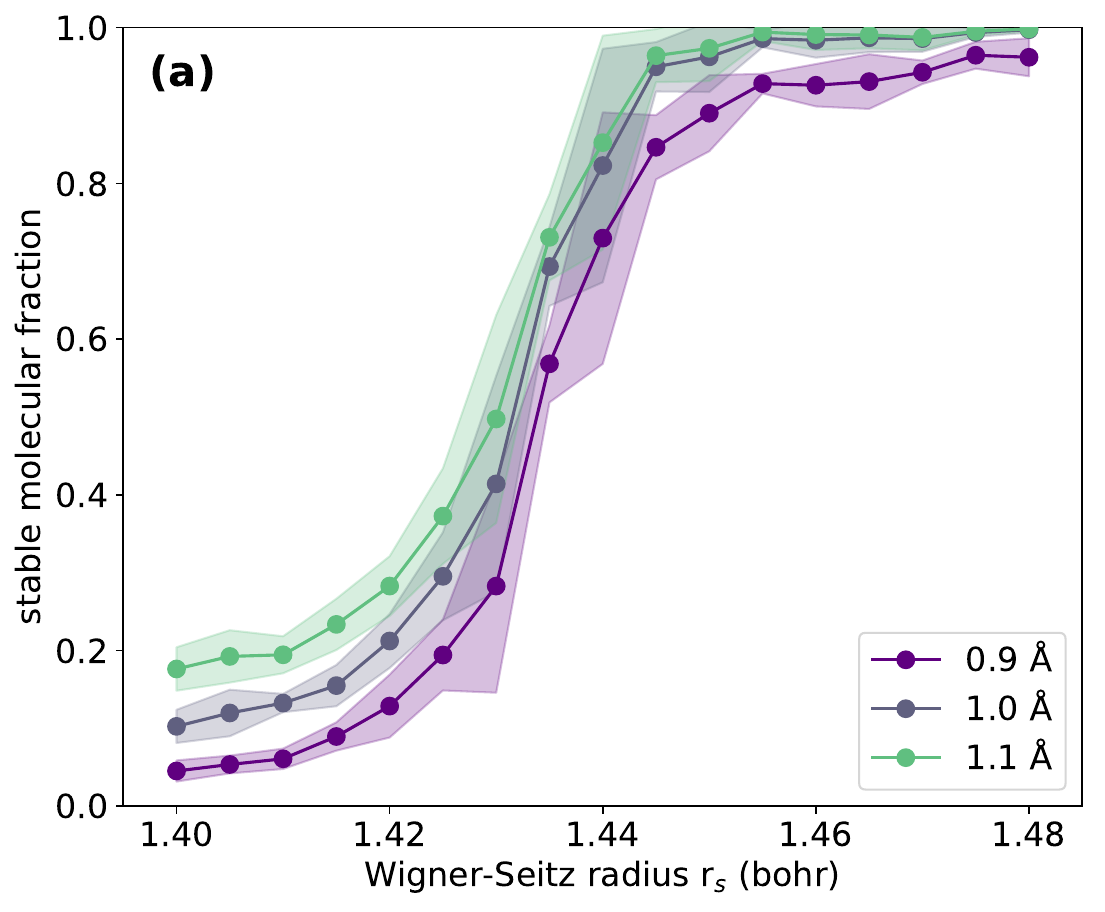}%
    \includegraphics[width=0.5\linewidth]{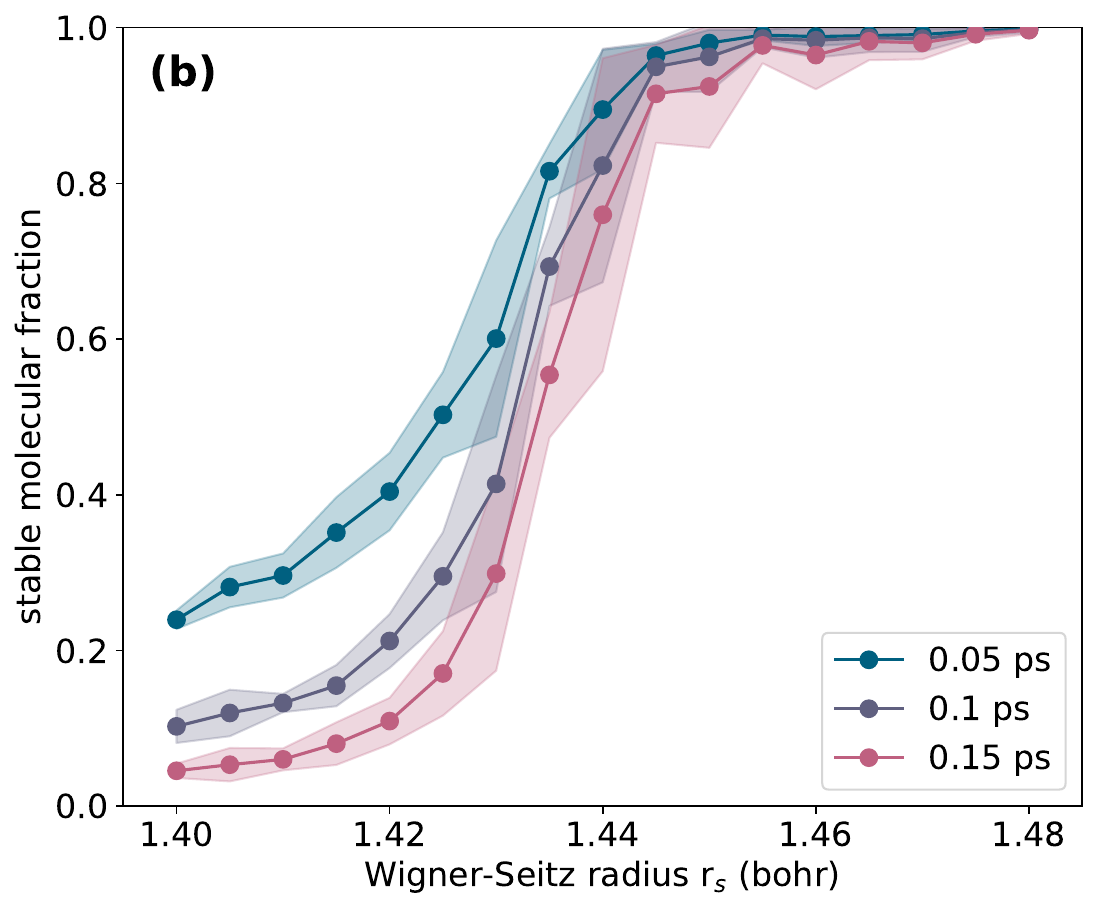}
    
    \caption{%
        \emph{Sensitivity of stable molecular fractions} to parameter choices.
        Shown are stable molecular fractions as a function of cell density for different choices of spatial (left) and temporal cutoff (right).
    }
    \label{si:fig:smfsensitivity}
\end{figure}

\Cref{si:fig:smfsensitivity} shows the influence of parameter choices on calculated stable molecular fractions~$\mu$.
Neither the spatial cutoff nor the temporal cutoff qualitatively change the isothermal $\mu$ over $r_s$ curves.

\subsection{Training of machine learning interatomic potentials}

The \textbf{Yukawa} and \textbf{Tersoff} traditional empirical potentials were optimised using the Nelder-Mead simplex algorithm \citesi{nm1965si} as implemented in the SciPy \citesi{vghv2020si} package. 
The \gls{md} simulations were performed with the standard implementations of these potentials in LAMMPS.

\textbf{Ultra-fast potentials} (UFP2, UFP3) were trained with the Python code available at \url{https://github.com/uf3/uf3}.
For two-body terms, we used a radial cutoff of 3.5\,\AA{} and six knots/\AA; for three-body terms, we used a cutoff of 2\,\AA{} and four knots/\AA.
The behaviour at the limits of the knot intervals we defined by setting the parameters \texttt{leading\_trim=0} and \texttt{trailing\_trim=3}. 
The energy/force weighting parameter we set to 0.25.
For regularisation, we set \texttt{ridge\_1b=1e-5}, \texttt{curvature\_1b=0}, \texttt{ridge\_2b=1e-6}, \texttt{curvature\_2b=1e-5}, \texttt{ridge\_3b=1e-6}, and \texttt{curvature\_3b=1e-8}.
The \gls{md} simulations were performed with the ML-UF3 plugin for LAMMPS, available at \url{https://github.com/monk-04/uf3/tree/lammps_implementation/lammps_plugin}.

\textbf{PACE} was trained using the PACEmaker code \citesi{lood2021qsi,blmd2022q} documented at \url{https://pacemaker.readthedocs.io} using the settings below.
The \gls{md} simulations were performed with the ML-PACE plugin for LAMMPS, available at \url{https://github.com/ICAMS/lammps-user-pace}.

\begin{minipage}{0.5\linewidth-\tabcolsep}
\begin{verbatim}
potential:
  deltaSplineBins: 0.001
  elements: [H]
  embeddings:
    ALL: { 
      npot: 'FinnisSinclairShiftedScaled', 
      fs_parameters: [ 1, 1, 1, 0.5], 
      ndensity: 2 
    }
  bonds:
    ALL: {
      radbase: ChebExpCos, 
      radparameters: [ 5.25 ],
      rcut: 3.0,
      dcut: 0.01,
      NameOfCutoffFunction: cos,
    }
  functions:
    ALL: {
      nradmax_by_orders: [150,20,4,2,2],
      lmax_by_orders: [ 0,5,3,2,2],
    }    
\end{verbatim}    
\end{minipage}%
\hfill\vline\hfill%
\begin{minipage}{0.5\linewidth-3\tabcolsep}
\begin{verbatim}
cutoff: 3.0 
seed: 7

fit:
  loss: 
    kappa: 0.001
    L1_coeffs: 0.00
    L2_coeffs: 0.00
    w1_coeffs: 0
    w2_coeffs: 0
    w0_rad: 0
    w1_rad: 0
    w2_rad: 0
  optimizer: BFGS
  maxiter: 10000
  fit_cycles: 2

backend:
  evaluator: tensorpot
  batch_size: 50   
\end{verbatim}
\end{minipage}

\bigskip

\textbf{MACE} was trained using the ACEsuit code available at \url{https://github.com/ACEsuit/mace}, using the command string

\texttt{%
python ./mace/scripts/run\_train.py    --name="ScaleShiftMACE"    --model=MACE \\   --train\_file="train.xyz"    --valid\_fraction=0.05    --test\_file="test.xyz" \\   --E0s='{1:-13.663181292231226}'    --model="ScaleShiftMACE"   \\ --hidden\_irreps='128x0e + 128x1o + 128x2e'    --r\_max=3.5    --batch\_size=8   \\ --max\_num\_epochs=80    --ema    --ema\_decay=0.99    --amsgrad    --default\_dtype="float64"    \\ --device=cuda    --seed=1 --swa}

The \gls{md} simulations were performed with the MACE plugin for LAMMPS described at \url{https://mace-docs.readthedocs.io}, section ``MACE in LAMMPS''.


\section{Analysis details}\label{si:sec:results}


\subsection{Derived properties}\label{si:sec:properties}

\Cref{si:fig:properties} presents \gls{md}-derived properties for all benchmarked \glspl{mlp}.

\begin{figure}[hbtp]
    \centering

    \begin{subfigure}{0.95\linewidth}
        \includegraphics[width=0.44\linewidth,trim={5pt 5pt 5pt 5pt},clip]{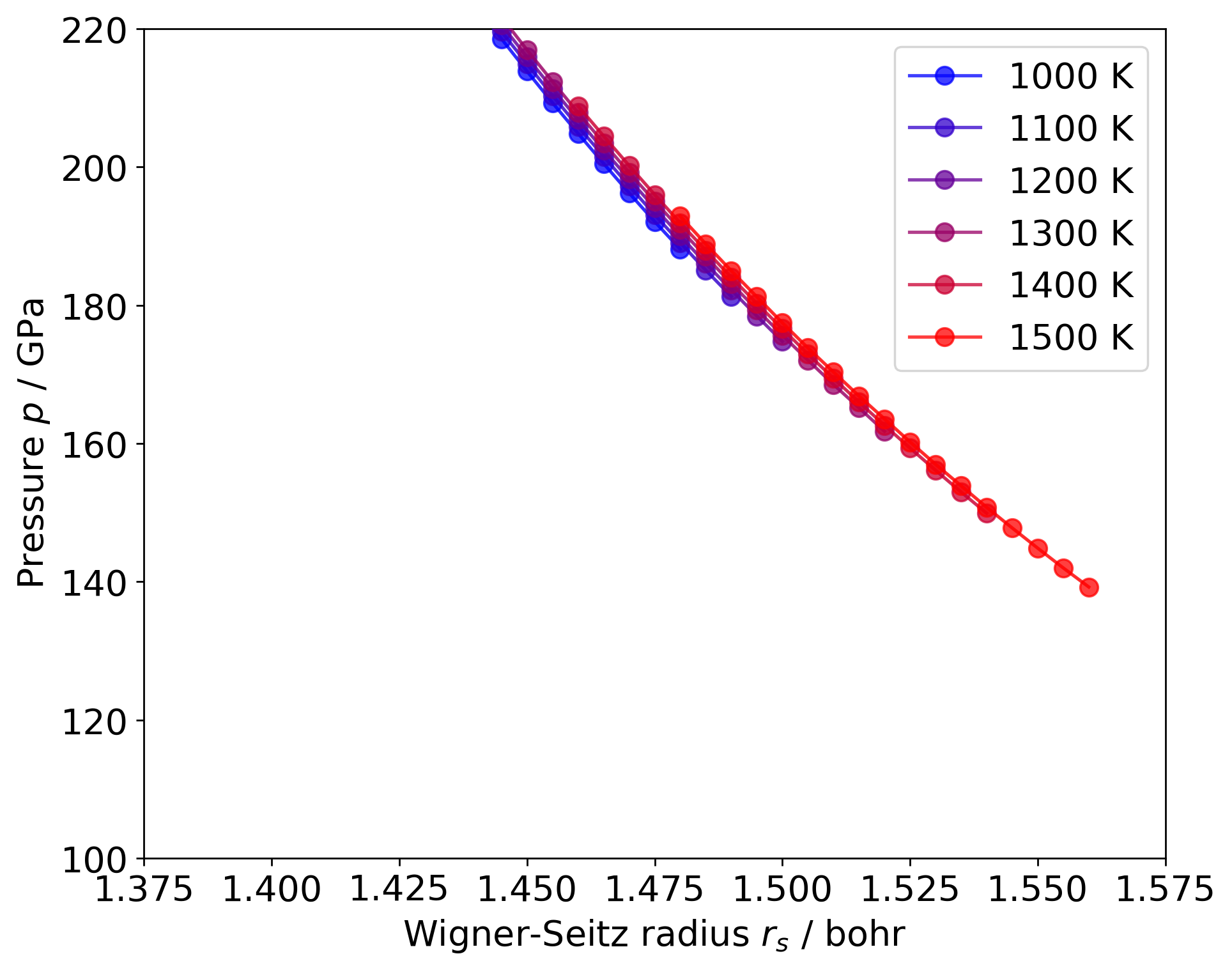}%
        \quad\!\!%
        \includegraphics[width=0.44\linewidth,trim={5pt 5pt 5pt 5pt},clip]{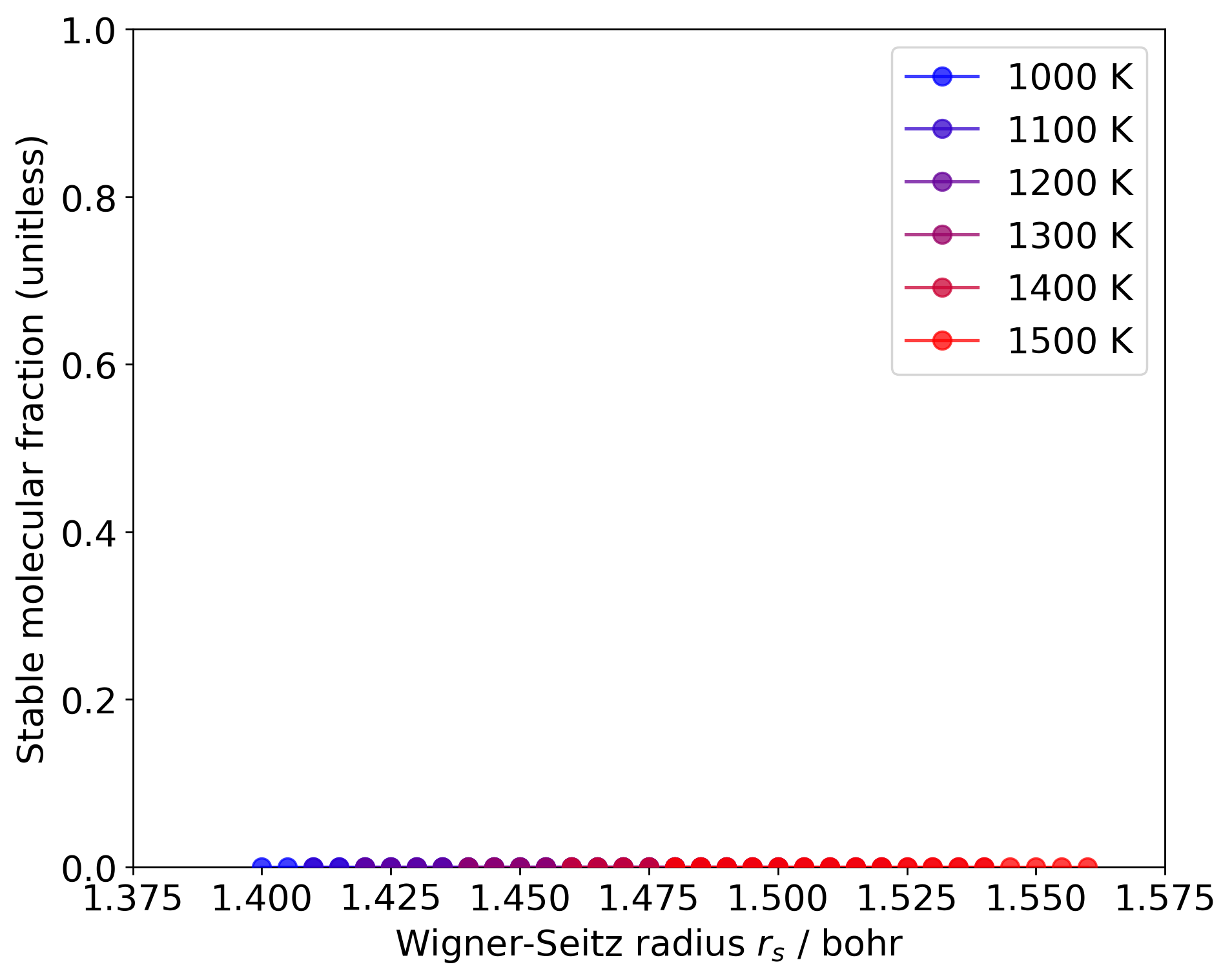}
    
        \includegraphics[width=0.44\linewidth,trim={5pt 5pt 5pt 5pt},clip]{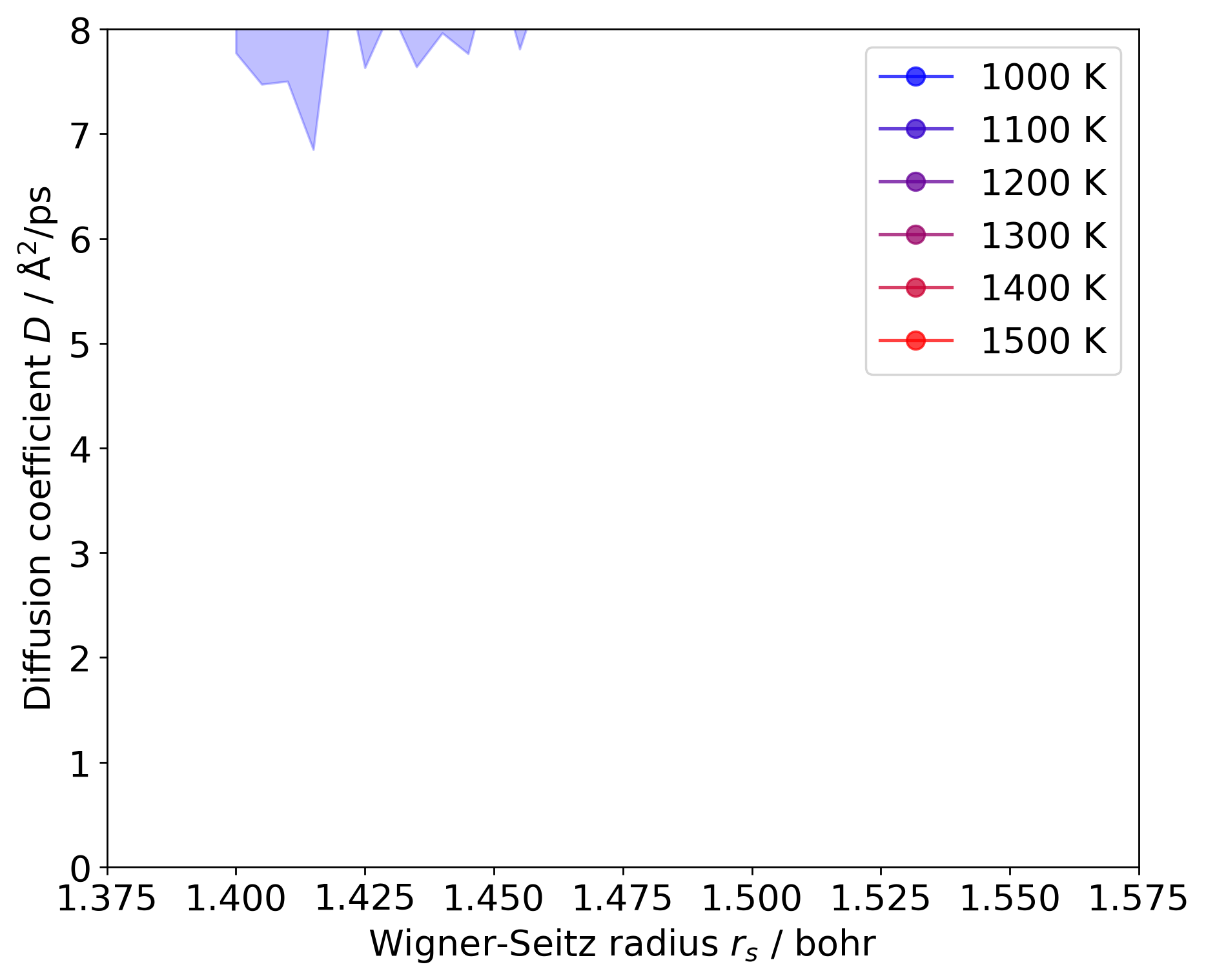}
        \includegraphics[width=0.55\linewidth,trim={5pt 5pt 5pt 5pt},clip]{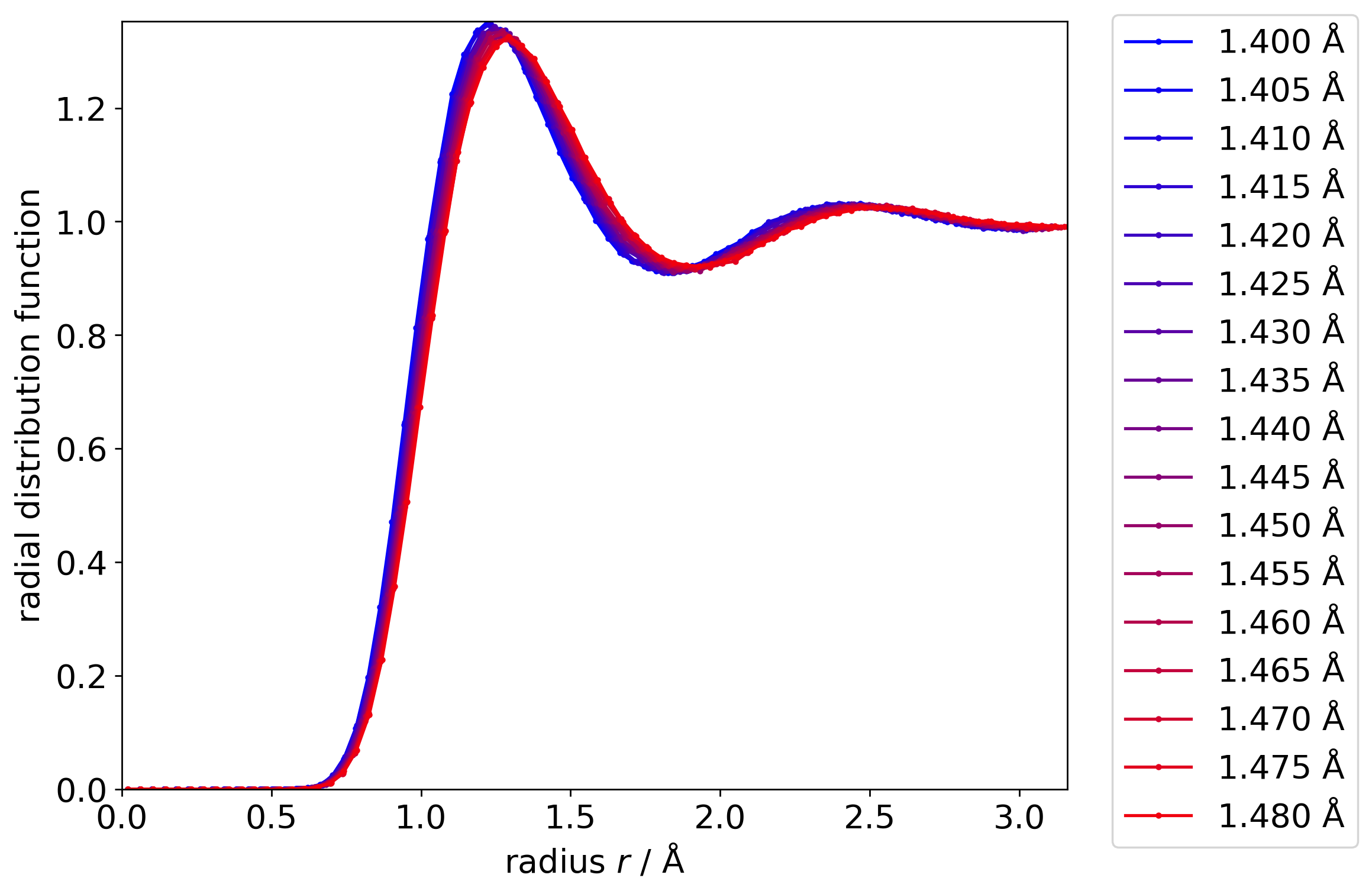}
        
        \subcaption{Yukawa potential.}
        \label{si:fig:properties:yukawa}
    \end{subfigure}

    \begin{subfigure}{0.95\linewidth}
        \includegraphics[width=0.44\linewidth,trim={5pt 5pt 5pt 5pt},clip]{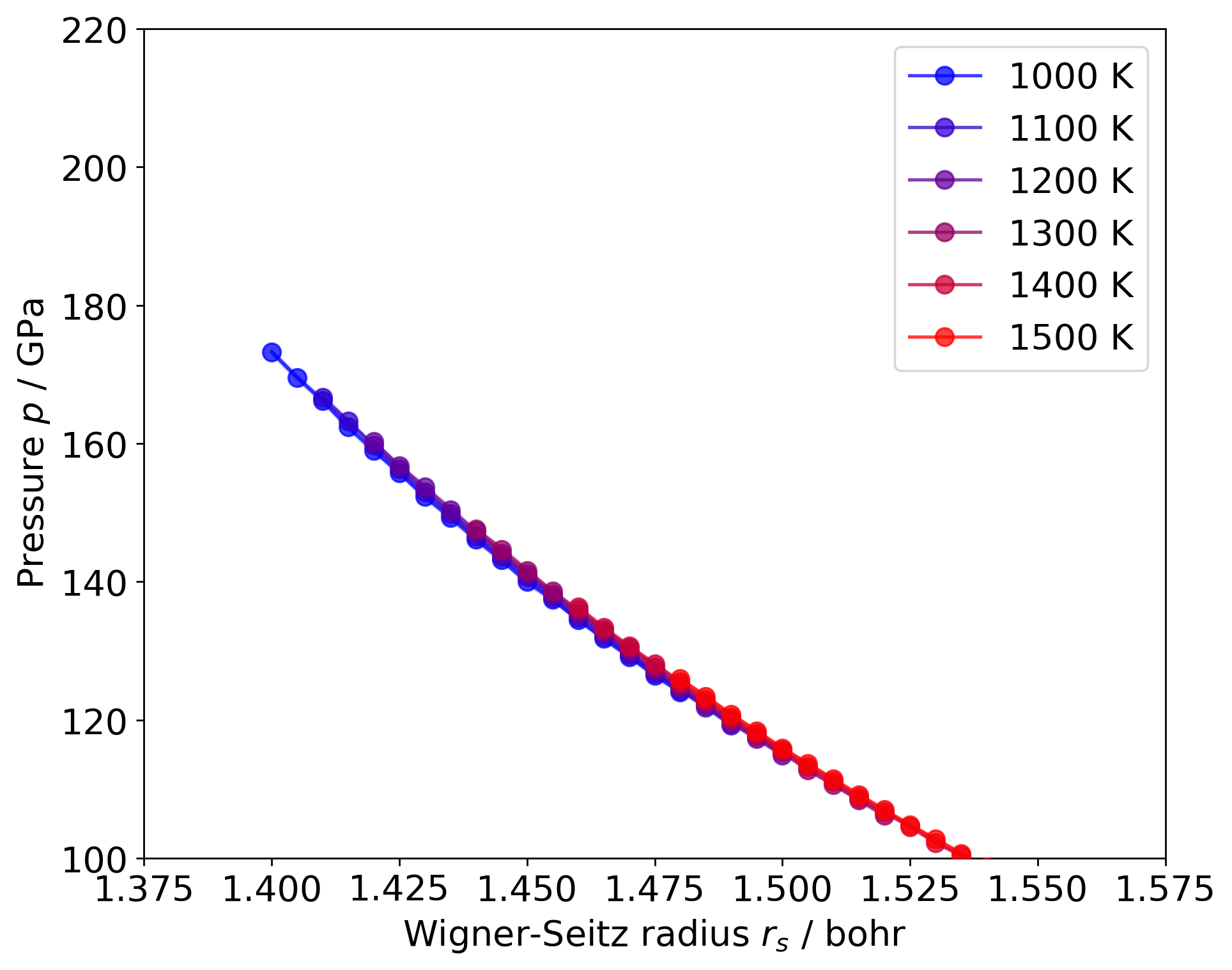}%
        \quad\!\!%
        \includegraphics[width=0.44\linewidth,trim={5pt 5pt 5pt 5pt},clip]{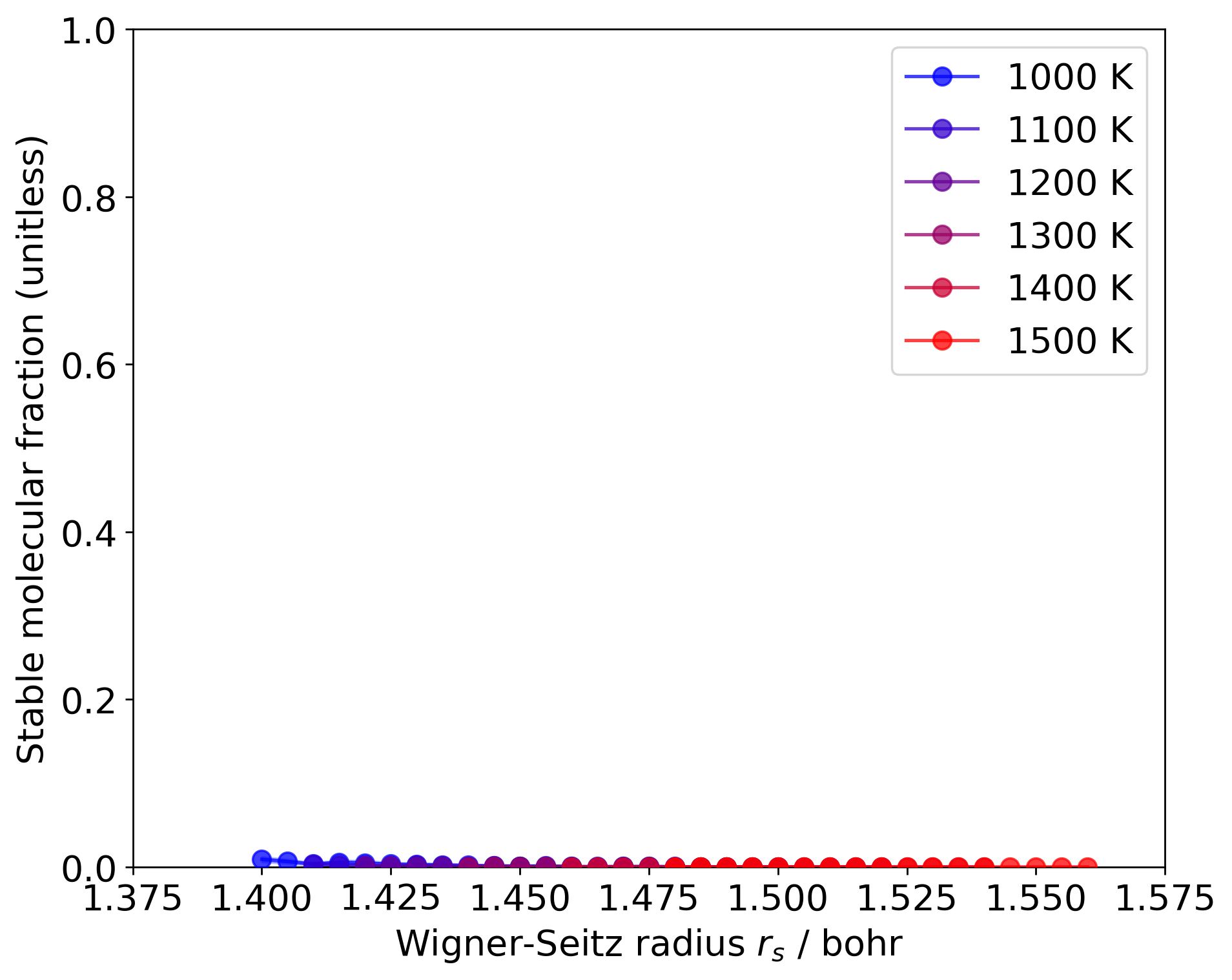}
    
        \includegraphics[width=0.44\linewidth,trim={5pt 5pt 5pt 5pt},clip]{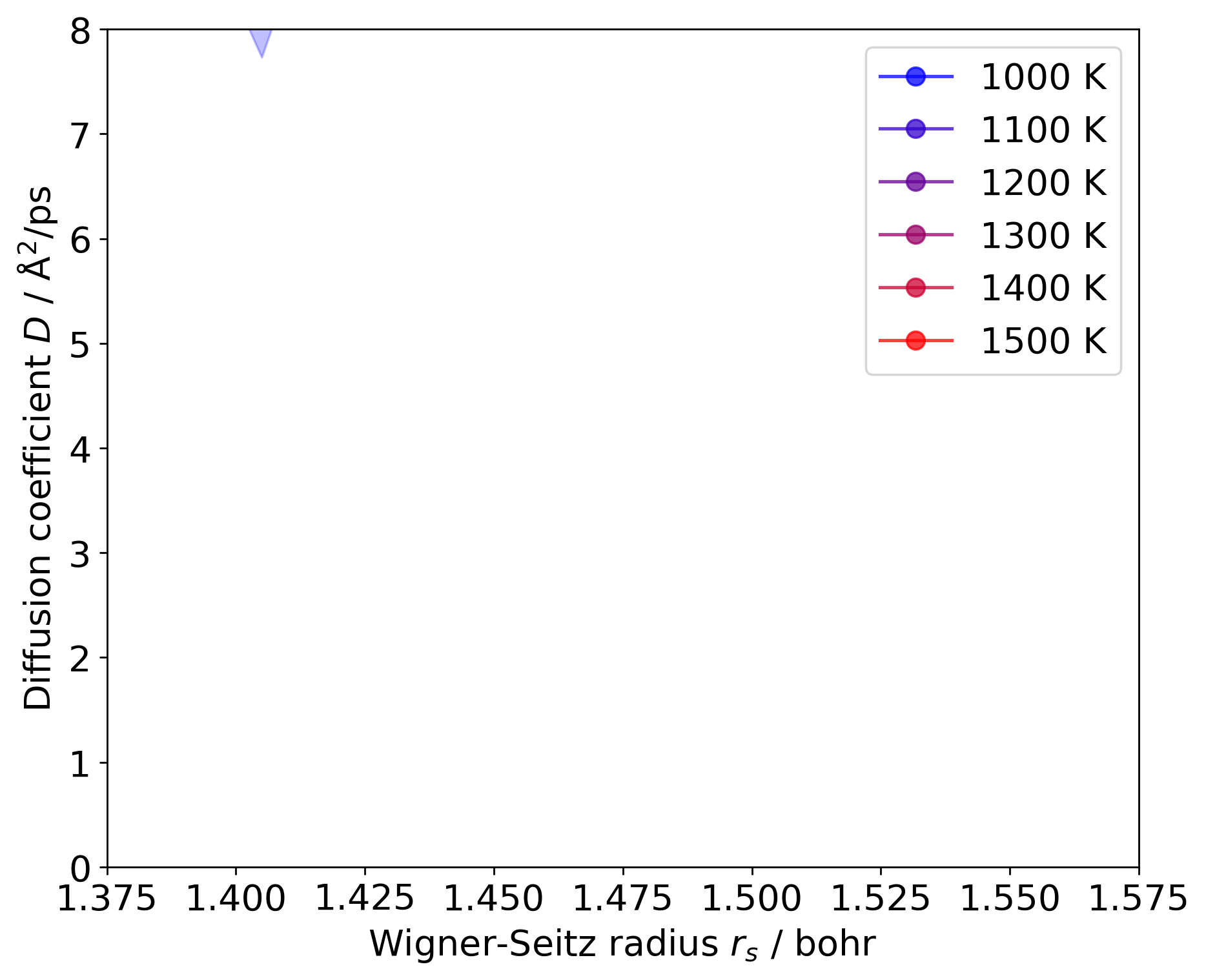}
        \includegraphics[width=0.55\linewidth,trim={5pt 5pt 5pt 5pt},clip]{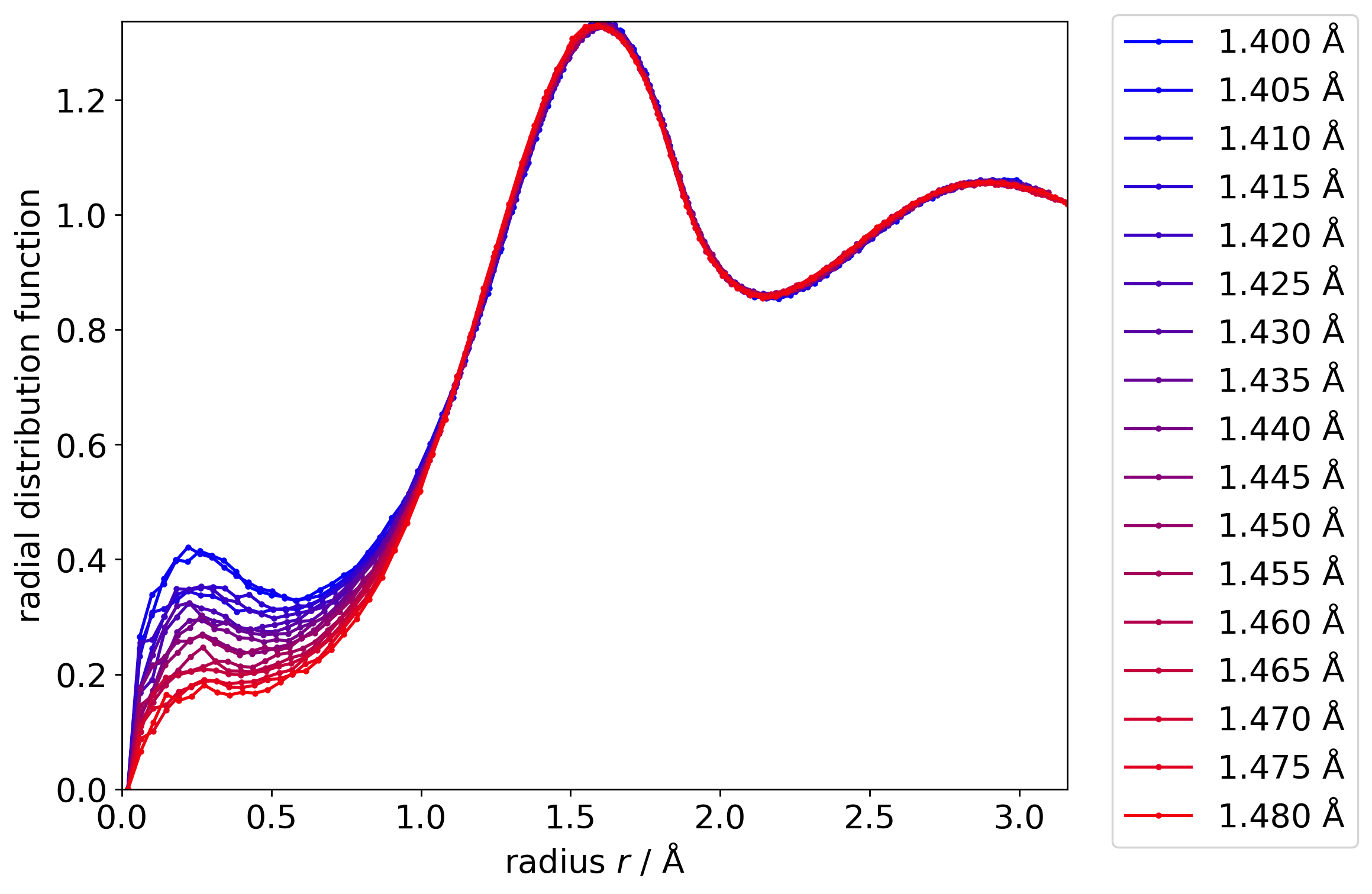}
        
        \subcaption{Tersoff potential.}
        \label{si:fig:properties:tersoff}
    \end{subfigure}

    \caption{%
        \emph{Derived properties}.
        Shown are pressure (top left), stable molecular fraction (top right), and diffusion coefficient (bottom left) as a function of the Wigner-Seitz radius~$r_s$ and for different temperatures (color-coded), as well as radial distribution functions (bottom right) at 1000\,K for different Wigner-Seitz radii (color-coded), for all benchmarked \glspl{mlp} (\subref{si:fig:properties:yukawa}--\subref{si:fig:properties:mace}).
        See \cref{sec:properties} for details.
    }
\end{figure}

\begin{figure}[hbtp]\ContinuedFloat
    \centering

    \begin{subfigure}{0.95\linewidth}
        \includegraphics[width=0.44\linewidth,trim={5pt 5pt 5pt 5pt},clip]{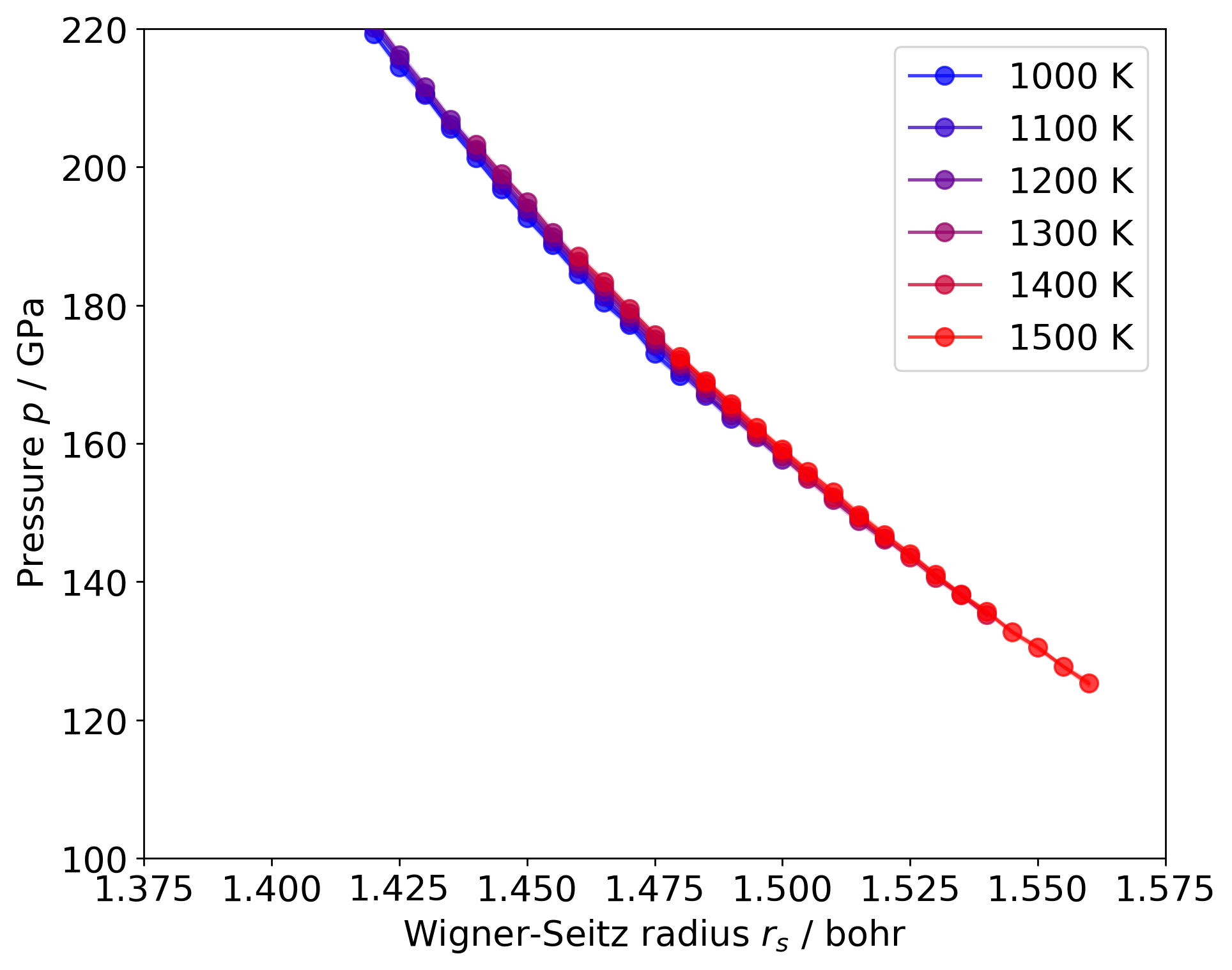}%
        \quad\!\!%
        \includegraphics[width=0.44\linewidth,trim={5pt 5pt 5pt 5pt},clip]{si_fig_properties_uf2-tb23_smf-rs}
    
        \includegraphics[width=0.44\linewidth,trim={5pt 5pt 5pt 5pt},clip]{si_fig_properties_uf2-tb23_dc-rs}
        \includegraphics[width=0.55\linewidth,trim={5pt 5pt 5pt 5pt},clip]{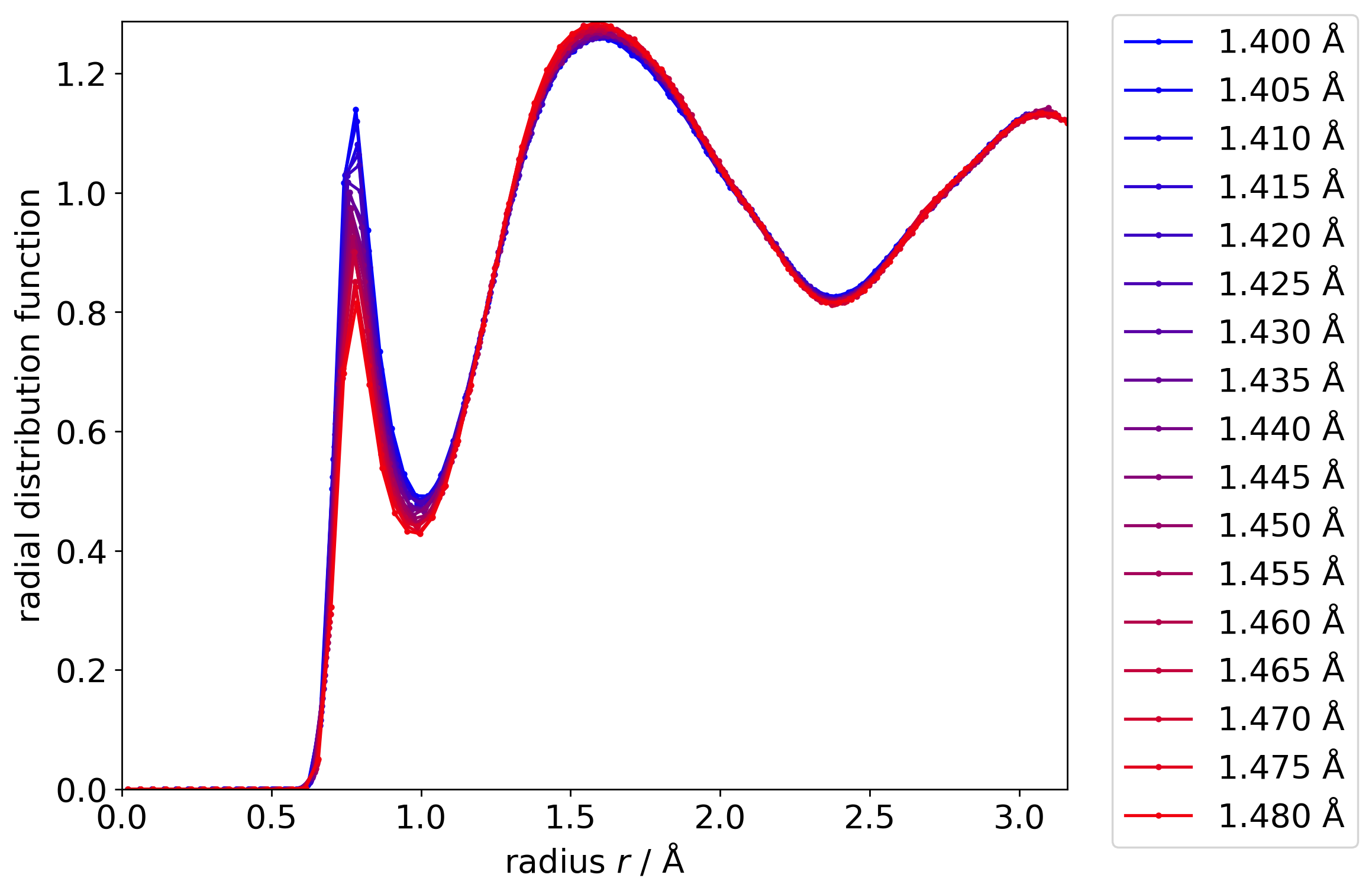}
        
        \subcaption{UFP2 potential.}
        \label{si:fig:properties:ufptwo}
    \end{subfigure}

    \begin{subfigure}{0.95\linewidth}
        \includegraphics[width=0.44\linewidth,trim={5pt 5pt 5pt 5pt},clip]{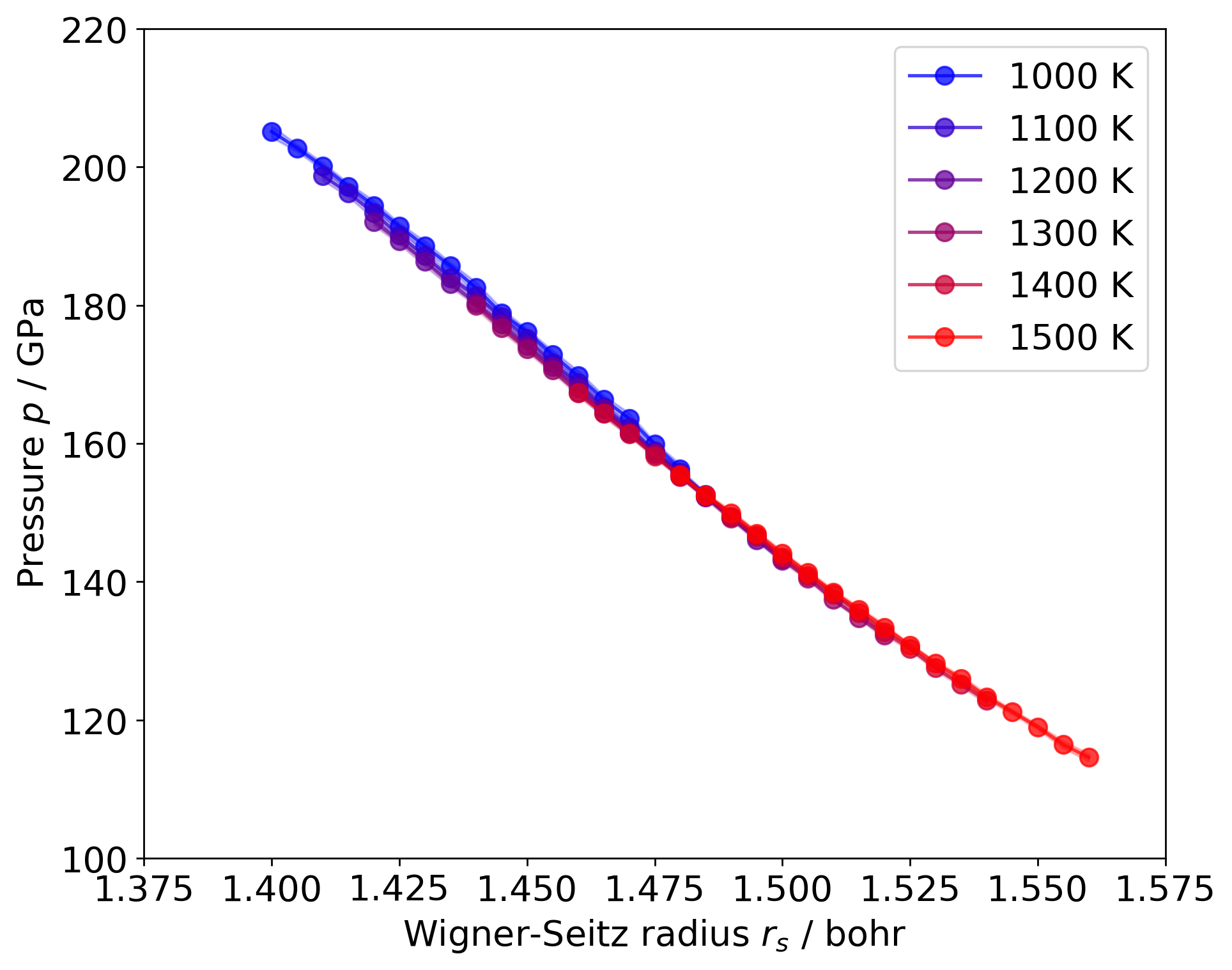}%
        \quad\!\!%
        \includegraphics[width=0.44\linewidth,trim={5pt 5pt 5pt 5pt},clip]{si_fig_properties_uf3-tb23_smf-rs}
    
        \includegraphics[width=0.44\linewidth,trim={5pt 5pt 5pt 5pt},clip]{si_fig_properties_uf3-tb23_dc-rs}
        \includegraphics[width=0.55\linewidth,trim={5pt 5pt 5pt 5pt},clip]{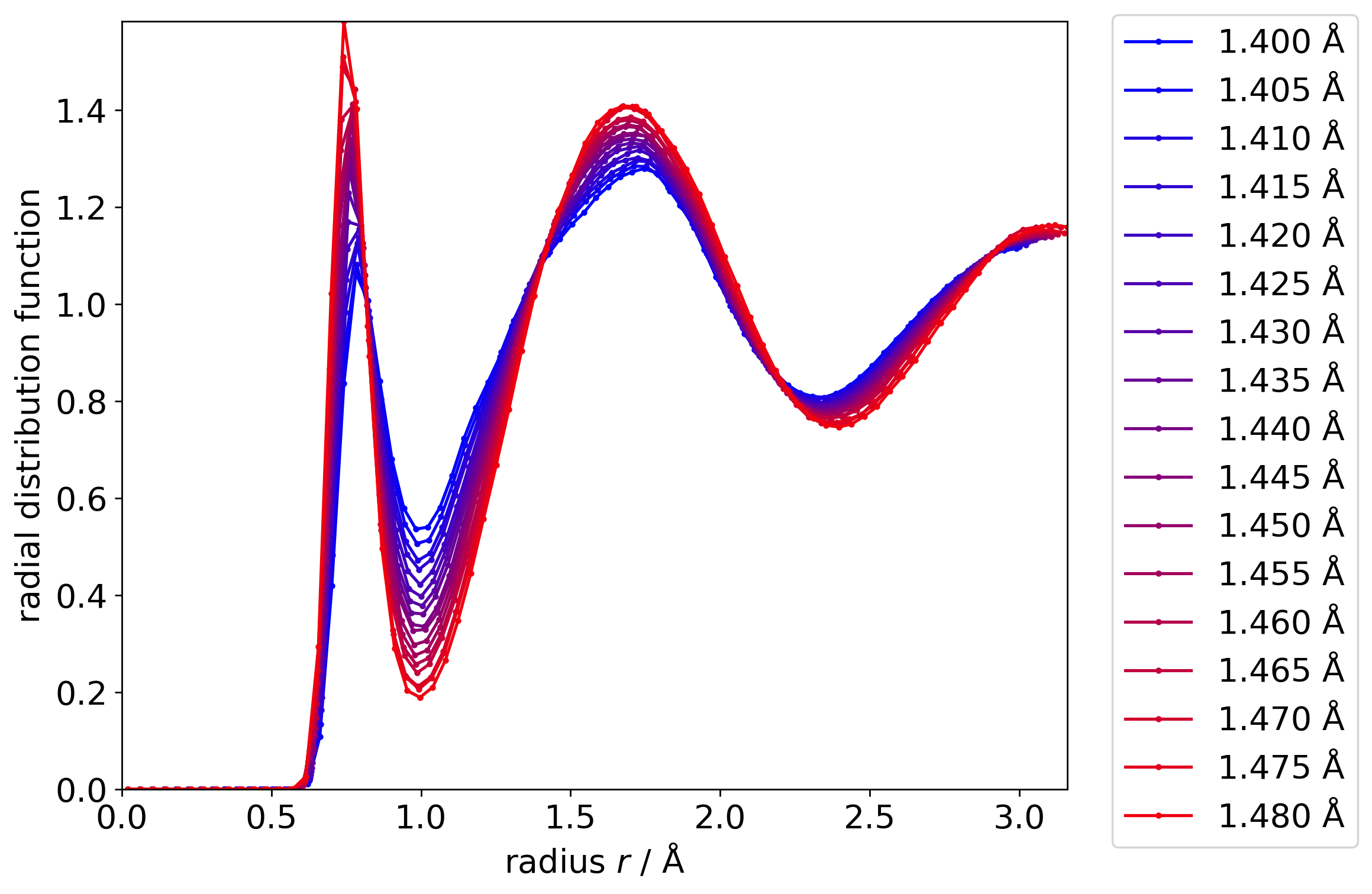}
        
        \subcaption{UFP3 potential.}
        \label{si:fig:properties:ufpthree}
    \end{subfigure}

    \caption{%
        \emph{Derived properties} (continued).
        Shown are pressure (top left), stable molecular fraction (top right), and diffusion coefficient (bottom left) as a function of the Wigner-Seitz radius~$r_s$ and for different temperatures (color-coded), as well as radial distribution functions (bottom right) at 1000\,K for different Wigner-Seitz radii (color-coded), for all benchmarked \glspl{mlp} (\subref{si:fig:properties:yukawa}--\subref{si:fig:properties:mace}).
        See \cref{sec:properties} for details.
    }
\end{figure}

\begin{figure}[hbtp]\ContinuedFloat
    \centering

    \begin{subfigure}{0.95\linewidth}
        \includegraphics[width=0.44\linewidth,trim={5pt 5pt 5pt 5pt},clip]{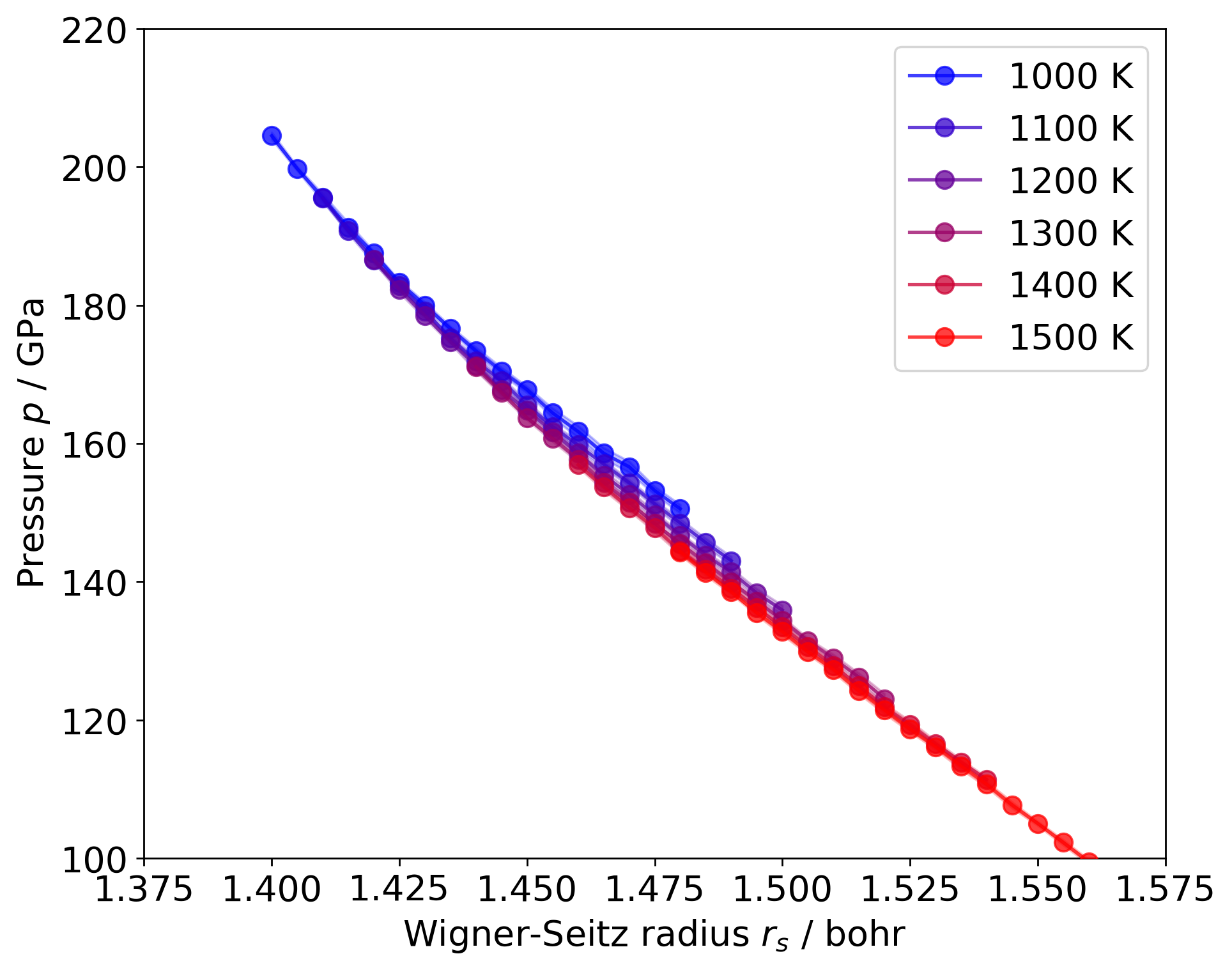}%
        \quad\!\!%
        \includegraphics[width=0.44\linewidth,trim={5pt 5pt 5pt 5pt},clip]{si_fig_properties_pace-bj23_smf-rs}
    
        \includegraphics[width=0.44\linewidth,trim={5pt 5pt 5pt 5pt},clip]{si_fig_properties_pace-bj23_dc-rs}
        \includegraphics[width=0.55\linewidth,trim={5pt 5pt 5pt 5pt},clip]{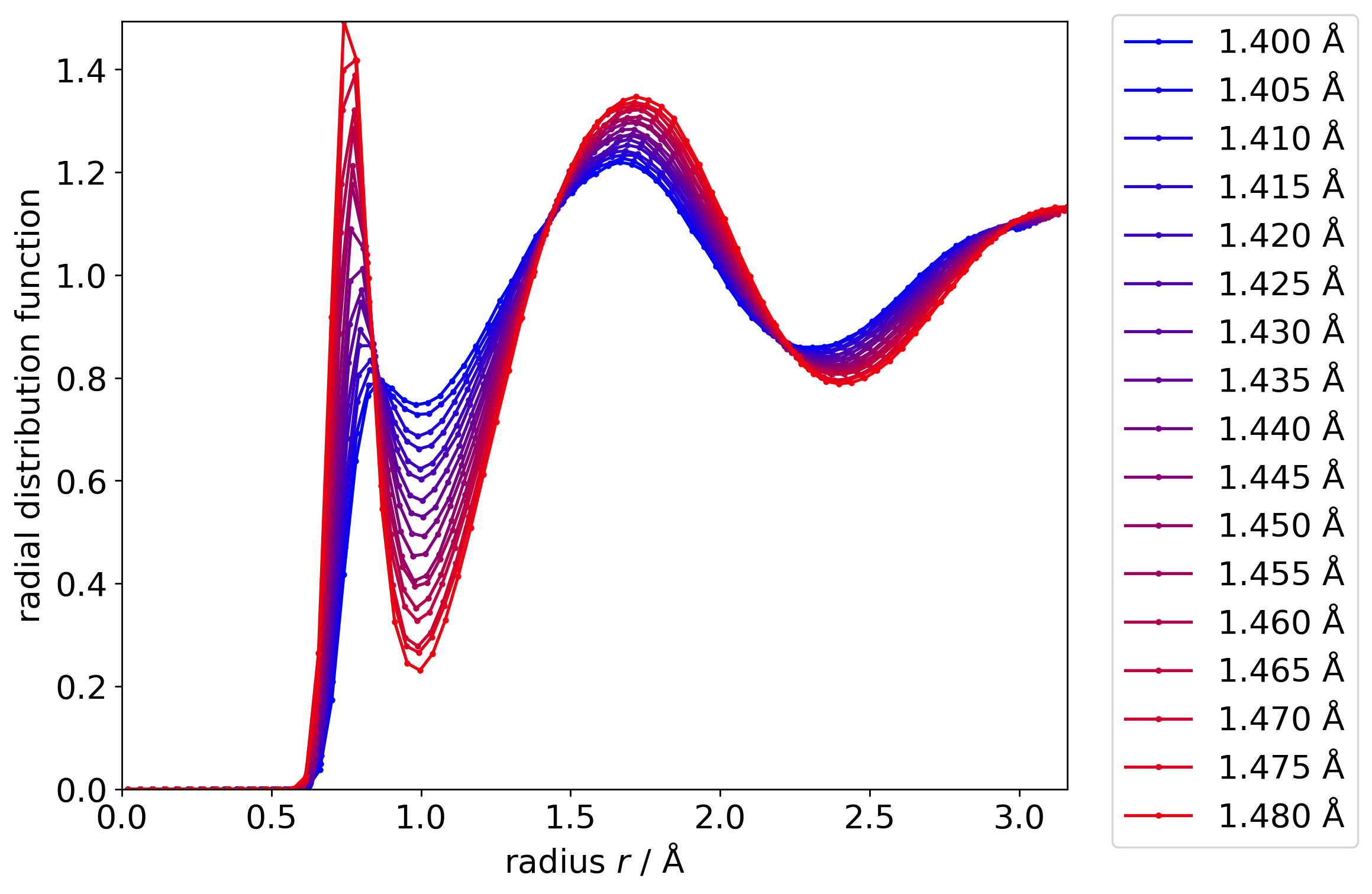}
        
        \subcaption{PACE potential.}
        \label{si:fig:properties:pace}
    \end{subfigure}

    \begin{subfigure}{0.95\linewidth}
        \includegraphics[width=0.44\linewidth,trim={5pt 5pt 5pt 5pt},clip]{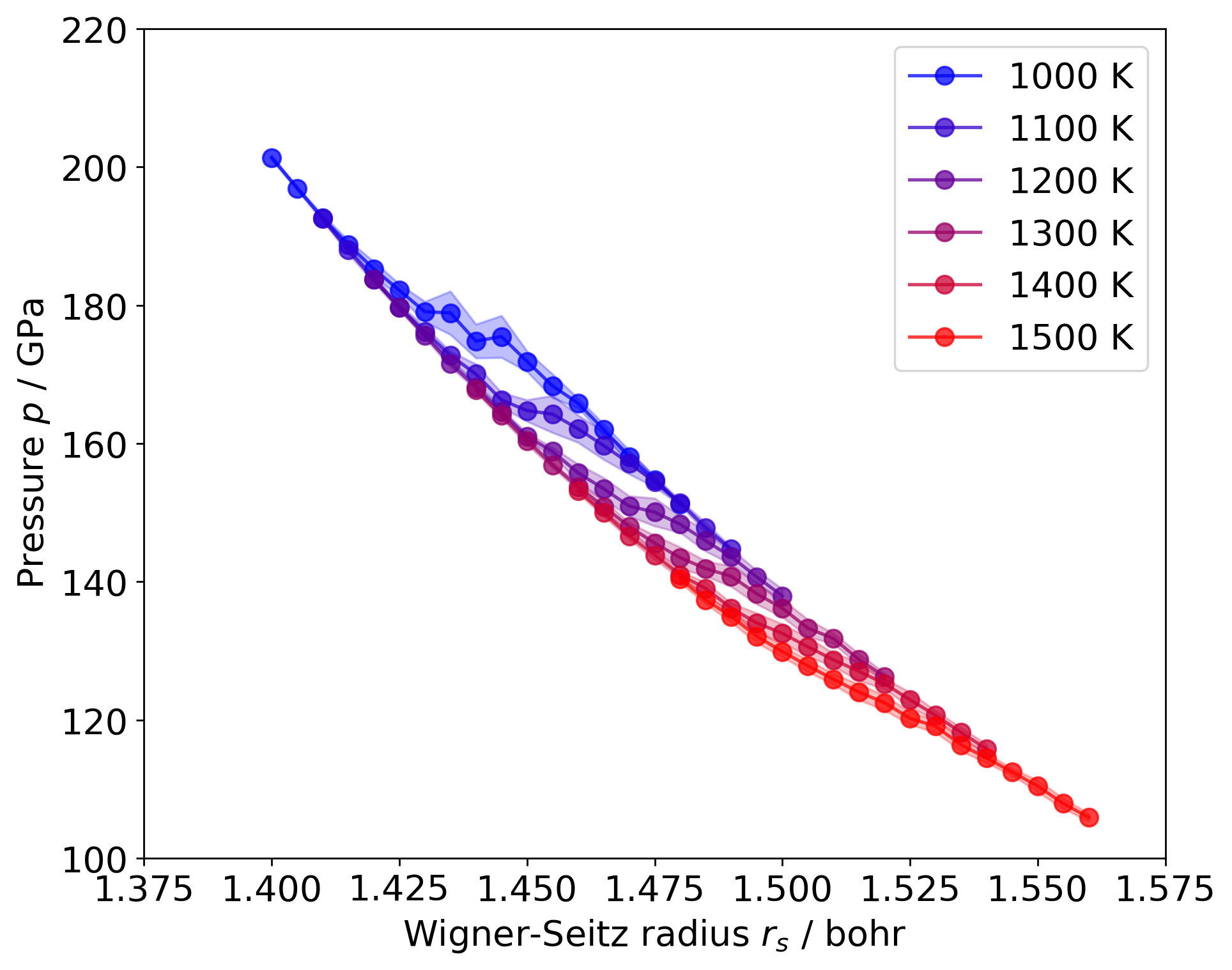}%
        \quad\!\!%
        \includegraphics[width=0.44\linewidth,trim={5pt 5pt 5pt 5pt},clip]{si_fig_properties_mace-bj23_smf-rs}
    
        \includegraphics[width=0.44\linewidth,trim={5pt 5pt 5pt 5pt},clip]{si_fig_properties_mace-bj23_dc-rs}
        \includegraphics[width=0.55\linewidth,trim={5pt 5pt 5pt 5pt},clip]{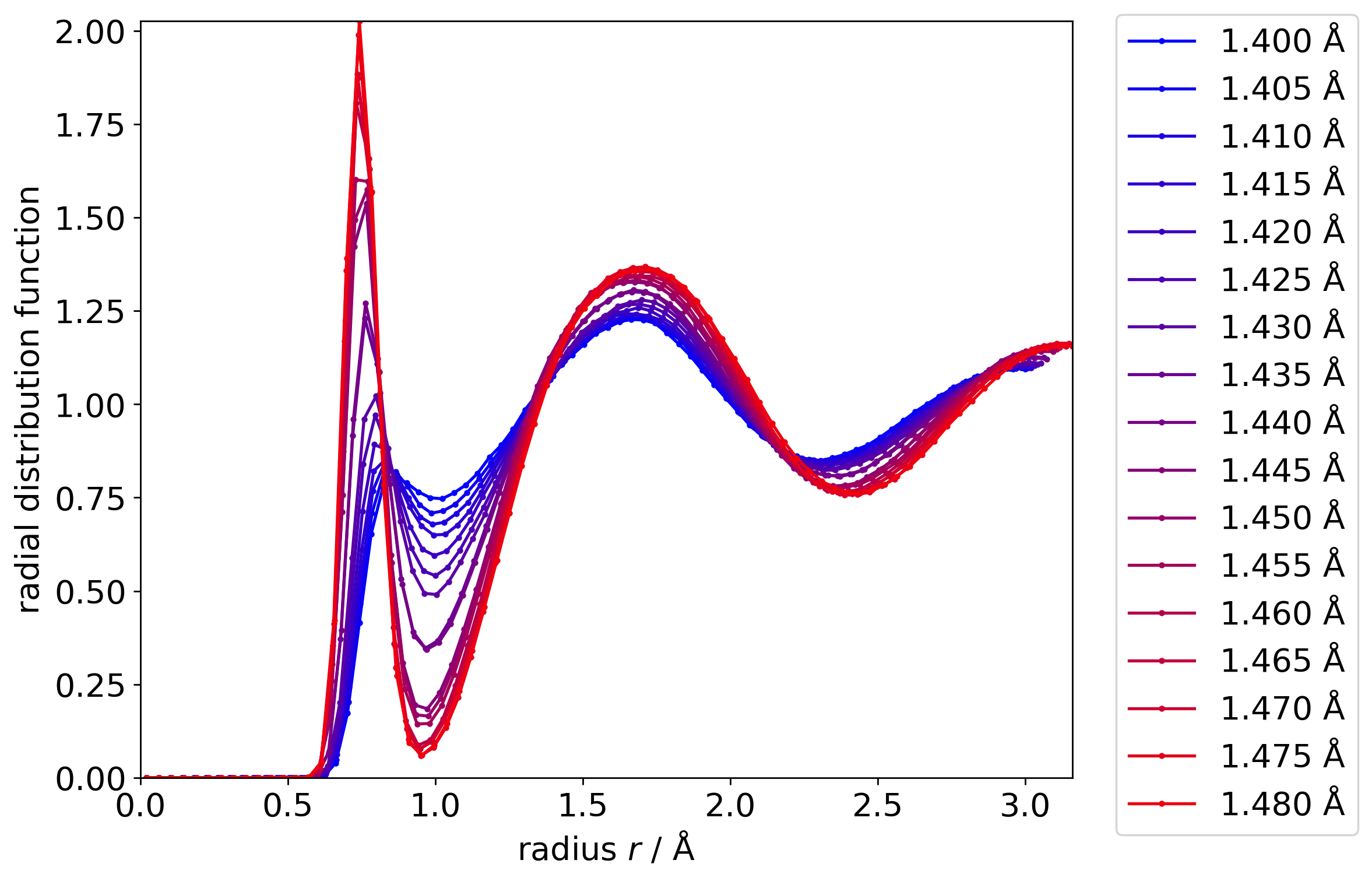}
        
        \subcaption{MACE potential.}
        \label{si:fig:properties:mace}
    \end{subfigure}

    \caption{%
        \emph{Derived properties} (continued).
        Shown are pressure (top left), stable molecular fraction (top right), and diffusion coefficient (bottom left) as a function of the Wigner-Seitz radius~$r_s$ and for different temperatures (color-coded), as well as radial distribution functions (bottom right) at 1000\,K for different Wigner-Seitz radii (color-coded), for all benchmarked \glspl{mlp} (\subref{si:fig:properties:yukawa}--\subref{si:fig:properties:mace}).
        See \cref{sec:properties} for details.
    }
    \label{si:fig:properties}
\end{figure}


\subsection{Scatter plots}\label{si:sec:scatter}

\Cref{si:fig:scatterplots} presents scatter plots showing test-set predictions for all benchmarked \glspl{mlp}.

\begin{figure}[hbtp]
    \centering

    \begin{subfigure}{0.95\linewidth}
        \includegraphics[width=\linewidth]{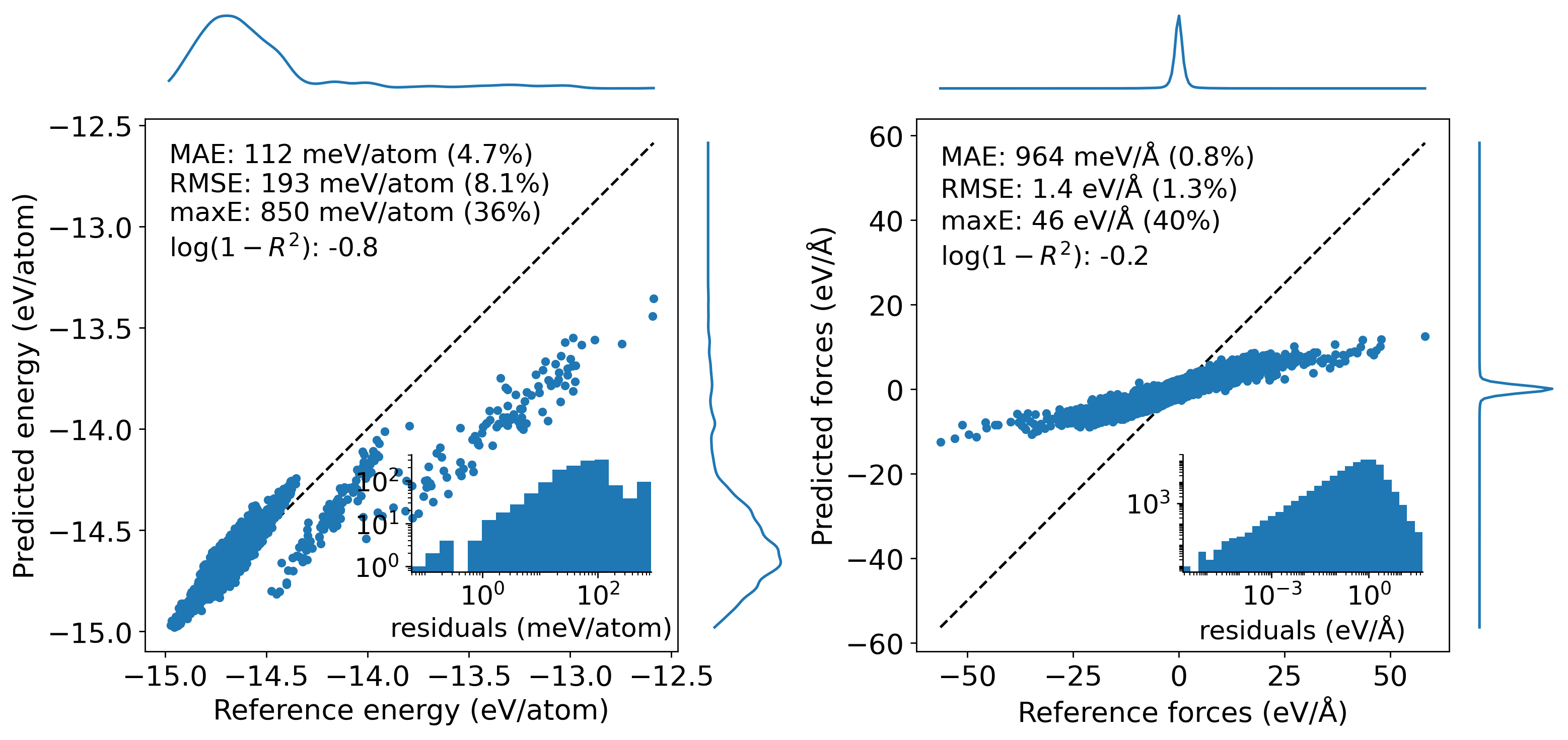}    
        \subcaption{Yukawa potential.}
        \label{si:fig:scatter:yukawa}
    \end{subfigure}

    \begin{subfigure}{0.95\linewidth}
        \includegraphics[width=\linewidth]{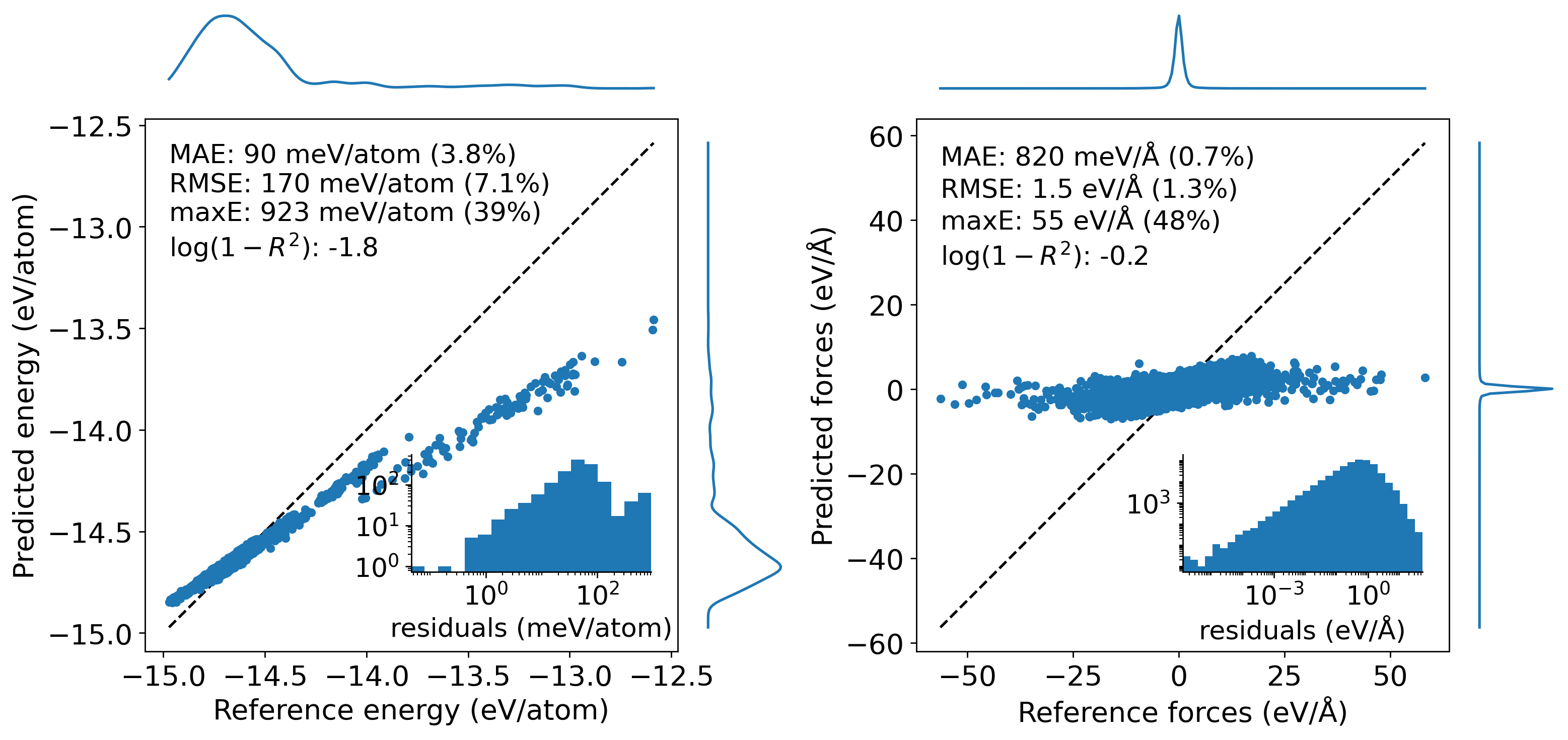}
        \subcaption{Tersoff potential.}
        \label{si:fig:scatter:tersoff}
    \end{subfigure}

    \begin{subfigure}{0.95\linewidth}
        \includegraphics[width=\linewidth]{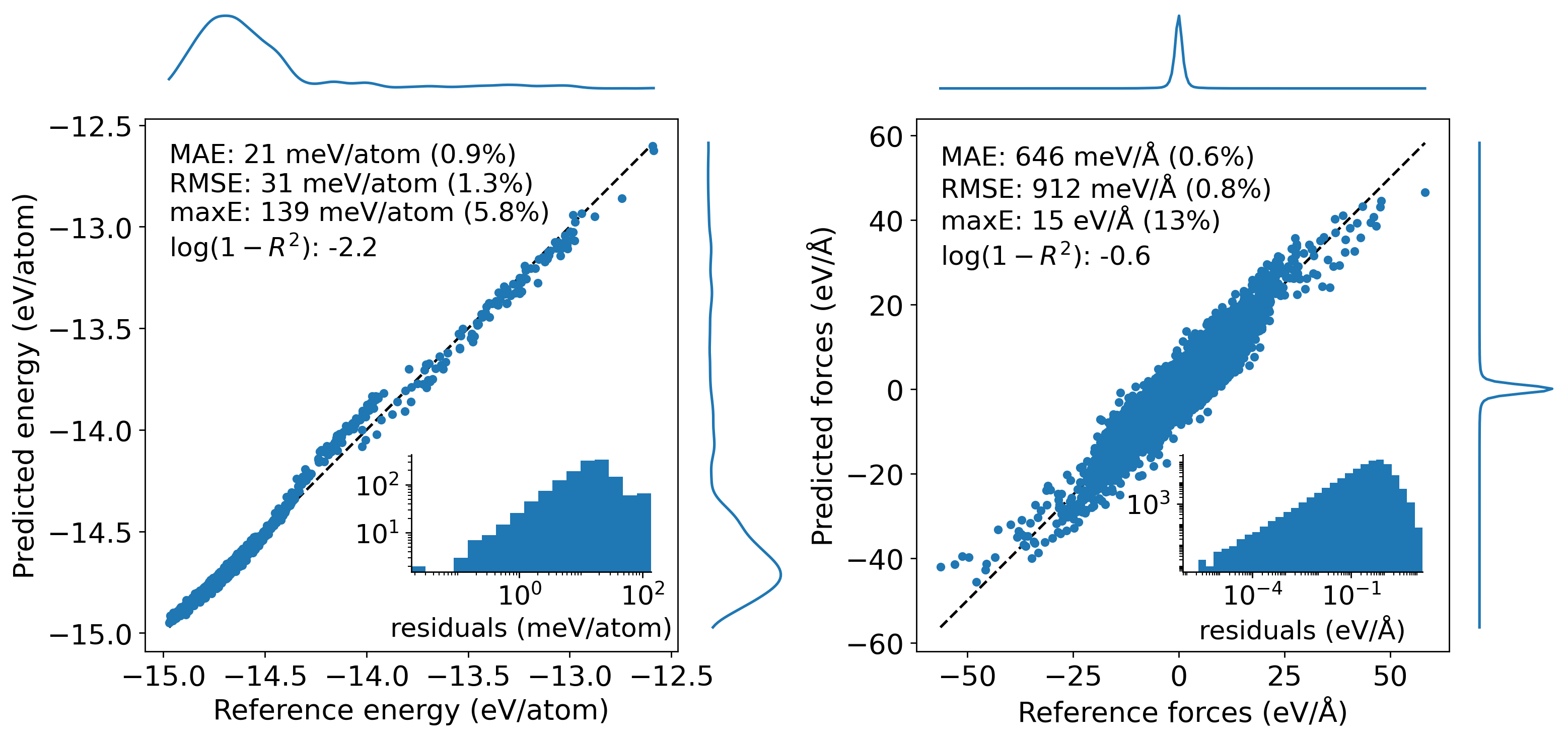}
        \subcaption{UFP2 potential.}
        \label{si:fig:scatter:uftwo}
    \end{subfigure}

    \caption{%
        \emph{Scatter plots}.
        See main text for explanation.
    }
\end{figure}

\begin{figure}[hbtp]\ContinuedFloat
    \centering

    \begin{subfigure}{0.95\linewidth}
        \includegraphics[width=\linewidth]{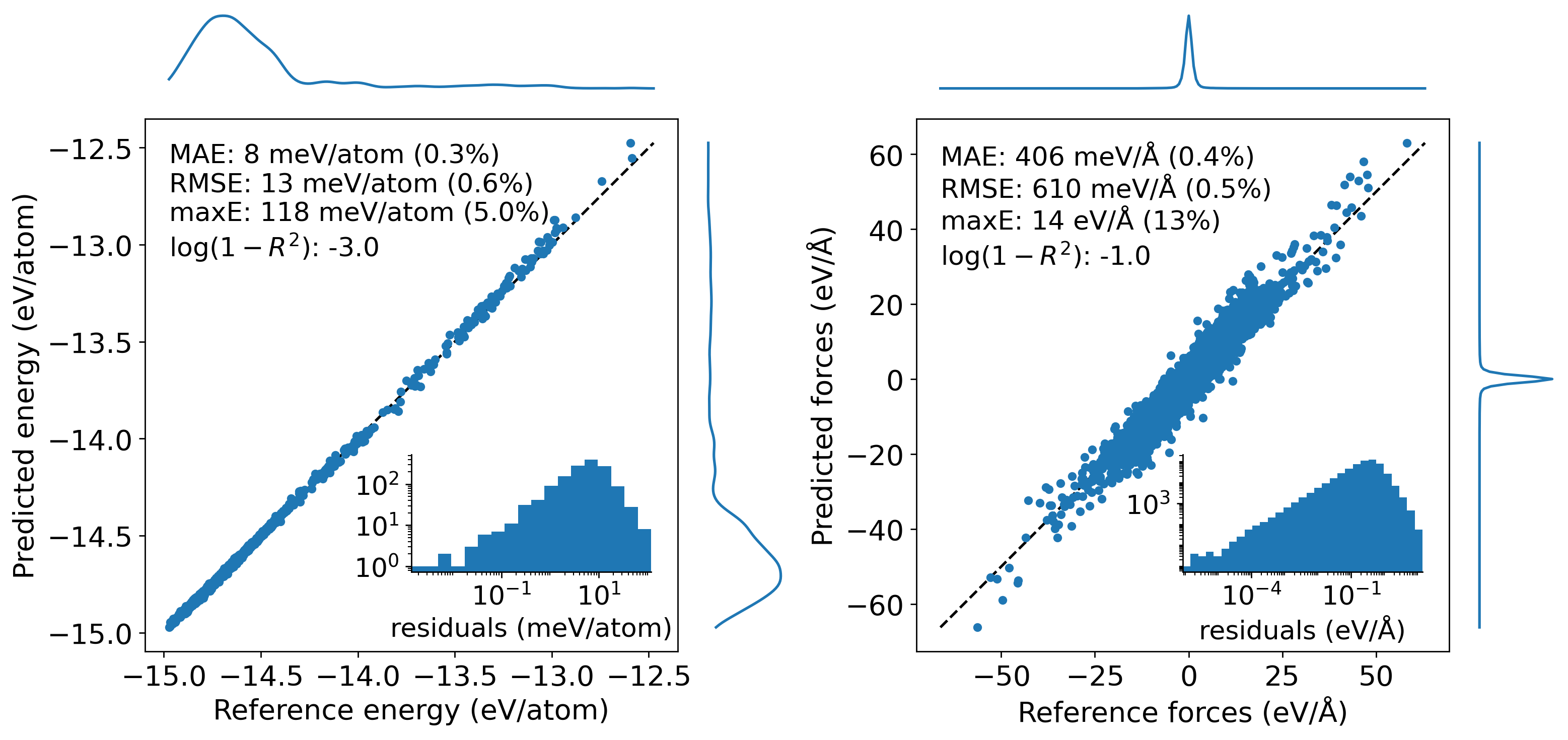}    
        \subcaption{UFP3 potential.}
        \label{si:fig:scatter:ufthree}
    \end{subfigure}

    \begin{subfigure}{0.95\linewidth}
        \includegraphics[width=\linewidth]{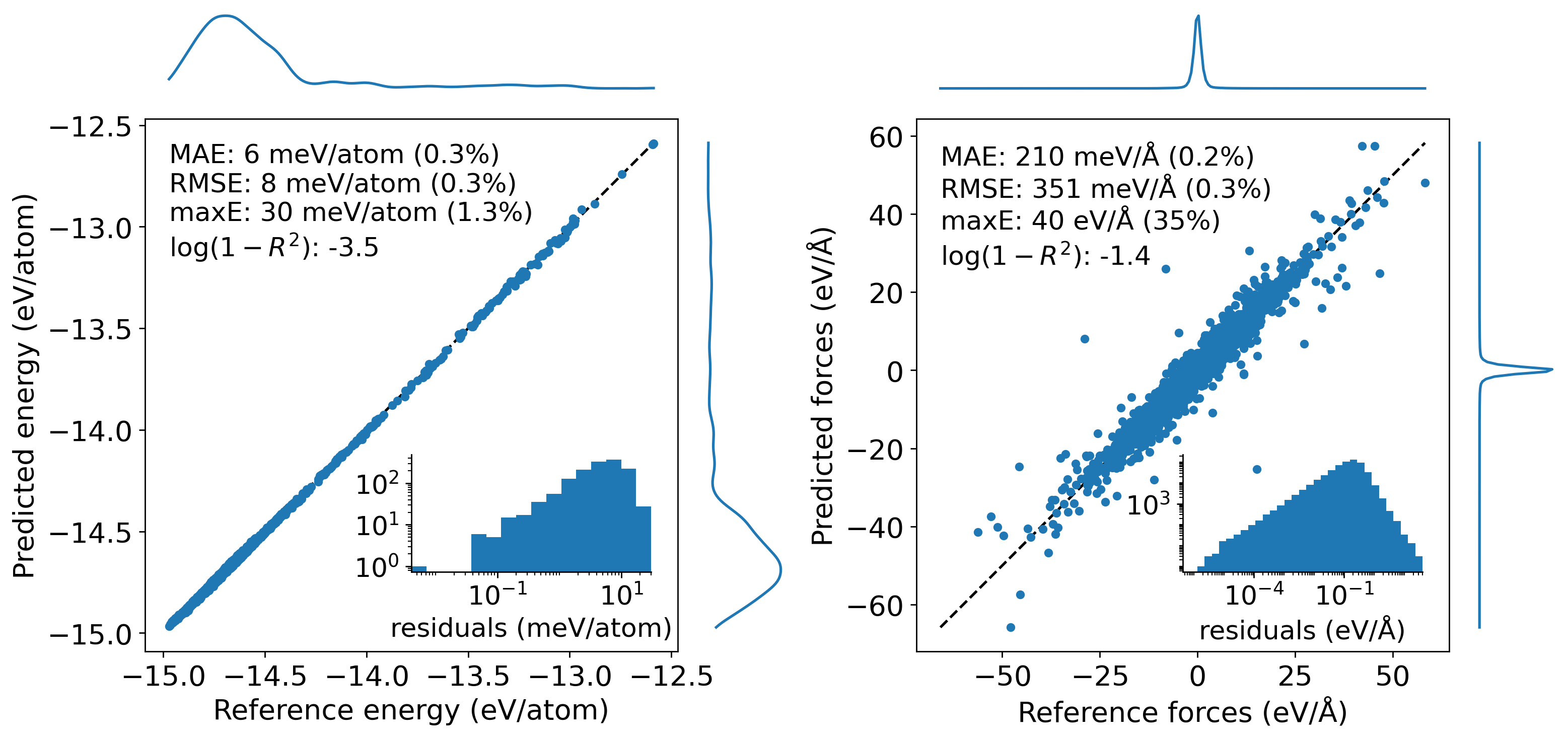}
        \subcaption{PACE potential.}
        \label{si:fig:scatter:pace}
    \end{subfigure}

    \begin{subfigure}{0.95\linewidth}
        \includegraphics[width=\linewidth]{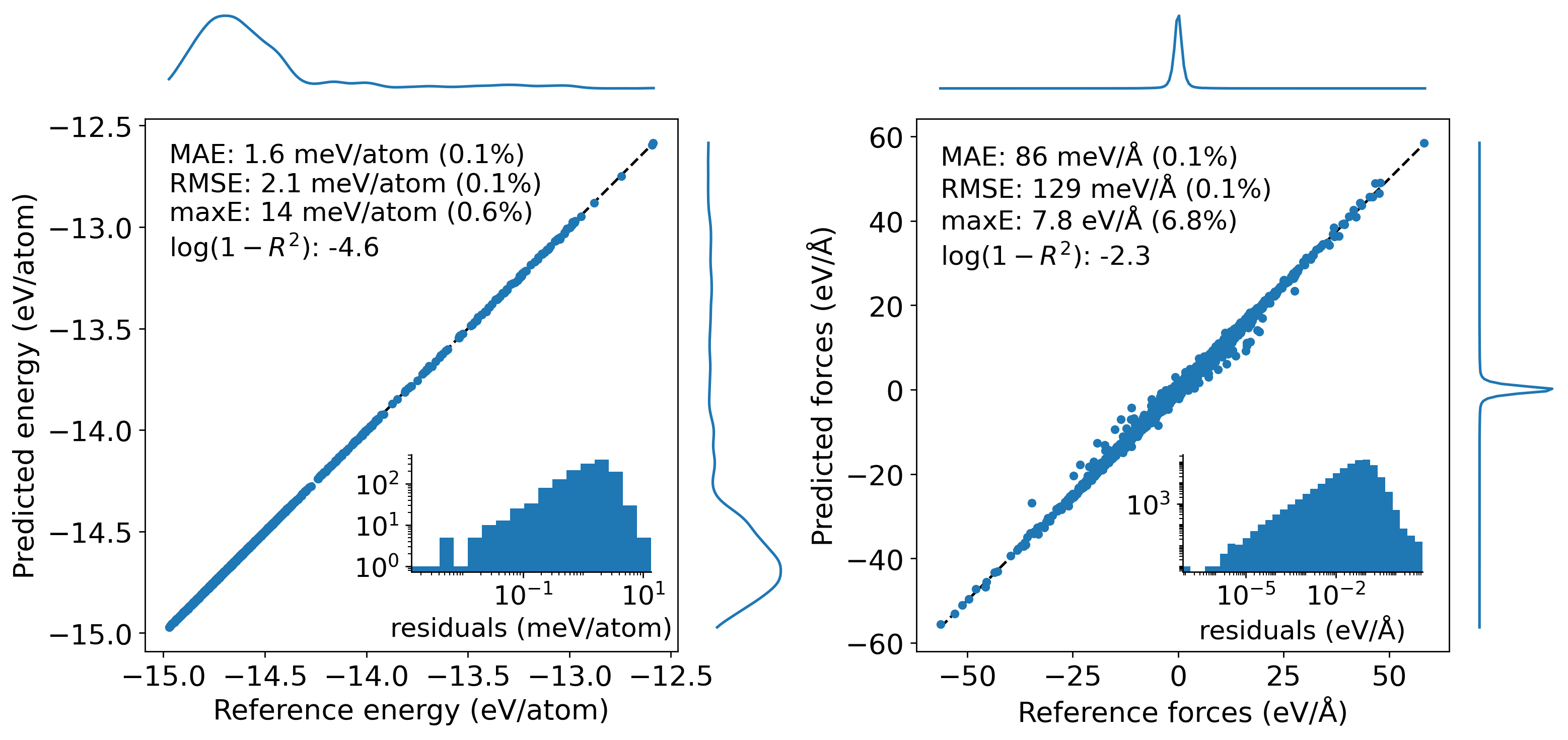}
        \subcaption{MACE potential.}
        \label{si:fig:scatter:mace}
    \end{subfigure}

    \caption{%
        \emph{Scatter plots} (continued).
        See main text for explanation.
    }
    \label{si:fig:scatterplots}
\end{figure}


\subsection{Hellinger distances}\label{si:sec:hellingerdists}

\Cref{si:tab:hellingerdists} presents \glspl{hd} for all \gls{md}-derived properties for all benchmarked \glspl{mlp}.
Wigner-Seitz radii are given as indices, specified by \cref{si:tab:rsindices}.

\begin{table}[hbtp]
    \begin{minipage}[t]{0.48\linewidth}
        \begin{tabular}[t]{@{}rcccccc@{}}
            \toprule
            & \multicolumn{6}{c}{Temperature / K} \\
            \cmidrule(lr){2-7} 
            I. & 1000 & 1100 & 1200 & 1300 & 1400 & 1500 \\
            \midrule
             1 & 1.400 & 1.410 & 1.420 & 1.440 & 1.460 & 1.480 \\
             2 & 1.405 & 1.415 & 1.425 & 1.445 & 1.465 & 1.485 \\
             3 & 1.410 & 1.420 & 1.430 & 1.450 & 1.470 & 1.490 \\
             4 & 1.415 & 1.425 & 1.435 & 1.455 & 1.475 & 1.495 \\
             5 & 1.420 & 1.430 & 1.440 & 1.460 & 1.480 & 1.500 \\
             6 & 1.425 & 1.435 & 1.445 & 1.465 & 1.485 & 1.505 \\
             7 & 1.430 & 1.440 & 1.450 & 1.470 & 1.490 & 1.510 \\
             8 & 1.435 & 1.445 & 1.455 & 1.475 & 1.495 & 1.515 \\
             9 & 1.440 & 1.450 & 1.460 & 1.480 & 1.500 & 1.520 \\
            10 & 1.445 & 1.455 & 1.465 & 1.485 & 1.505 & 1.525 \\
            11 & 1.450 & 1.460 & 1.470 & 1.490 & 1.510 & 1.530 \\
            12 & 1.455 & 1.465 & 1.475 & 1.495 & 1.515 & 1.535 \\
            13 & 1.460 & 1.470 & 1.480 & 1.500 & 1.520 & 1.540 \\
            14 & 1.465 & 1.475 & 1.485 & 1.505 & 1.525 & 1.545 \\
            15 & 1.470 & 1.480 & 1.490 & 1.510 & 1.530 & 1.550 \\
            16 & 1.475 & 1.485 & 1.495 & 1.515 & 1.535 & 1.555 \\
            17 & 1.480 & 1.490 & 1.500 & 1.520 & 1.540 & 1.560 \\
            \bottomrule
        \end{tabular}
        \end{minipage}%
    \hfill%
    \begin{minipage}[t]{0.52\linewidth-2\tabcolsep}
        \caption{%
            \emph{Wigner-Seitz radii indices} for \cref{si:tab:hellingerdists}.
            For each index (I.), the Wigner-Seitz radius~$r_s$ is given for each temperature.
        }
        \label{si:tab:rsindices}
    \end{minipage}
\end{table}

\begin{table}[p]
    \caption{%
        \emph{Hellinger distances} for all \gls{md}-derived properties and benchmarked \glspl{mlp}.
    }
    \label{si:tab:hellingerdists}
    
    \centering

    \begin{subtable}{\linewidth}
        \caption{%
            \emph{Yukawa potential}.
            Shown are \glspl{hd} of pressures, stable molecular fractions, diffusion coefficients, and radial distribution functions (top to bottom) as functions of Wigner-Seitz radius (see \cref{si:tab:rsindices}).
        }

        \bigskip

        \begingroup
        \small\setlength{\tabcolsep}{0.5em}

        Pressures:

        \smallskip

        \begin{tabular}{@{}lccccccccccccccccc@{}}
            \toprule
            & \multicolumn{17}{c}{Index of Wigner-Seitz radius rs} \\
            \cmidrule(lr){2-18} 
            T/K & 1 & 2 & 3 & 4 & 5 & 6 & 7 & 8 & 9 & 10 & 11 & 12 & 13 & 14 & 15 & 16 & 17 \\
            \midrule
            1000 & 1.00 & 1.00 & 1.00 & 1.00 & 1.00 & 1.00 & 1.00 & 1.00 & 1.00 & 1.00 & 1.00 & 1.00 & 1.00 & 1.00 & 1.00 & 1.00 & 1.00 \\
            1100 & 1.00 & 1.00 & 1.00 & 1.00 & 1.00 & 1.00 & 1.00 & 1.00 & 1.00 & 1.00 & 1.00 & 1.00 & 1.00 & 1.00 & 1.00 & 1.00 & 1.00 \\
            1200 & 1.00 & 1.00 & 1.00 & 1.00 & 1.00 & 1.00 & 1.00 & 1.00 & 1.00 & 1.00 & 1.00 & 1.00 & 1.00 & 1.00 & 1.00 & 1.00 & 1.00 \\
            1300 & 1.00 & 1.00 & 1.00 & 1.00 & 1.00 & 1.00 & 1.00 & 1.00 & 1.00 & 1.00 & 1.00 & 1.00 & 1.00 & 1.00 & 1.00 & 1.00 & 1.00 \\
            1400 & 1.00 & 1.00 & 1.00 & 1.00 & 1.00 & 1.00 & 1.00 & 1.00 & 1.00 & 1.00 & 1.00 & 1.00 & 1.00 & 1.00 & 1.00 & 1.00 & 1.00 \\
            1500 & 1.00 & 1.00 & 1.00 & 1.00 & 1.00 & 1.00 & 1.00 & 1.00 & 1.00 & 1.00 & 1.00 & 1.00 & 1.00 & 1.00 & 1.00 & 1.00 & 1.00 \\
            \bottomrule
        \end{tabular}

        \bigskip

        Stable molecular fractions:

        \smallskip

        \begin{tabular}{@{}lccccccccccccccccc@{}}
            \toprule
            & \multicolumn{17}{c}{Index of Wigner-Seitz radius rs} \\
            \cmidrule(lr){2-18} 
            T/K & 1 & 2 & 3 & 4 & 5 & 6 & 7 & 8 & 9 & 10 & 11 & 12 & 13 & 14 & 15 & 16 & 17 \\
            \midrule
            1000 & 1.00 & 1.00 & 1.00 & 1.00 & 1.00 & 1.00 & 1.00 & 1.00 & 1.00 & 1.00 & 1.00 & 1.00 & 1.00 & 1.00 & 1.00 & 1.00 & 1.00 \\
            1100 & 1.00 & 1.00 & 1.00 & 1.00 & 1.00 & 1.00 & 1.00 & 1.00 & 1.00 & 1.00 & 1.00 & 1.00 & 1.00 & 1.00 & 1.00 & 1.00 & 1.00 \\
            1200 & 1.00 & 1.00 & 1.00 & 1.00 & 1.00 & 1.00 & 1.00 & 1.00 & 1.00 & 1.00 & 1.00 & 1.00 & 1.00 & 1.00 & 1.00 & 1.00 & 1.00 \\
            1300 & 1.00 & 1.00 & 1.00 & 1.00 & 1.00 & 1.00 & 1.00 & 1.00 & 1.00 & 1.00 & 1.00 & 1.00 & 1.00 & 1.00 & 1.00 & 1.00 & 1.00 \\
            1400 & 1.00 & 1.00 & 1.00 & 1.00 & 1.00 & 1.00 & 1.00 & 1.00 & 1.00 & 1.00 & 1.00 & 1.00 & 1.00 & 1.00 & 1.00 & 1.00 & 1.00 \\
            1500 & 1.00 & 1.00 & 1.00 & 1.00 & 1.00 & 1.00 & 1.00 & 1.00 & 1.00 & 1.00 & 1.00 & 1.00 & 1.00 & 1.00 & 1.00 & 1.00 & 1.00 \\
            \bottomrule
        \end{tabular}        

        \bigskip

        Diffusion coefficients:

        \smallskip

        \begin{tabular}{@{}lccccccccccccccccc@{}}
            \toprule
            & \multicolumn{17}{c}{Index of Wigner-Seitz radius rs} \\
            \cmidrule(lr){2-18} 
            T/K & 1 & 2 & 3 & 4 & 5 & 6 & 7 & 8 & 9 & 10 & 11 & 12 & 13 & 14 & 15 & 16 & 17 \\
            \midrule
            1000 & 0.91 & 0.97 & 0.98 & 0.91 & 1.00 & 1.00 & 1.00 & 1.00 & 1.00 & 1.00 & 1.00 & 1.00 & 1.00 & 1.00 & 1.00 & 1.00 & 1.00 \\
            1100 & 0.99 & 0.96 & 0.91 & 0.98 & 0.99 & 1.00 & 1.00 & 1.00 & 1.00 & 1.00 & 1.00 & 1.00 & 1.00 & 1.00 & 1.00 & 1.00 & 1.00 \\
            1200 & 0.95 & 0.78 & 1.00 & 0.96 & 0.97 & 1.00 & 1.00 & 1.00 & 1.00 & 1.00 & 1.00 & 1.00 & 1.00 & 1.00 & 1.00 & 1.00 & 1.00 \\
            1300 & 0.97 & 0.98 & 0.99 & 0.99 & 1.00 & 1.00 & 1.00 & 1.00 & 1.00 & 1.00 & 1.00 & 1.00 & 1.00 & 1.00 & 1.00 & 1.00 & 1.00 \\
            1400 & 0.94 & 1.00 & 0.99 & 1.00 & 0.98 & 0.99 & 1.00 & 0.99 & 1.00 & 1.00 & 1.00 & 1.00 & 1.00 & 1.00 & 1.00 & 1.00 & 1.00 \\
            1500 & 1.00 & 0.96 & 1.00 & 0.98 & 1.00 & 1.00 & 1.00 & 1.00 & 0.99 & 1.00 & 1.00 & 1.00 & 1.00 & 1.00 & 1.00 & 1.00 & 1.00 \\
            \bottomrule
        \end{tabular}

        \bigskip

        Radial distribution functions:

        \smallskip

        \begin{tabular}{@{}lccccccccccccccccc@{}}
            \toprule
            & \multicolumn{17}{c}{Index of Wigner-Seitz radius rs} \\
            \cmidrule(lr){2-18} 
            T/K & 1 & 2 & 3 & 4 & 5 & 6 & 7 & 8 & 9 & 10 & 11 & 12 & 13 & 14 & 15 & 16 & 17 \\
            \midrule
            1000 & 0.78 & 0.79 & 0.79 & 0.79 & 0.77 & 0.78 & 0.76 & 0.78 & 0.78 & 0.79 & 0.79 & 0.79 & 0.81 & 0.80 & 0.79 & 0.79 & 0.79 \\
            1100 & 0.78 & 0.80 & 0.79 & 0.78 & 0.79 & 0.80 & 0.78 & 0.76 & 0.79 & 0.79 & 0.78 & 0.80 & 0.80 & 0.81 & 0.80 & 0.81 & 0.79 \\
            1200 & 0.78 & 0.79 & 0.79 & 0.78 & 0.79 & 0.79 & 0.81 & 0.77 & 0.80 & 0.79 & 0.80 & 0.79 & 0.80 & 0.80 & 0.82 & 0.81 & 0.79 \\
            1300 & 0.78 & 0.79 & 0.81 & 0.81 & 0.81 & 0.77 & 0.78 & 0.77 & 0.79 & 0.79 & 0.80 & 0.81 & 0.79 & 0.80 & 0.78 & 0.78 & 0.79 \\
            1400 & 0.80 & 0.80 & 0.80 & 0.82 & 0.77 & 0.79 & 0.79 & 0.78 & 0.79 & 0.80 & 0.81 & 0.80 & 0.81 & 0.79 & 0.80 & 0.81 & 0.80 \\
            1500 & 0.79 & 0.79 & 0.78 & 0.80 & 0.79 & 0.82 & 0.80 & 0.81 & 0.80 & 0.81 & 0.81 & 0.81 & 0.82 & 0.80 & 0.80 & 0.81 & 0.82 \\
            \bottomrule
        \end{tabular}
                \endgroup
    \end{subtable}
    
\end{table}

\begin{table}[p]\ContinuedFloat
    \caption{%
        \emph{Hellinger distances} (continued) for all \gls{md}-derived properties and benchmarked \glspl{mlp}.
    }
    
    \centering

    \begin{subtable}{\linewidth}
        \caption{%
            \emph{Tersoff potential}.
            Shown are \glspl{hd} of pressures, stable molecular fractions, diffusion coefficients, and radial distribution functions (top to bottom) as functions of Wigner-Seitz radius (see \cref{si:tab:rsindices}).
        }

        \bigskip

        \begingroup
        \small\setlength{\tabcolsep}{0.5em}

        Pressures:

        \smallskip

        \begin{tabular}{@{}lccccccccccccccccc@{}}
            \toprule
            & \multicolumn{17}{c}{Index of Wigner-Seitz radius rs} \\
            \cmidrule(lr){2-18} 
            T/K & 1 & 2 & 3 & 4 & 5 & 6 & 7 & 8 & 9 & 10 & 11 & 12 & 13 & 14 & 15 & 16 & 17 \\
            \midrule
            1000 & 1.00 & 1.00 & 1.00 & 1.00 & 1.00 & 1.00 & 1.00 & 1.00 & 1.00 & 1.00 & 1.00 & 1.00 & 1.00 & 1.00 & 1.00 & 1.00 & 1.00 \\
            1100 & 1.00 & 1.00 & 1.00 & 1.00 & 1.00 & 1.00 & 1.00 & 1.00 & 1.00 & 1.00 & 1.00 & 1.00 & 1.00 & 1.00 & 1.00 & 1.00 & 1.00 \\
            1200 & 1.00 & 1.00 & 1.00 & 1.00 & 1.00 & 1.00 & 1.00 & 1.00 & 1.00 & 1.00 & 1.00 & 1.00 & 1.00 & 1.00 & 1.00 & 1.00 & 1.00 \\
            1300 & 1.00 & 1.00 & 1.00 & 1.00 & 1.00 & 1.00 & 1.00 & 1.00 & 1.00 & 1.00 & 1.00 & 1.00 & 1.00 & 1.00 & 1.00 & 1.00 & 1.00 \\
            1400 & 1.00 & 1.00 & 1.00 & 1.00 & 1.00 & 1.00 & 1.00 & 1.00 & 1.00 & 1.00 & 1.00 & 1.00 & 1.00 & 1.00 & 1.00 & 1.00 & 1.00 \\
            1500 & 1.00 & 1.00 & 1.00 & 1.00 & 1.00 & 1.00 & 1.00 & 1.00 & 1.00 & 1.00 & 1.00 & 1.00 & 1.00 & 1.00 & 1.00 & 1.00 & 1.00 \\
            \bottomrule
        \end{tabular}
        
        \bigskip

        Stable molecular fractions:

        \smallskip

        \begin{tabular}{@{}lccccccccccccccccc@{}}
            \toprule
            & \multicolumn{17}{c}{Index of Wigner-Seitz radius rs} \\
            \cmidrule(lr){2-18} 
            T/K & 1 & 2 & 3 & 4 & 5 & 6 & 7 & 8 & 9 & 10 & 11 & 12 & 13 & 14 & 15 & 16 & 17 \\
            \midrule
            1000 & 0.20 & 0.52 & 0.90 & 0.88 & 0.88 & 0.97 & 0.90 & 1.00 & 0.99 & 1.00 & 1.00 & 1.00 & 1.00 & 1.00 & 1.00 & 1.00 & 1.00 \\
            1100 & 0.30 & 0.92 & 0.76 & 0.98 & 0.94 & 0.99 & 0.96 & 0.93 & 0.99 & 1.00 & 1.00 & 1.00 & 1.00 & 1.00 & 1.00 & 1.00 & 1.00 \\
            1200 & 0.74 & 0.54 & 0.73 & 0.91 & 0.91 & 0.93 & 1.00 & 0.95 & 0.97 & 1.00 & 1.00 & 1.00 & 1.00 & 1.00 & 1.00 & 1.00 & 1.00 \\
            1300 & 0.95 & 0.99 & 1.00 & 0.83 & 0.99 & 0.99 & 0.98 & 0.98 & 0.99 & 0.99 & 1.00 & 1.00 & 1.00 & 1.00 & 1.00 & 1.00 & 1.00 \\
            1400 & 0.90 & 0.93 & 1.00 & 0.99 & 0.99 & 1.00 & 1.00 & 0.99 & 1.00 & 1.00 & 1.00 & 1.00 & 1.00 & 1.00 & 1.00 & 1.00 & 1.00 \\
            1500 & 0.96 & 1.00 & 1.00 & 1.00 & 1.00 & 1.00 & 1.00 & 1.00 & 1.00 & 1.00 & 1.00 & 1.00 & 1.00 & 1.00 & 1.00 & 1.00 & 1.00 \\
            \bottomrule
        \end{tabular}
        
        \bigskip

        Diffusion coefficients:

        \smallskip

        \begin{tabular}{@{}lccccccccccccccccc@{}}
            \toprule
            & \multicolumn{17}{c}{Index of Wigner-Seitz radius rs} \\
            \cmidrule(lr){2-18} 
            T/K & 1 & 2 & 3 & 4 & 5 & 6 & 7 & 8 & 9 & 10 & 11 & 12 & 13 & 14 & 15 & 16 & 17 \\
            \midrule
            1000 & 0.99 & 0.97 & 1.00 & 0.98 & 1.00 & 1.00 & 1.00 & 1.00 & 1.00 & 1.00 & 1.00 & 1.00 & 1.00 & 1.00 & 1.00 & 1.00 & 1.00 \\
            1100 & 0.99 & 0.96 & 0.99 & 0.98 & 0.99 & 1.00 & 1.00 & 1.00 & 1.00 & 1.00 & 1.00 & 1.00 & 1.00 & 1.00 & 1.00 & 1.00 & 1.00 \\
            1200 & 0.97 & 0.95 & 0.94 & 1.00 & 1.00 & 0.98 & 1.00 & 1.00 & 1.00 & 1.00 & 1.00 & 1.00 & 1.00 & 1.00 & 1.00 & 1.00 & 1.00 \\
            1300 & 0.99 & 1.00 & 0.99 & 1.00 & 0.98 & 1.00 & 0.92 & 1.00 & 1.00 & 1.00 & 1.00 & 1.00 & 1.00 & 1.00 & 1.00 & 1.00 & 1.00 \\
            1400 & 1.00 & 0.99 & 1.00 & 1.00 & 1.00 & 0.99 & 1.00 & 1.00 & 1.00 & 1.00 & 1.00 & 1.00 & 1.00 & 1.00 & 1.00 & 1.00 & 1.00 \\
            1500 & 0.99 & 0.97 & 1.00 & 0.98 & 1.00 & 1.00 & 1.00 & 0.99 & 1.00 & 1.00 & 1.00 & 1.00 & 1.00 & 1.00 & 1.00 & 1.00 & 1.00 \\
            \bottomrule
        \end{tabular}
        
        \bigskip

        Radial distribution functions:

        \smallskip

        \begin{tabular}{@{}lccccccccccccccccc@{}}
            \toprule
            & \multicolumn{17}{c}{Index of Wigner-Seitz radius rs} \\
            \cmidrule(lr){2-18} 
            T/K & 1 & 2 & 3 & 4 & 5 & 6 & 7 & 8 & 9 & 10 & 11 & 12 & 13 & 14 & 15 & 16 & 17 \\
            \midrule
            1000 & 0.86 & 0.87 & 0.87 & 0.85 & 0.84 & 0.90 & 0.85 & 0.89 & 0.87 & 0.88 & 0.89 & 0.91 & 0.92 & 0.91 & 0.91 & 0.92 & 0.94 \\
            1100 & 0.87 & 0.88 & 0.87 & 0.87 & 0.87 & 0.90 & 0.89 & 0.88 & 0.91 & 0.90 & 0.92 & 0.91 & 0.93 & 0.92 & 0.93 & 0.91 & 0.93 \\
            1200 & 0.88 & 0.89 & 0.87 & 0.87 & 0.90 & 0.90 & 0.92 & 0.89 & 0.90 & 0.92 & 0.91 & 0.91 & 0.94 & 0.94 & 0.93 & 0.94 & 0.94 \\
            1300 & 0.88 & 0.89 & 0.91 & 0.91 & 0.90 & 0.89 & 0.92 & 0.93 & 0.92 & 0.91 & 0.92 & 0.93 & 0.93 & 0.93 & 0.95 & 0.94 & 0.94 \\
            1400 & 0.90 & 0.90 & 0.90 & 0.92 & 0.90 & 0.92 & 0.90 & 0.90 & 0.92 & 0.93 & 0.94 & 0.93 & 0.93 & 0.94 & 0.94 & 0.95 & 0.96 \\
            1500 & 0.91 & 0.91 & 0.92 & 0.92 & 0.91 & 0.93 & 0.94 & 0.93 & 0.93 & 0.92 & 0.93 & 0.95 & 0.94 & 0.95 & 0.95 & 0.96 & 0.96 \\
            \bottomrule
        \end{tabular}
        
        \endgroup
    \end{subtable}
    
\end{table}

\begin{table}[p]\ContinuedFloat
    \caption{%
        \emph{Hellinger distances} (continued) for all \gls{md}-derived properties and benchmarked \glspl{mlp}.
    }
    
    \centering

    \begin{subtable}{\linewidth}
        \caption{%
            \emph{UFP2 potential}.
            Shown are \glspl{hd} of pressures, stable molecular fractions, diffusion coefficients, and radial distribution functions (top to bottom) as functions of Wigner-Seitz radius (see \cref{si:tab:rsindices}).
        }

        \bigskip

        \begingroup
        \small\setlength{\tabcolsep}{0.5em}

        Pressures:

        \smallskip

        \begin{tabular}{@{}lccccccccccccccccc@{}}
            \toprule
            & \multicolumn{17}{c}{Index of Wigner-Seitz radius rs} \\
            \cmidrule(lr){2-18} 
            T/K & 1 & 2 & 3 & 4 & 5 & 6 & 7 & 8 & 9 & 10 & 11 & 12 & 13 & 14 & 15 & 16 & 17 \\
            \midrule
            1000 & 1.00 & 1.00 & 1.00 & 1.00 & 1.00 & 1.00 & 1.00 & 1.00 & 1.00 & 1.00 & 1.00 & 1.00 & 1.00 & 1.00 & 1.00 & 1.00 & 1.00 \\
            1100 & 1.00 & 1.00 & 1.00 & 1.00 & 1.00 & 1.00 & 1.00 & 1.00 & 1.00 & 1.00 & 1.00 & 1.00 & 1.00 & 1.00 & 1.00 & 1.00 & 1.00 \\
            1200 & 1.00 & 1.00 & 1.00 & 1.00 & 1.00 & 1.00 & 1.00 & 1.00 & 1.00 & 1.00 & 1.00 & 1.00 & 1.00 & 1.00 & 1.00 & 1.00 & 1.00 \\
            1300 & 1.00 & 1.00 & 1.00 & 1.00 & 1.00 & 1.00 & 1.00 & 1.00 & 1.00 & 1.00 & 1.00 & 1.00 & 1.00 & 1.00 & 1.00 & 1.00 & 1.00 \\
            1400 & 1.00 & 1.00 & 1.00 & 1.00 & 1.00 & 1.00 & 1.00 & 1.00 & 1.00 & 1.00 & 1.00 & 1.00 & 1.00 & 1.00 & 1.00 & 1.00 & 1.00 \\
            1500 & 1.00 & 1.00 & 1.00 & 1.00 & 1.00 & 1.00 & 1.00 & 1.00 & 1.00 & 1.00 & 1.00 & 1.00 & 1.00 & 1.00 & 1.00 & 1.00 & 1.00 \\
            \bottomrule
        \end{tabular}
        
        \bigskip

        Stable molecular fractions:

        \smallskip

        \begin{tabular}{@{}lccccccccccccccccc@{}}
            \toprule
            & \multicolumn{17}{c}{Index of Wigner-Seitz radius rs} \\
            \cmidrule(lr){2-18} 
            T/K & 1 & 2 & 3 & 4 & 5 & 6 & 7 & 8 & 9 & 10 & 11 & 12 & 13 & 14 & 15 & 16 & 17 \\
            \midrule
            1000 & 0.86 & 0.66 & 0.15 & 0.25 & 0.42 & 0.88 & 0.80 & 1.00 & 0.98 & 1.00 & 1.00 & 1.00 & 1.00 & 1.00 & 1.00 & 1.00 & 1.00 \\
            1100 & 0.88 & 0.37 & 0.13 & 0.04 & 0.58 & 0.87 & 0.87 & 0.82 & 0.97 & 0.99 & 1.00 & 1.00 & 1.00 & 1.00 & 1.00 & 1.00 & 1.00 \\
            1200 & 0.56 & 0.45 & 0.08 & 0.39 & 0.63 & 0.72 & 0.97 & 0.88 & 0.90 & 0.98 & 1.00 & 1.00 & 1.00 & 1.00 & 1.00 & 1.00 & 1.00 \\
            1300 & 0.10 & 0.77 & 0.92 & 0.65 & 0.96 & 0.95 & 0.91 & 0.94 & 0.97 & 0.98 & 0.99 & 1.00 & 1.00 & 1.00 & 1.00 & 1.00 & 1.00 \\
            1400 & 0.69 & 0.66 & 0.88 & 0.97 & 0.97 & 1.00 & 0.96 & 0.95 & 0.97 & 0.99 & 1.00 & 1.00 & 1.00 & 1.00 & 1.00 & 1.00 & 1.00 \\
            1500 & 0.81 & 0.99 & 0.92 & 0.90 & 0.94 & 1.00 & 1.00 & 0.99 & 1.00 & 1.00 & 1.00 & 1.00 & 1.00 & 1.00 & 1.00 & 1.00 & 1.00 \\
            \bottomrule
        \end{tabular}
               
        \bigskip

        Diffusion coefficients:

        \smallskip

        \begin{tabular}{@{}lccccccccccccccccc@{}}
            \toprule
            & \multicolumn{17}{c}{Index of Wigner-Seitz radius rs} \\
            \cmidrule(lr){2-18} 
            T/K & 1 & 2 & 3 & 4 & 5 & 6 & 7 & 8 & 9 & 10 & 11 & 12 & 13 & 14 & 15 & 16 & 17 \\
            \midrule
            1000 & 0.10 & 0.35 & 0.45 & 0.37 & 0.70 & 0.88 & 0.76 & 1.00 & 1.00 & 1.00 & 1.00 & 1.00 & 1.00 & 1.00 & 1.00 & 1.00 & 1.00 \\
            1100 & 0.04 & 0.34 & 0.40 & 0.66 & 0.69 & 0.77 & 0.70 & 0.93 & 0.98 & 1.00 & 1.00 & 1.00 & 1.00 & 1.00 & 1.00 & 1.00 & 1.00 \\
            1200 & 0.22 & 0.28 & 0.32 & 0.55 & 0.63 & 0.46 & 0.74 & 0.94 & 0.79 & 1.00 & 1.00 & 1.00 & 1.00 & 0.99 & 1.00 & 1.00 & 1.00 \\
            1300 & 0.46 & 0.47 & 0.69 & 0.55 & 0.74 & 0.75 & 0.93 & 0.97 & 0.97 & 0.99 & 1.00 & 1.00 & 1.00 & 1.00 & 1.00 & 1.00 & 1.00 \\
            1400 & 0.46 & 0.79 & 0.81 & 0.97 & 0.78 & 0.92 & 0.99 & 0.93 & 1.00 & 1.00 & 1.00 & 1.00 & 1.00 & 1.00 & 1.00 & 1.00 & 1.00 \\
            1500 & 0.85 & 0.75 & 0.91 & 0.74 & 0.99 & 1.00 & 1.00 & 1.00 & 0.99 & 1.00 & 1.00 & 1.00 & 1.00 & 1.00 & 1.00 & 1.00 & 1.00 \\
            \bottomrule
        \end{tabular}
        
        \bigskip

        Radial distribution functions:

        \smallskip

        \begin{tabular}{@{}lccccccccccccccccc@{}}
            \toprule
            & \multicolumn{17}{c}{Index of Wigner-Seitz radius rs} \\
            \cmidrule(lr){2-18} 
            T/K & 1 & 2 & 3 & 4 & 5 & 6 & 7 & 8 & 9 & 10 & 11 & 12 & 13 & 14 & 15 & 16 & 17 \\
            \midrule
            1000 & 0.65 & 0.62 & 0.61 & 0.60 & 0.58 & 0.52 & 0.45 & 0.57 & 0.56 & 0.63 & 0.69 & 0.68 & 0.68 & 0.65 & 0.69 & 0.71 & 0.69 \\
            1100 & 0.60 & 0.59 & 0.59 & 0.61 & 0.55 & 0.50 & 0.45 & 0.44 & 0.58 & 0.60 & 0.66 & 0.67 & 0.70 & 0.68 & 0.71 & 0.71 & 0.71 \\
            1200 & 0.58 & 0.54 & 0.54 & 0.50 & 0.48 & 0.47 & 0.47 & 0.50 & 0.57 & 0.59 & 0.65 & 0.67 & 0.70 & 0.70 & 0.71 & 0.73 & 0.74 \\
            1300 & 0.49 & 0.52 & 0.55 & 0.58 & 0.54 & 0.63 & 0.65 & 0.65 & 0.66 & 0.68 & 0.70 & 0.71 & 0.75 & 0.77 & 0.78 & 0.78 & 0.78 \\
            1400 & 0.64 & 0.65 & 0.66 & 0.69 & 0.71 & 0.70 & 0.69 & 0.68 & 0.72 & 0.73 & 0.74 & 0.76 & 0.77 & 0.78 & 0.78 & 0.80 & 0.80 \\
            1500 & 0.69 & 0.71 & 0.73 & 0.73 & 0.72 & 0.76 & 0.76 & 0.73 & 0.77 & 0.77 & 0.78 & 0.79 & 0.80 & 0.79 & 0.81 & 0.82 & 0.80 \\
            \bottomrule
        \end{tabular}        
        
        \endgroup
    \end{subtable}
    
\end{table}

\begin{table}[p]\ContinuedFloat
    \caption{%
        \emph{Hellinger distances} (continued) for all \gls{md}-derived properties and benchmarked \glspl{mlp}.
    }
    
    \centering

    \begin{subtable}{\linewidth}
        \caption{%
            \emph{UFP3 potential}.
            Shown are \glspl{hd} of pressures, stable molecular fractions, diffusion coefficients, and radial distribution functions (top to bottom) as functions of Wigner-Seitz radius (see \cref{si:tab:rsindices}).
        }

        \bigskip

        \begingroup
        \small\setlength{\tabcolsep}{0.5em}

        Pressures:

        \smallskip

        \begin{tabular}{@{}lccccccccccccccccc@{}}
            \toprule
            & \multicolumn{17}{c}{Index of Wigner-Seitz radius rs} \\
            \cmidrule(lr){2-18} 
            T/K & 1 & 2 & 3 & 4 & 5 & 6 & 7 & 8 & 9 & 10 & 11 & 12 & 13 & 14 & 15 & 16 & 17 \\
            \midrule
            1000 & 0.88 & 1.00 & 1.00 & 1.00 & 1.00 & 1.00 & 0.90 & 0.95 & 0.57 & 0.89 & 0.99 & 1.00 & 1.00 & 1.00 & 1.00 & 1.00 & 1.00 \\
            1100 & 1.00 & 1.00 & 1.00 & 1.00 & 1.00 & 1.00 & 1.00 & 1.00 & 0.99 & 0.93 & 0.91 & 1.00 & 1.00 & 1.00 & 1.00 & 1.00 & 1.00 \\
            1200 & 1.00 & 1.00 & 1.00 & 1.00 & 1.00 & 1.00 & 1.00 & 1.00 & 1.00 & 1.00 & 0.94 & 0.98 & 1.00 & 1.00 & 1.00 & 1.00 & 1.00 \\
            1300 & 1.00 & 1.00 & 1.00 & 1.00 & 1.00 & 1.00 & 1.00 & 1.00 & 1.00 & 1.00 & 0.98 & 0.98 & 1.00 & 1.00 & 1.00 & 1.00 & 1.00 \\
            1400 & 1.00 & 1.00 & 1.00 & 1.00 & 1.00 & 1.00 & 1.00 & 0.99 & 0.98 & 1.00 & 1.00 & 1.00 & 1.00 & 1.00 & 1.00 & 1.00 & 1.00 \\
            1500 & 1.00 & 1.00 & 1.00 & 1.00 & 1.00 & 1.00 & 1.00 & 1.00 & 1.00 & 1.00 & 1.00 & 1.00 & 1.00 & 1.00 & 1.00 & 1.00 & 1.00 \\
            \bottomrule
        \end{tabular}
        
        \bigskip

        Stable molecular fractions:

        \smallskip

        \begin{tabular}{@{}lccccccccccccccccc@{}}
            \toprule
            & \multicolumn{17}{c}{Index of Wigner-Seitz radius rs} \\
            \cmidrule(lr){2-18} 
            T/K & 1 & 2 & 3 & 4 & 5 & 6 & 7 & 8 & 9 & 10 & 11 & 12 & 13 & 14 & 15 & 16 & 17 \\
            \midrule
            1000 & 1.00 & 1.00 & 1.00 & 0.97 & 0.94 & 0.50 & 0.44 & 0.83 & 0.71 & 1.00 & 1.00 & 1.00 & 1.00 & 1.00 & 1.00 & 1.00 & 1.00 \\
            1100 & 1.00 & 1.00 & 1.00 & 1.00 & 0.98 & 0.99 & 0.54 & 0.43 & 0.52 & 0.81 & 0.94 & 1.00 & 1.00 & 1.00 & 1.00 & 1.00 & 1.00 \\
            1200 & 0.99 & 1.00 & 1.00 & 1.00 & 0.99 & 1.00 & 1.00 & 0.46 & 0.46 & 0.54 & 0.87 & 0.99 & 1.00 & 1.00 & 1.00 & 1.00 & 1.00 \\
            1300 & 1.00 & 0.97 & 1.00 & 1.00 & 0.99 & 0.93 & 0.48 & 0.33 & 0.59 & 0.64 & 0.84 & 0.94 & 1.00 & 1.00 & 1.00 & 1.00 & 1.00 \\
            1400 & 1.00 & 1.00 & 0.99 & 0.98 & 0.96 & 0.82 & 0.29 & 0.57 & 0.71 & 0.87 & 0.97 & 1.00 & 1.00 & 1.00 & 1.00 & 1.00 & 1.00 \\
            1500 & 0.99 & 0.86 & 0.83 & 0.54 & 0.17 & 0.24 & 0.92 & 0.83 & 0.92 & 1.00 & 1.00 & 1.00 & 1.00 & 1.00 & 1.00 & 1.00 & 1.00 \\
            \bottomrule
        \end{tabular}
        
        \bigskip

        Diffusion coefficients:

        \smallskip

        \begin{tabular}{@{}lccccccccccccccccc@{}}
            \toprule
            & \multicolumn{17}{c}{Index of Wigner-Seitz radius rs} \\
            \cmidrule(lr){2-18} 
            T/K & 1 & 2 & 3 & 4 & 5 & 6 & 7 & 8 & 9 & 10 & 11 & 12 & 13 & 14 & 15 & 16 & 17 \\
            \midrule
            1000 & 0.56 & 0.66 & 0.95 & 0.38 & 0.54 & 0.20 & 0.27 & 0.80 & 0.71 & 0.95 & 0.92 & 0.99 & 0.95 & 1.00 & 0.65 & 0.99 & 0.63 \\
            1100 & 0.80 & 0.55 & 0.67 & 0.79 & 0.47 & 0.40 & 0.08 & 0.02 & 0.35 & 0.38 & 0.78 & 0.98 & 1.00 & 1.00 & 0.94 & 0.88 & 0.93 \\
            1200 & 0.60 & 0.64 & 0.53 & 0.61 & 0.54 & 0.41 & 0.26 & 0.01 & 0.01 & 0.34 & 0.84 & 0.97 & 0.99 & 1.00 & 1.00 & 0.94 & 0.91 \\
            1300 & 0.70 & 0.04 & 0.41 & 0.99 & 0.40 & 0.39 & 0.13 & 0.22 & 0.54 & 0.44 & 0.62 & 0.93 & 0.95 & 1.00 & 1.00 & 1.00 & 0.99 \\
            1400 & 0.47 & 0.58 & 0.51 & 0.28 & 0.16 & 0.13 & 0.11 & 0.29 & 0.28 & 0.52 & 0.84 & 1.00 & 0.99 & 0.99 & 1.00 & 1.00 & 1.00 \\
            1500 & 0.26 & 0.13 & 0.32 & 0.04 & 0.05 & 0.32 & 0.86 & 0.80 & 0.83 & 0.97 & 1.00 & 1.00 & 1.00 & 1.00 & 1.00 & 0.98 & 1.00 \\
            \bottomrule
        \end{tabular}
        
        \bigskip

        Radial distribution functions:

        \smallskip

        \begin{tabular}{@{}lccccccccccccccccc@{}}
            \toprule
            & \multicolumn{17}{c}{Index of Wigner-Seitz radius rs} \\
            \cmidrule(lr){2-18} 
            T/K & 1 & 2 & 3 & 4 & 5 & 6 & 7 & 8 & 9 & 10 & 11 & 12 & 13 & 14 & 15 & 16 & 17 \\
            \midrule
            1000 & 0.60 & 0.57 & 0.59 & 0.59 & 0.58 & 0.55 & 0.36 & 0.30 & 0.26 & 0.38 & 0.36 & 0.35 & 0.47 & 0.46 & 0.52 & 0.50 & 0.42 \\
            1100 & 0.60 & 0.61 & 0.65 & 0.62 & 0.61 & 0.63 & 0.57 & 0.55 & 0.36 & 0.32 & 0.37 & 0.38 & 0.49 & 0.50 & 0.50 & 0.55 & 0.54 \\
            1200 & 0.67 & 0.64 & 0.66 & 0.66 & 0.67 & 0.66 & 0.68 & 0.60 & 0.52 & 0.49 & 0.44 & 0.48 & 0.52 & 0.54 & 0.52 & 0.56 & 0.58 \\
            1300 & 0.66 & 0.65 & 0.67 & 0.71 & 0.67 & 0.66 & 0.66 & 0.59 & 0.55 & 0.51 & 0.53 & 0.55 & 0.60 & 0.60 & 0.66 & 0.65 & 0.68 \\
            1400 & 0.69 & 0.69 & 0.72 & 0.71 & 0.69 & 0.73 & 0.59 & 0.58 & 0.52 & 0.57 & 0.59 & 0.63 & 0.64 & 0.70 & 0.69 & 0.74 & 0.74 \\
            1500 & 0.71 & 0.70 & 0.71 & 0.69 & 0.66 & 0.67 & 0.61 & 0.59 & 0.63 & 0.71 & 0.70 & 0.73 & 0.74 & 0.76 & 0.78 & 0.77 & 0.77 \\
            \bottomrule
        \end{tabular}        
        
        \endgroup
    \end{subtable}
    
\end{table}

\begin{table}[p]\ContinuedFloat
    \caption{%
        \emph{Hellinger distances} (continued) for all \gls{md}-derived properties and benchmarked \glspl{mlp}.
    }
    
    \centering

    \begin{subtable}{\linewidth}
        \caption{%
            \emph{PACE potential}.
            Shown are \glspl{hd} of pressures, stable molecular fractions, diffusion coefficients, and radial distribution functions (top to bottom) as functions of Wigner-Seitz radius (see \cref{si:tab:rsindices}).
        }

        \bigskip

        \begingroup
        \small\setlength{\tabcolsep}{0.5em}

        Pressures:

        \smallskip

        \begin{tabular}{@{}lccccccccccccccccc@{}}
            \toprule
            & \multicolumn{17}{c}{Index of Wigner-Seitz radius rs} \\
            \cmidrule(lr){2-18} 
            T/K & 1 & 2 & 3 & 4 & 5 & 6 & 7 & 8 & 9 & 10 & 11 & 12 & 13 & 14 & 15 & 16 & 17 \\
            \midrule
            1000 & 0.90 & 0.97 & 0.93 & 0.69 & 0.92 & 0.34 & 0.47 & 0.88 & 0.68 & 1.00 & 1.00 & 1.00 & 0.98 & 0.96 & 0.64 & 0.62 & 0.22 \\
            1100 & 1.00 & 0.99 & 1.00 & 0.99 & 1.00 & 0.97 & 0.43 & 0.49 & 0.36 & 0.71 & 0.75 & 1.00 & 1.00 & 1.00 & 0.99 & 0.90 & 0.67 \\
            1200 & 1.00 & 1.00 & 1.00 & 1.00 & 0.93 & 0.98 & 1.00 & 0.77 & 0.57 & 0.16 & 0.54 & 0.80 & 0.92 & 1.00 & 0.99 & 0.99 & 0.96 \\
            1300 & 1.00 & 1.00 & 1.00 & 1.00 & 1.00 & 0.94 & 1.00 & 0.61 & 0.37 & 0.23 & 0.46 & 0.70 & 1.00 & 0.93 & 1.00 & 1.00 & 1.00 \\
            1400 & 1.00 & 1.00 & 1.00 & 1.00 & 1.00 & 1.00 & 0.55 & 0.59 & 0.45 & 0.40 & 0.78 & 0.96 & 1.00 & 1.00 & 1.00 & 1.00 & 1.00 \\
            1500 & 1.00 & 1.00 & 1.00 & 1.00 & 0.78 & 1.00 & 0.12 & 0.40 & 0.81 & 0.99 & 1.00 & 1.00 & 1.00 & 1.00 & 1.00 & 1.00 & 1.00 \\
            \bottomrule
        \end{tabular}
                       
        \bigskip

        Stable molecular fractions:

        \smallskip

        \begin{tabular}{@{}lccccccccccccccccc@{}}
            \toprule
            & \multicolumn{17}{c}{Index of Wigner-Seitz radius rs} \\
            \cmidrule(lr){2-18} 
            T/K & 1 & 2 & 3 & 4 & 5 & 6 & 7 & 8 & 9 & 10 & 11 & 12 & 13 & 14 & 15 & 16 & 17 \\
            \midrule
            1000 & 0.25 & 0.21 & 0.47 & 0.33 & 0.49 & 0.81 & 0.73 & 0.99 & 0.93 & 1.00 & 1.00 & 1.00 & 1.00 & 1.00 & 1.00 & 1.00 & 1.00 \\
            1100 & 0.03 & 0.18 & 0.04 & 0.04 & 0.21 & 0.35 & 0.57 & 0.66 & 0.89 & 0.94 & 0.98 & 1.00 & 1.00 & 1.00 & 1.00 & 1.00 & 1.00 \\
            1200 & 0.12 & 0.28 & 0.03 & 0.06 & 0.02 & 0.18 & 0.10 & 0.33 & 0.50 & 0.78 & 0.94 & 1.00 & 1.00 & 1.00 & 1.00 & 1.00 & 1.00 \\
            1300 & 0.14 & 0.29 & 0.37 & 0.27 & 0.66 & 0.04 & 0.15 & 0.61 & 0.79 & 0.75 & 0.83 & 0.92 & 1.00 & 1.00 & 1.00 & 1.00 & 1.00 \\
            1400 & 0.42 & 0.84 & 0.71 & 0.45 & 0.26 & 0.09 & 0.40 & 0.62 & 0.68 & 0.76 & 0.92 & 0.99 & 1.00 & 1.00 & 0.99 & 1.00 & 1.00 \\
            1500 & 0.59 & 0.21 & 0.54 & 0.42 & 0.15 & 0.03 & 0.73 & 0.71 & 0.80 & 0.98 & 1.00 & 1.00 & 1.00 & 1.00 & 1.00 & 1.00 & 1.00 \\
            \bottomrule
        \end{tabular}
        
        \bigskip

        Diffusion coefficients:

        \smallskip

        \begin{tabular}{@{}lccccccccccccccccc@{}}
            \toprule
            & \multicolumn{17}{c}{Index of Wigner-Seitz radius rs} \\
            \cmidrule(lr){2-18} 
            T/K & 1 & 2 & 3 & 4 & 5 & 6 & 7 & 8 & 9 & 10 & 11 & 12 & 13 & 14 & 15 & 16 & 17 \\
            \midrule
            1000 & 0.19 & 0.48 & 0.40 & 0.51 & 0.53 & 0.64 & 0.41 & 0.97 & 0.98 & 1.00 & 1.00 & 1.00 & 1.00 & 1.00 & 0.92 & 0.99 & 0.85 \\
            1100 & 0.27 & 0.23 & 0.19 & 0.22 & 0.45 & 0.38 & 0.40 & 0.31 & 0.72 & 0.80 & 0.96 & 1.00 & 1.00 & 1.00 & 0.98 & 0.97 & 0.90 \\
            1200 & 0.04 & 0.01 & 0.04 & 0.20 & 0.08 & 0.06 & 0.18 & 0.42 & 0.39 & 0.67 & 1.00 & 1.00 & 0.86 & 0.98 & 0.99 & 0.93 & 0.79 \\
            1300 & 0.31 & 0.18 & 0.30 & 0.16 & 0.01 & 0.15 & 0.06 & 0.67 & 0.76 & 0.58 & 0.83 & 0.81 & 0.99 & 0.98 & 0.98 & 0.99 & 0.95 \\
            1400 & 0.04 & 0.13 & 0.10 & 0.20 & 0.10 & 0.09 & 0.29 & 0.39 & 0.46 & 0.66 & 0.77 & 0.95 & 0.95 & 0.95 & 0.96 & 0.97 & 0.96 \\
            1500 & 0.20 & 0.23 & 0.19 & 0.01 & 0.06 & 0.24 & 0.70 & 0.94 & 0.59 & 1.00 & 0.96 & 0.85 & 0.95 & 0.88 & 0.85 & 0.75 & 0.98 \\
            \bottomrule
        \end{tabular}
        
        \bigskip

        Radial distribution functions:

        \smallskip

        \begin{tabular}{@{}lccccccccccccccccc@{}}
            \toprule
            & \multicolumn{17}{c}{Index of Wigner-Seitz radius rs} \\
            \cmidrule(lr){2-18} 
            T/K & 1 & 2 & 3 & 4 & 5 & 6 & 7 & 8 & 9 & 10 & 11 & 12 & 13 & 14 & 15 & 16 & 17 \\
            \midrule
            1000 & 0.10 & 0.19 & 0.10 & 0.13 & 0.12 & 0.27 & 0.27 & 0.54 & 0.51 & 0.58 & 0.63 & 0.60 & 0.59 & 0.51 & 0.44 & 0.59 & 0.44 \\
            1100 & 0.10 & 0.06 & 0.07 & 0.08 & 0.07 & 0.12 & 0.14 & 0.12 & 0.39 & 0.45 & 0.55 & 0.61 & 0.56 & 0.51 & 0.53 & 0.46 & 0.43 \\
            1200 & 0.17 & 0.12 & 0.16 & 0.12 & 0.13 & 0.12 & 0.14 & 0.08 & 0.14 & 0.23 & 0.41 & 0.49 & 0.49 & 0.54 & 0.48 & 0.47 & 0.45 \\
            1300 & 0.21 & 0.20 & 0.19 & 0.27 & 0.25 & 0.15 & 0.12 & 0.12 & 0.20 & 0.23 & 0.31 & 0.41 & 0.49 & 0.41 & 0.50 & 0.38 & 0.33 \\
            1400 & 0.27 & 0.32 & 0.29 & 0.26 & 0.23 & 0.17 & 0.11 & 0.14 & 0.19 & 0.19 & 0.29 & 0.38 & 0.39 & 0.37 & 0.38 & 0.39 & 0.31 \\
            1500 & 0.25 & 0.23 & 0.33 & 0.29 & 0.18 & 0.18 & 0.12 & 0.21 & 0.29 & 0.43 & 0.41 & 0.45 & 0.39 & 0.39 & 0.42 & 0.40 & 0.35 \\
            \bottomrule
        \end{tabular}
        
        \endgroup
    \end{subtable}
    
\end{table}

\begin{table}[p]\ContinuedFloat
    \caption{%
        \emph{Hellinger distances} (continued) for all \gls{md}-derived properties and benchmarked \glspl{mlp}.
    }
    
    \centering

    \begin{subtable}{\linewidth}
        \caption{%
            \emph{MACE potential}.
            Shown are \glspl{hd} of pressures, stable molecular fractions, diffusion coefficients, and radial distribution functions (top to bottom) as functions of Wigner-Seitz radius (see \cref{si:tab:rsindices}).
        }

        \bigskip

        \begingroup
        \small\setlength{\tabcolsep}{0.5em}

        Pressures:

        \smallskip

        \begin{tabular}{@{}lccccccccccccccccc@{}}
            \toprule
            & \multicolumn{17}{c}{Index of Wigner-Seitz radius rs} \\
            \cmidrule(lr){2-18} 
            T/K & 1 & 2 & 3 & 4 & 5 & 6 & 7 & 8 & 9 & 10 & 11 & 12 & 13 & 14 & 15 & 16 & 17 \\
            \midrule
            1000 & 0.41 & 0.49 & 0.15 & 0.05 & 0.06 & 0.02 & 0.14 & 0.17 & 0.17 & 0.32 & 0.16 & 0.40 & 0.39 & 0.37 & 0.41 & 0.01 & 0.02 \\
            1100 & 0.03 & 0.11 & 0.13 & 0.01 & 0.02 & 0.01 & 0.00 & 0.12 & 0.23 & 0.08 & 0.09 & 0.30 & 0.22 & 0.31 & 0.33 & 0.17 & 0.06 \\
            1200 & 0.18 & 0.16 & 0.13 & 0.06 & 0.19 & 0.28 & 0.00 & 0.17 & 0.06 & 0.13 & 0.61 & 0.35 & 0.25 & 0.63 & 0.27 & 0.30 & 0.38 \\
            1300 & 0.36 & 0.10 & 0.10 & 0.18 & 0.21 & 0.00 & 0.03 & 0.09 & 0.17 & 0.12 & 0.05 & 0.17 & 0.78 & 0.22 & 0.44 & 0.38 & 0.49 \\
            1400 & 0.52 & 0.59 & 0.25 & 0.28 & 0.24 & 0.17 & 0.13 & 0.13 & 0.19 & 0.12 & 0.34 & 0.21 & 0.40 & 0.44 & 0.27 & 0.61 & 0.50 \\
            1500 & 0.11 & 0.14 & 0.41 & 0.07 & 0.04 & 0.23 & 0.41 & 0.17 & 0.34 & 0.73 & 0.74 & 0.77 & 0.80 & 0.85 & 0.80 & 0.88 & 0.95 \\
            \bottomrule
        \end{tabular}
        
        \bigskip

        Stable molecular fractions:

        \smallskip

        \begin{tabular}{@{}lccccccccccccccccc@{}}
            \toprule
            & \multicolumn{17}{c}{Index of Wigner-Seitz radius rs} \\
            \cmidrule(lr){2-18} 
            T/K & 1 & 2 & 3 & 4 & 5 & 6 & 7 & 8 & 9 & 10 & 11 & 12 & 13 & 14 & 15 & 16 & 17 \\
            \midrule
            1000 & 0.21 & 0.11 & 0.16 & 0.17 & 0.00 & 0.05 & 0.26 & 0.18 & 0.26 & 0.34 & 0.34 & 0.35 & 0.16 & 0.26 & 0.37 & 0.00 & 0.00 \\
            1100 & 0.03 & 0.23 & 0.03 & 0.19 & 0.19 & 0.03 & 0.03 & 0.22 & 0.39 & 0.07 & 0.11 & 0.29 & 0.24 & 0.24 & 0.12 & 0.12 & 0.01 \\
            1200 & 0.16 & 0.12 & 0.06 & 0.13 & 0.26 & 0.11 & 0.11 & 0.34 & 0.22 & 0.22 & 0.67 & 0.41 & 0.32 & 0.64 & 0.23 & 0.20 & 0.33 \\
            1300 & 0.06 & 0.26 & 0.09 & 0.21 & 0.02 & 0.16 & 0.20 & 0.33 & 0.21 & 0.27 & 0.07 & 0.17 & 0.66 & 0.23 & 0.46 & 0.13 & 0.27 \\
            1400 & 0.09 & 0.23 & 0.04 & 0.02 & 0.00 & 0.17 & 0.32 & 0.30 & 0.27 & 0.18 & 0.28 & 0.20 & 0.30 & 0.28 & 0.07 & 0.24 & 0.50 \\
            1500 & 0.01 & 0.16 & 0.01 & 0.03 & 0.01 & 0.07 & 0.41 & 0.17 & 0.31 & 0.63 & 0.62 & 0.49 & 0.39 & 0.27 & 0.30 & 0.42 & 0.42 \\
            \bottomrule
        \end{tabular}
        
        \bigskip

        Diffusion coefficients:

        \smallskip

        \begin{tabular}{@{}lccccccccccccccccc@{}}
            \toprule
            & \multicolumn{17}{c}{Index of Wigner-Seitz radius rs} \\
            \cmidrule(lr){2-18} 
            T/K & 1 & 2 & 3 & 4 & 5 & 6 & 7 & 8 & 9 & 10 & 11 & 12 & 13 & 14 & 15 & 16 & 17 \\
            \midrule
            1000 & 0.20 & 0.21 & 0.33 & 0.07 & 0.13 & 0.10 & 0.05 & 0.25 & 0.24 & 0.38 & 0.04 & 0.24 & 0.16 & 0.20 & 0.13 & 0.03 & 0.10 \\
            1100 & 0.09 & 0.13 & 0.24 & 0.13 & 0.26 & 0.22 & 0.03 & 0.14 & 0.16 & 0.01 & 0.05 & 0.40 & 0.36 & 0.13 & 0.17 & 0.13 & 0.03 \\
            1200 & 0.09 & 0.01 & 0.01 & 0.16 & 0.07 & 0.01 & 0.26 & 0.11 & 0.10 & 0.42 & 0.62 & 0.32 & 0.22 & 0.17 & 0.19 & 0.04 & 0.18 \\
            1300 & 0.38 & 0.22 & 0.16 & 0.28 & 0.05 & 0.23 & 0.14 & 0.57 & 0.47 & 0.20 & 0.02 & 0.15 & 0.25 & 0.32 & 0.11 & 0.38 & 0.09 \\
            1400 & 0.03 & 0.25 & 0.06 & 0.11 & 0.10 & 0.00 & 0.02 & 0.11 & 0.07 & 0.11 & 0.06 & 0.41 & 0.20 & 0.11 & 0.15 & 0.23 & 0.27 \\
            1500 & 0.15 & 0.28 & 0.14 & 0.06 & 0.07 & 0.24 & 0.69 & 0.17 & 0.12 & 0.72 & 0.59 & 0.33 & 0.30 & 0.11 & 0.11 & 0.07 & 0.31 \\
            \bottomrule
        \end{tabular}
        
        \bigskip

        Radial distribution functions:

        \smallskip

        \begin{tabular}{@{}lccccccccccccccccc@{}}
            \toprule
            & \multicolumn{17}{c}{Index of Wigner-Seitz radius rs} \\
            \cmidrule(lr){2-18} 
            T/K & 1 & 2 & 3 & 4 & 5 & 6 & 7 & 8 & 9 & 10 & 11 & 12 & 13 & 14 & 15 & 16 & 17 \\
            \midrule
            1000 & 0.08 & 0.08 & 0.07 & 0.07 & 0.06 & 0.05 & 0.08 & 0.11 & 0.15 & 0.19 & 0.09 & 0.15 & 0.22 & 0.20 & 0.19 & 0.08 & 0.07 \\
            1100 & 0.09 & 0.06 & 0.07 & 0.08 & 0.05 & 0.07 & 0.05 & 0.07 & 0.15 & 0.05 & 0.09 & 0.18 & 0.15 & 0.15 & 0.11 & 0.16 & 0.04 \\
            1200 & 0.06 & 0.06 & 0.06 & 0.05 & 0.07 & 0.07 & 0.10 & 0.06 & 0.11 & 0.16 & 0.31 & 0.17 & 0.13 & 0.17 & 0.11 & 0.08 & 0.11 \\
            1300 & 0.05 & 0.06 & 0.07 & 0.06 & 0.06 & 0.10 & 0.09 & 0.11 & 0.14 & 0.12 & 0.07 & 0.12 & 0.22 & 0.11 & 0.10 & 0.10 & 0.09 \\
            1400 & 0.05 & 0.08 & 0.05 & 0.05 & 0.04 & 0.07 & 0.07 & 0.09 & 0.11 & 0.08 & 0.11 & 0.10 & 0.10 & 0.09 & 0.06 & 0.11 & 0.09 \\
            1500 & 0.07 & 0.12 & 0.07 & 0.05 & 0.07 & 0.09 & 0.12 & 0.11 & 0.13 & 0.27 & 0.16 & 0.18 & 0.16 & 0.09 & 0.12 & 0.14 & 0.13 \\
            \bottomrule
        \end{tabular}
        
        \endgroup
    \end{subtable}
    
\end{table}


\clearpage
\addcontentsline{toc}{section}{References}


\providecommand{\BAN}{\,}\providecommand{\BAP}{.\,}\providecommand{\BANE}{}\providecommand{\BAPE}{.}

\end{document}